\documentclass[aps,prb,twocolumn,showpacs,superscriptaddress,longbibliography]{revtex4-1}
\usepackage{graphicx}  
\usepackage{dcolumn}   
\usepackage{bm}        
\usepackage{amssymb}   
\usepackage{multirow}
\usepackage{amsmath}
\usepackage{natbib}
\usepackage{xcolor}
\usepackage[unicode=true,colorlinks=true]{hyperref}
\newcommand{\affA}{Tyndall National Institute, Lee Maltings, Dyke Parade, Cork, Ireland}
\newcommand{\affB}{Department of Physics, University College Cork, Cork, Ireland}

\newcommand{\etal}{\textit{et al}. }

\begin{document}

\title{Theoretical analysis of influence of random alloy fluctuations on the opto-electronic properties of site-controlled (111)-oriented InGaAs/GaAs quantum dots}

\author{R.~Benchamekh}\affiliation{\affA}
\author{S.~Schulz}
\affiliation{\affA}
\author{E.~P.~O'Reilly}
\affiliation{\affA}\affiliation{\affB}

\date{\today}
\begin{abstract}
We use an $sp^3d^5s^* $ tight-binding model to
investigate the electronic and optical properties of realistic
site-controlled (111)-oriented InGaAs/GaAs quantum dots. Special
attention is paid to the impact of random alloy fluctuations on key
factors that determine the fine-structure splitting in these
systems. Using a pure InAs/GaAs quantum dot as a reference system,
we show that the combination of spin-orbit coupling and biaxial
strain effects can lead to sizeable spin-splitting effects in these
systems. Then, a realistic alloyed InGaAs/GaAs quantum dot with 25\%
InAs content is studied. Our analysis reveals that the impact of
random alloy fluctuations on the electronic and optical properties
of (111)-oriented InGaAs/GaAs quantum dots reduces strongly as the
lateral size of the dot increases and approaches realistic sizes.
For instance the optical matrix element shows an almost vanishing
anisotropy in the (111)-growth plane. Furthermore, conduction and
valence band mixing effects in the system under consideration are
strongly reduced compared to standard (100)-oriented InGaAs/GaAs
systems. All these factors strongly indicate a reduced fine
structure splitting in site-controlled (111)-oriented InGaAs/GaAs
quantum dots. Thus, we conclude that quantum dots with realistic
(50-80~nm) base length represent promising candidates for
polarization entangled photon generation, consistent with recent
experimental data.
\end{abstract}
\maketitle

\section{Introduction}

Much effort from the scientific community is dedicated to the design
of quantum information devices using non-classical light emitters.
One of the main challenges towards achieving quantum information
applications is the realization of on demand entangled photon
sources.~\cite{BoEk2000,MuBo2014} Semiconductor quantum dots (QDs)
offer the possibility to generate polarization entangled photons via
a biexciton-exciton cascade.~\cite{Benson,Akopian} Zinc-blende (ZB)
InAs-based QDs grown along the [001]-direction, due to their well
established growth procedures, have attracted significant interest
for these kind of applications.~\cite{MuBo2014,22726724} However, in
these systems the degenerate bright excitonic ground states in an
ideal QD are split by the so-called fine-structure splitting (FSS)
preventing therefore entangled photon emission. This FSS arises from
the underlying $C_{2v}$ symmetry of the combined system of QD
geometry and ZB crystal structure.~\cite{SeSc2005,BeNa2003} Several
different approaches, such as applying external electric or strain
fields, have been discussed in the literature to reduce or ideally
eliminate the
FSS.~\cite{LuSi2012,RaDi2012,PoBe2014,HuWi2014,22726724} Of
particular interest to this study, a vanishing FSS can be achieved
by growing ZB-based QDs along the [111]-direction, since in this
case the system has ideally $C_{3v}$
symmetry.~\cite{Schliwa,Singh,JuDi2013} It has been shown both in
theory~\cite{Schliwa,Singh,Karlsson} and
experiment~\cite{JuDi2013,JuMu2015,Wang2015} that InGaAs/GaAs QDs
grown along the [111]-direction are promising candidates to achieve
entangled-photon emission. Despite the incontestable importance
of these devices, there is a lack of detailed theoretical
investigation on realistically sized and shaped (111)-oriented
InGaAs QD systems. Indeed, previous theoretical studies are mainly
related to model systems with dot geometries and dimensions carried
over from the (001)-oriented systems. However, the geometric
dimensions and shape of these site-controlled (111)-oriented
InGaAs-based QDs are very different from the structures assumed in
previous theoretical studies.~\cite{Schliwa,Singh} Site-controlled
InGaAs/GaAs QDs exhibit typically a base length of 50-80 nm and a
height of approximately 2~nm.~\cite{HeYo2010} This raises the
question how these very different geometrical features affect the
electronic and optical properties of 
(111)-oriented site-controlled InGaAs/GaAs QDs.

Experimental studies on site-controlled (111)-oriented InGaAs/GaAs
QDs have demonstrated the generation of polarization entangled
photons.~\cite{JuDi2013,JuMu2015} Due to their high structural
uniformity and spectral purity,~\cite{MeDi2009} these systems are of
particular interest for device design. Moreover, experimental
studies reveal that these dots seem to be very robust against
composition fluctuations, since a very high percentage of the dots
in the samples investigated emit entangled photons.\cite{JuDi2013}
This asks for a detailed atomistic analysis of the experimentally
relevant site-controlled InGaAs/GaAs QDs to shed more light on the
underlying physics which enables robust entangled photon generation.
However, so far no such theoretical study on realistic
(111)-oriented InGaAs/GaAs QDs has been presented.

Theoretical studies addressing the electronic and optical properties
of realistic site-controlled structures use mainly continuum-based
$\mathbf{k}\cdot\mathbf{p}$ models.~\cite{MaORe2014} These
approaches allow for insights into the general electronic and
optical properties of these structures but neglect atomistic effects
arising for example from random alloy fluctuations. It has been
shown that in (001)-oriented InGaAs/GaAs QDs, alloy fluctuations can
have a significant effect on the electronic structure, especially
when studying the FSS.~\cite{BeNa2003} However, since the
site-controlled QDs considered here are grown along a different
crystallographic direction plus they have much larger base length
than standard (001)-oriented Stranski-Krastanov InGaAs/GaAs QDs, it
is not immediately clear how alloy fluctuations will affect the
electronic structure of realistic site-controlled InGaAs/GaAs
(111)-oriented QDs. We show here that the geometric features,
such as dot size and aspect ratio, significantly impact electronic
and optical properties of site-controlled (111)-oriented
InGaAs/GaAs QDs. This is especially true for key factors that affect
the magnitude of the FSS. In fact our analysis reveals that the dot
dimensions are critical for achieving entangled photon emission
from these structures. Our analysis also highlights that the
properties of (111)-oriented site-controlled dots are very different
from standard (001)-oriented systems or (111)-oriented structures
using geometries and dot sizes carried over from previous (001)-plane analysis.

We present here a detailed atomistic $sp^3d^5s^*$ tight-binding (TB)
analysis to investigate the key factors that enable the generation
of entangled photons from a high fraction of the site-controlled QDs
investigated in Ref.~\citenum{JuDi2013}. We calculate that the
impact of alloy fluctuations decreases with increasing dot base size
both for electrons and for holes, while the low dot height leads to
the electron being only weakly confined in the dot, with a wave
function which then has predominantly GaAs character. In fact
we find a strong asymmetry in the number of bound electron and hole
states. A detailed investigation of how different contributions,
namely strain field, first- and second-order piezoelectricity and
spin-orbit coupling effects, impact the properties of the QD system
under consideration is presented. We find that the mixing
between valence and conduction states also decreases significantly
with increasing dot base length, thereby significantly reducing the
size of several terms that can contribute to a finite
FSS.~\cite{krapek} The electron $p$-state splitting (\textit{i.e.}
the splitting between the first and second excited electron state)
has previously been used as an indication of the ability of a given
system to generate polarization entangled photons. In (001)-oriented
InAs/GaAs QDs, even in the absence of alloy fluctuations, the
underlying atomistic symmetry of the system leads to the situation
that the electron $p$-states are no longer degenerate. Strain and
piezoelectric effects usually increase the splitting of the
$p$-states.~\cite{Bester2005} Thus, for (001)-oriented
InGaAs/GaAs QD systems, the $p$-state splitting arises mainly due to
geometrical effects (lack of inversion symmetry due to dot
geometries). In this work we consider both $p$-state splitting and
the anisotropy of the in-plane optical momentum matrix element $E_p$
between the electron and hole ground states as a measure of the
symmetry reduction of the QDs due to random alloy fluctuations.

Even though the FSS provide a direct measure of the ability to
generate polarization entangled photons, we highlight here the
challenges and difficulties of carrying out an FSS calculation using the TB
method for realistic site-controlled (111)-oriented InGaAs/GaAs QDs.
This originates in particular from the small number of bound
electron and hole states arising from the geometrical features of
(111)-oriented site-controlled InGaAs/GaAs QDs.

Our results show that while the in-plane optical momentum matrix
element is perfectly isotropic for a pure InAs/GaAs QD, the
$p$-state splitting does not vanish even when neglecting strain and
built-in potential effects. We show here that this feature
originates from spin-orbit coupling effects. This is a very
different effect when compared to (001)-oriented InGaAs/GaAs QDs
where geometrical aspects are important, as discussed above. In the
case of an alloyed InGaAs/GaAs QD, the random alloy fluctuations
increase the $p$-state splitting and cause an anisotropy of the
optical momentum matrix element. Our results reveal that the impact
of random alloy fluctuations on the electronic structure decreases
as the lateral size of the QDs increases and that for realistic
site-controlled (111)-oriented InGaAs/GaAs QDs the anisotropy of the
optical momentum matrix element is negligible and the $p$-state
splitting is mainly determined by spin-splitting effects.   We
conclude that the combination of all of these factors then leads to
site-controlled (111)-oriented InGaAs/GaAs QDs with large base
length and low dot height being particularly promising candidates
for devices delivering entangled-photon emission on demand.

This paper is organized as follows.  The details of the supporting theoretical
framework are discussed in Sec.~\ref{sec:Theory}. We review in
Sec.~\ref{sec:Exp_SC} available experimental literature data on the
structural properties of site-controlled (111)-oriented InGaAs/GaAs QDs and discuss how
this data is incorporated in our theoretical framework. Our results
are presented in Sec.~\ref{sec:Results}, where we undertake a complete analysis
of a (111)-oriented InGaAs/GaAs model system to study the impact of
different contributions, namely the spin-orbit coupling, strain
effects, built-in fields and random alloy fluctuations on the
electronic structure. Finally, we summarize our results in
Sec.~\ref{sec:Conclusion}.

\section{Theory}
\label{sec:Theory}

In this section, we introduce the theoretical framework applied here to
describe the electronic structure of site-controlled (111)-oriented
InGaAs/GaAs QDs. The generation of the underlying atomistic grid for
a (111)-oriented ZB system is presented in Sec.~\ref{sec:Theory_SC}.
The applied valence force field (VFF) model, to
obtain the relaxed atomic position in a mixed InAs and GaAs system,
is discussed in Sec.~\ref{sec:Theory_Strain}, followed by the
calculation of the arising first- and second-order piezoelectric
built-in potentials (Sec.~\ref{sec:Theory_Piezo}). Subsequently, in
Sec.~\ref{sec:Theory_TB}, we focus on the $sp^{3}d^{5}s^{*}$ TB
model that forms the basis of our electronic structure calculations.

\subsection{Supercell}
\label{sec:Theory_SC}
\begin{figure}
\includegraphics[width=0.49\columnwidth]{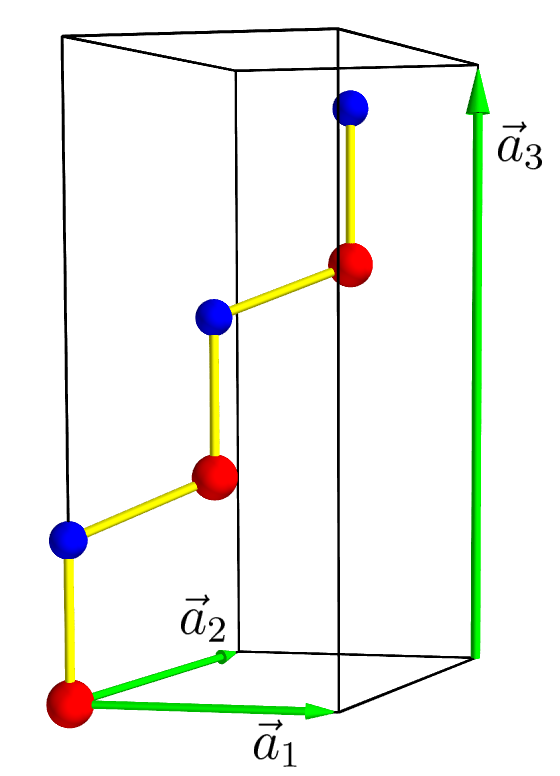}
\includegraphics[width=0.49\columnwidth]{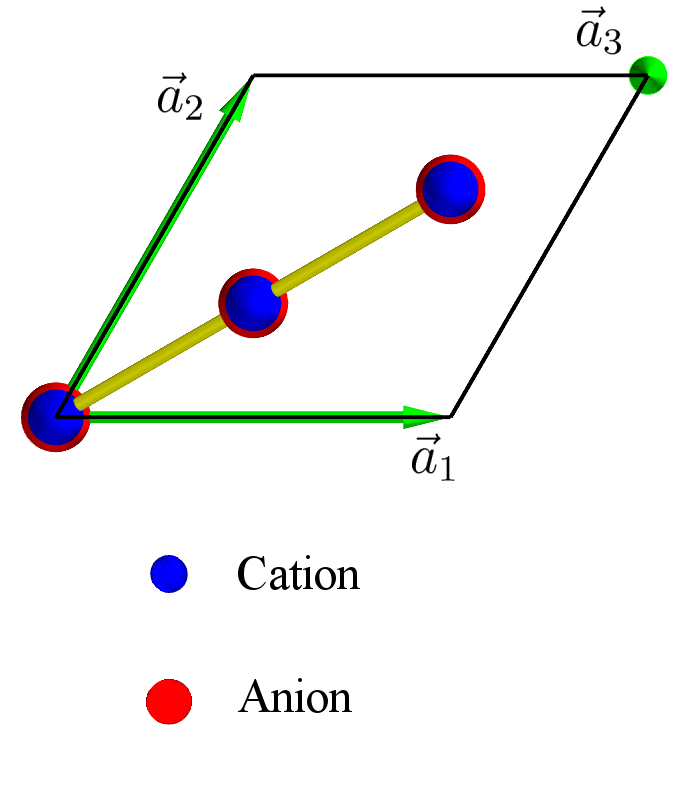}
\caption{Perspective view (left) and top view (right) of the six atom unit cell.}\label{fig:unit_cell}
\end{figure}

Since we are using an atomistic approach to describe the electronic
properties of site-controlled (111)-oriented InGaAs/GaAs QDs, the grid
underlying our calculations must reflect the atomic arrangement
along this direction. This can be achieved by considering a unit
cell of a ZB material defined by the lattice vectors:
\begin{equation}
\begin{array}{l}
\mathbf{a}_1=(-\frac{1}{2},\frac{1}{2},0) a_0  \\
\mathbf{a}_2=(-\frac{1}{2},0,\frac{1}{2}) a_0 \\
\mathbf{a}_3=(1,1,1) a_0\,\, .
\end{array} \nonumber
\end{equation}
Here, $a_0$ is the lattice constant of the barrier material, in our
case GaAs. In doing so we  account for the $C_{3v}$ symmetry of the
ZB lattice when oriented along the [111]-direction. The unit cell
contains six atoms (three anions $\{A\}_{i=1,..,3}$ and three
cations $\{C\}_{i=1,..,3}$) placed at:
\begin{equation}
\begin{array}{cc}
A_1=(0,0,0) ;   & C_1=(\frac{1}{4},\frac{1}{4},\frac{1}{4})\\
A_2=(0,\frac{1}{2},\frac{1}{2}) ; & C_2=(\frac{1}{4},\frac{3}{4},\frac{3}{4})\\
A_3=(0,1,1) ;& C_3=(\frac{1}{4},\frac{5}{4},\frac{5}{4}).
\end{array} \nonumber
\end{equation}
Translating the unit cell by the vectors:
\begin{equation}
\begin{array}{l}
\mathbf{T}_{ijk}=i\; \mathbf{a}_1+ j\; \mathbf{a}_2+ k\;\mathbf{a}_3 ; \\
0 \leq i \leq N_1 ,  0 \leq j \leq N_2,  0 \leq k \leq N_3\, ,
\end{array} \nonumber
\end{equation}
allows us to construct a supercell (SC) of a (111)-oriented ZB lattice. A
schematic illustration of the SC is displayed in
Fig.~\ref{fig:unit_cell}. Thus we obtain a matrix of GaAs defined by
the lattice vectors $\mathbf{b}_i= N_i \mathbf{a}_i, i=1,2,3 $. Note that
the vectors and atomic positions are expressed in the conventional
coordinate system. All following expression are therefore also given
in the standard (001)-oriented frame. Choosing $N_2=N_1 $, the
lateral dimension of the SC is $l=N_1 a_0/\sqrt{2} $ while its
height is given by $h= N_3 a_0\sqrt{3} $.

\subsection{Valence force field model}
\label{sec:Theory_Strain}

The difference in the lattice constants between InAs and GaAs gives
rise to a strain field in the nanostructure. This strain field
corresponds to a displacement of the atoms from their equilibrium
positions. To find the relaxed atomic positions in the SC, we use
the Keating VFF model.~\cite{Keating} In this model the total energy of an atom $i$ is given
by:
\begin{equation}
\begin{array}{rl}
U_i= & \sum_{j}^N \frac{3 \alpha_{ij}}{16 d_{ij}^2}\left[ \mathbf{r}_{ij}^2- d_{ij}^2 \right]^2 \\
  + & \sum_{j,k>j}^N \frac{3\beta_{ijk}}{16 d_{ij}d_{ik}} \left[
\mathbf{r}_{ij}\cdot \mathbf{r}_{ik} +\frac{d_{ij}d_{ik}}{3} \right]^2\, ,
\end{array}
\end{equation}
where $\mathbf{r}_{ij}$ is the vector between the atoms $i$ and $j$,
and $d_{ij}$ is the bond length between atoms $i$ and $j$. Bond
stretching and bending constants are denoted by $\alpha_{ij}$ and
$\beta_{ijk}$, respectively. The parameters for $\alpha_{ij}$ and
$\beta_{ijk}$ are chosen to fit the macroscopic elastic constants
$C_{11}$ and $C_{12}$ of GaAs and InAs.\cite{Ram-mohan} The total
energy of the system is minimized with respect to the atomic
coordinates, yielding the new and relaxed atomic positions of the
atoms in the SC. In and around the nanostructure, the relaxed atomic
positions deviate slightly from their equilibrium positions,
inducing therefore local deformations. These local strain effects
lead also to the appearance of piezoelectric built-in fields in
InGaAs heterostructures. We discuss the calculation of the built-in
potential on our irregular atomistic grid in the next section.

\subsection{Piezoelectric potential}
\label{sec:Theory_Piezo}

Depending on the crystal structure, semiconductor materials can
exhibit an electric polarization.~\cite{Nye85,Grim2007} This
electric polarization can be divided into strain independent
(spontaneous) and strain dependent (piezoelectric)
contributions.~\cite{Nye85} ZB semiconductors exhibit only strain
dependent piezoelectric polarization fields.~\cite{Nye85} In an
(001)-oriented ZB case, the piezoelectric polarization vector field
is connected to shear strain contributions.~\cite{Smith90,Niquet06} It has been shown by
several authors that these piezoelectric fields are important for a
realistic modeling of the electronic and optical properties of a
[111]-grown ZB
nanostructure.~\cite{Bester2011,Smith86,SmithJVST87,SmithPRB87,SmithPRL87,Smith89,Smith90,Smith88}
In (001)-oriented ZB materials, first- and second-order
piezoelectric polarization vector fields
$\mathbf{P}_\text{pz}^\text{FO}$ and
$\mathbf{P}_\text{pz}^\text{SO}$, respectively, are given
by:~\cite{Bester06,Niquet06,ScWi2007,Schulz11}
\begin{equation}
\begin{array}{lll}

\mathbf{P}_\text{pz}^\text{FO} &= & 2 e_{14}\left(
\begin{array}{c}
\epsilon_{yz}\\
\epsilon_{xz}\\
\epsilon_{xy} \end{array}\right), \\
 & & \\

\mathbf{P}_\text{pz}^\text{SO}& = &2 A_1\left(
\begin{array}{c}
Tr(\epsilon)\epsilon_{yz}\\
Tr(\epsilon)\epsilon_{xz}\\
Tr(\epsilon)\epsilon_{xy} \end{array}\right) +2 A_2 \left( \begin{array}{c}
\epsilon_{B,x}\epsilon_{yz}\\
\epsilon_{B,y}\epsilon_{xz}\\
\epsilon_{B,z}\epsilon_{xy} \end{array}\right) \\
&  +& 4 B_{156}\left( \begin{array}{c}
\epsilon_{xy}\epsilon_{xz}\\
\epsilon_{yz}\epsilon_{xy}\\
\epsilon_{yz}\epsilon_{xz} \end{array}\right).

\end{array}
\end{equation}
The first-order piezoelectric coefficient is denoted by $e_{14}$,
while $A_1=\frac{1}{3}B_{114}+\frac{2}{3}B_{124}$,
$A_2=\frac{2}{3}B_{114}-\frac{2}{3}B_{124}$ and $B_{156}$ are the
second-order piezoelectric coefficient. $B_{114} $, $B_{124} $ and
$B_{156} $ are the three independent elements of the second-order
piezoelectric tensor $B_{i\alpha\beta}$.~\cite{Grim2007}
The trace of the strain tensor $\epsilon$ is denoted by $Tr(\epsilon)$ while
$\epsilon_{B,i}=\frac{3}{2}\epsilon_{ii}- \frac{1}{2}Tr(\epsilon)$
are the biaxial strain components ($i = x,y,z$). First- and second-order
piezoelectric coefficients have been taken from
Ref.~\citenum{Bester2011}, which are similar to the recent hybrid functional density functional theory (DFT)
results of Caro~\etal\cite{CaSc2015} The main
difference between the two parameter sets is the value of $B_{156}$,
which has been shown to be of secondary importance for the QD system
considered here.~\cite{MaORe2014}

The charge density $\rho_{pz}$ arising from discontinuities in the polarization
$\mathbf{P}_\text{pz}=\mathbf{P}^\text{FO}_\text{pz}+\mathbf{P}^\text{SO}_\text{pz}$
is given by:
\begin{equation}
\rho_\text{pz}(\mathbf{r})= -\mathbf{\nabla} \cdot
\mathbf{P}_{pz}(\mathbf{r})\, .
\end{equation}
The corresponding electrostatic built-in potential $V_p$ is obtained
from solving Poisson's equation:
\begin{equation}
\nabla \cdot \left(\epsilon_0 \kappa_0(\mathbf{r}) \nabla \cdot
V_p(\mathbf{r})\right) = -\rho_\text{pz}(\mathbf{r})\,\, ,
\label{eq:Poisson}
\end{equation}
where $\epsilon_0$ is the vacuum permittivity and
$\kappa_0(\mathbf{r}) $ is the position dependent static dielectric constant.

To solve Eq.~(\ref{eq:Poisson}) we use a modified finite-difference method.
We cannot apply the standard finite-difference method since we are
dealing here with a ZB grid plus the fact that the atoms are
displaced from their equilibrium positions. To calculate derivatives
on this non-uniform grid we proceed in the following way. As an
example we take the differential $\frac{\partial \phi}{\partial x}$,
where we calculate the derivative of a quantity $\phi $ with respect
to the direction $x$. Similar considerations can be made for
$\frac{\partial \phi}{\partial y}$ and $\frac{\partial
\phi}{\partial z}$. In an (001)-oriented ZB system, the four vectors
linking a cation to its four nearest neighbor anions read:
\begin{eqnarray*}
\mathbf{r}_1 &=& a (1/4,1,4,1/4)\, , \\
\mathbf{r}_2 &=& a (-1/4,-1,4,1/4)\, , \\
\mathbf{r}_3 &=& a (1/4,-1,4,-1/4)\, , \\
\mathbf{r}_4 &=& a (-1/4,1,4,-1/4)\, ,
\end{eqnarray*}
where $a$ denotes the lattice constant. The atoms $\mathbf{r}_1$ and
$\mathbf{r}_3$ are in the half space $x>0$ while the atoms
$\mathbf{r}_2$ and $\mathbf{r}_4$ are in the half space $x<0$. Then
$\frac{\partial \phi}{\partial x} $ can be expressed as:
\begin{equation}\label{eq:partial}
\frac{\partial \phi}{\partial x}=\frac{ \left[ \phi(\mathbf{r}_1)+\phi(\mathbf{r}_3) \right] /2 - \left[ \phi(\mathbf{r}_2)+\phi(\mathbf{r}_4) \right] /2 }{\left| \left[\mathbf{x}_1+\mathbf{x}_3 \right]/2-\left[ \mathbf{x}_2+\mathbf{x}_4 \right] /2  \right|}.
\end{equation}
Using Eq.~(\ref{eq:partial}) allows us to solve Eq.~(\ref{eq:Poisson}) and thus to obtain the
piezoelectric potential $V_p$ at each atomic site. The built-in potential $V_p$ can then be included
in the TB Hamiltonian as we describe in the following section.

\subsection{Tight-binding model}
\label{sec:Theory_TB}

To achieve an atomistic description of the electronic structure of
site-controlled (111)-oriented InGaAs/GaAs QDs we apply an
$sp^3d^5s^* $ TB model. Our calculations are based on the two centre
parameters of Jancu \textit{et al.},~\cite{Jancu98} which give
excellent agreement with experimental and DFT data of the unstrained
binary materials. However, we reworked the strain parameters, to
reproduce the bulk deformation potentials of GaAs and InAs
recommended in Ref.~\citenum{Ram-mohan}. Our parameters are
summarized in Table.~\ref{tab:strain}. The
valence band offset of 0.23 eV between unstrained InAs and GaAs is used in our
calculations.~\cite{Raouafi,Soucail} Strain effects are included
into the TB Hamiltonian following the approach given in
Ref.~\citenum{Raouafi}, which is a generalization of earlier
strain models to an arbitrary local strain
tensor.~\cite{Jancu07,Niquet,Boykin,Zielinski} The strain dependence
of the interatomic matrix elements is included via a generalized
Harrison bond length scaling law.~\cite{Ren} On-site strain
corrections for the $p$ and $d$ orbitals with $\Gamma_{15}$ symmetry
are introduced by adding the following strain Hamiltonian:~\cite{Raouafi}
\begin{equation}\label{ham_str_n_int}
\delta\hat{H}=
\left(\begin{array}{ccc}
4\lambda_1 \varepsilon_{B,x}
   &\lambda_2\varepsilon_{xy} + \xi u_z
       &\lambda_2\varepsilon_{zx} + \xi u_y\\
\lambda_2\varepsilon_{xy}+ \xi u_z
   & 4\lambda_1 \varepsilon_{B,y}
       &\lambda_2\varepsilon_{yz} + \xi u_x\\
\lambda_2 \varepsilon_{zx}+ \xi u_y
   &\lambda_2\varepsilon_{yz}+ \xi u_x
       &4\lambda_1\varepsilon_{B,z}
\end{array}\right)\,\, . \nonumber
\end{equation}
Here, $\epsilon$ denotes the local strain tensor and $\mathbf{u}$ the
internal strain vector.
For $p$-orbitals we use $\lambda_1=\frac12E_p\pi_{001}$ and
$\lambda_2=\frac83E_p\pi_{111}$~\cite{Jancu07} while for $d$-orbitals
$\lambda_1=\frac12E_d\delta_{001}$ and
$\lambda_2=\frac83E_d\delta_{111}$;~\cite{Jancu07} we also use $\xi=\pm\lambda_2$, with ``$+$'' for anion and ``$-$'' for cation. The values for $\pi_{001}$, $\pi_{111}$, $\delta_{001}$ and $\delta_{111}$ are given it Table~\ref{tab:strain}.

\begin{table}
\caption{\label{tab:strain}Strain parameters used in the calculations. Notations from Ref.~\citenum{Jancu07} are applied.}
\begin{tabular*}{\columnwidth}{@{\extracolsep{\fill}}lll}

\hline\hline
                    & GaAs         & InAs     \\
\hline
$     n_s         $& $  0.8490$& $  1.0500$ \\
$     n_p         $& $  1.1914$& $  0.9275$ \\
$     n_d         $& $  1.9580$& $  1.6620$ \\
$     n_{s^*}     $& $  2.0000$& $  2.0000$ \\
$n_{ss\sigma}     $& $  3.3000$& $  2.2480$ \\
$n_{sp\sigma}     $& $  3.6100$& $  3.7700$ \\
$n_{sd\sigma}     $& $  2.7380$& $  1.4080$ \\
$n_{ss^*\sigma}   $& $  0.0000$& $  0.0000$ \\
$n_{s^*s^*\sigma} $& $  0.0000$& $  0.0000$ \\
$n_{s^*p\sigma}   $& $  2.5640$& $  2.1240$ \\
$n_{s^*d\sigma}   $& $  2.0000$& $  2.0000$ \\
$n_{pp\sigma}     $& $  3.0070$& $  2.6980$ \\
$n_{pp\pi}        $& $  3.1880$& $  3.7580$ \\
$n_{pd\sigma}     $& $  2.2470$& $  2.1890$ \\
$n_{pd\pi}        $& $  1.4289$& $  1.6069$ \\
$n_{dd\sigma}     $& $  1.0000$& $  2.4020$ \\
$n_{dd\pi}        $& $  2.2500$& $  1.2140$ \\
$n_{dd\delta}     $& $  2.9480$& $  2.5360$ \\
$\pi_{001}        $& $  0.2990$& $  0.0820$ \\
$\pi_{111}        $& $  0.4740$& $  0.4790$ \\
$\delta_{001}     $& $  0.0000$& $  0.0000$ \\
$\delta_{111}     $& $  0.0000$& $  0.0000$ \\

\hline\hline

\end{tabular*}
\end{table}
\begin{figure}
\centering
\includegraphics[width=0.6\columnwidth]{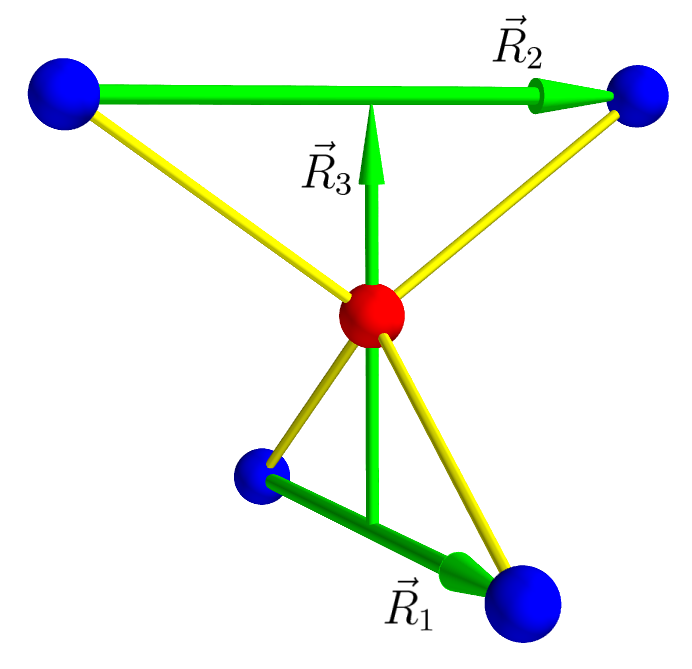}
\caption{Definition of the basis used in local strain field
calculation.}\label{fig:neighbors}
\end{figure}

We calculate the local strain tensor components $\epsilon_{ij}$ and
the internal strain vector $\mathbf{u}=(u_x,u_y,u_z)$ at each lattice site
following Ref.~\citenum{Raouafi}. As an example we consider for
instance an atom $C$ in the SC. The four nearest neighbor atoms are
labeled $\{A_i\}_{i=1..4}$. We introduce two sets of vectors. The
first set is denoted by $\{\mathbf{r}_i^0\}_{i=1..4}$ that
describes the unstrained bond lengths from atom $C$ to the four
nearest neighbor atoms $\{A_i\}_{i=1..4}$. The second set of vectors
$\{\mathbf{r}_i\}_{i=1..4} $ describes the same situation but
this time after the VFF minimization/relaxation procedure. Thus
$\{\mathbf{r}_i\}_{i=1..4} $ reflect the strained bond length.
The centre of the tetrahedron formed by atoms $A_i$ corresponds to
the centre of a sphere that touches all vertices of the tetrahedron
and its shape can be defined by three vectors
$\{\mathbf{R}_i\}_{i=1..3}$ chosen as:
\begin{eqnarray*}
\mathbf{R}_1 &=& \mathbf{r}_2 - \mathbf{r}_1,\\
\mathbf{R}_2 &=& \mathbf{r}_4 - \mathbf{r}_3,\\
\mathbf{R}_3 &=& 1/2(\mathbf{r}_4 + \mathbf{r}_3 -
\mathbf{r}_2 -\mathbf{r}_1).
\end{eqnarray*}
A schematic illustration of these vectors is given in Fig.~\ref{fig:neighbors}.

The matrix $T$ that connects the equilibrium tetrahedron vectors
$\{\mathbf{R}_i^0\}_{i=1..3}$ and the strained vectors
$\{\mathbf{R}_i\}_{i=1..3}$ is calculated from:
\begin{equation}
T\mathbf{R}_i^0=\mathbf{R}_i\, , \quad i=1..3 \quad .
\end{equation}
The local strain tensor $\epsilon$ for the atom $C$ is defined by the polar decomposition:
\begin{equation}
T=(1+\epsilon)P\, ,
\end{equation}
where $P$ is an orthogonal matrix that rotates the unstrained
tetrahedron vectors to the strained ones. The internal strain vector
$\mathbf{u}$ corresponds to the difference between the
tetrahedron centre and the atom $C$ position given after the VFF
minimization scaled by the equilibrium bond length.

In addition to including strain effects into our TB model, we need
also to take the piezoelectric contributions into account. We
incorporate the piezoelectric potential $V_p$, calculated from
Eq.~(\ref{eq:Poisson}), as a site-diagonal correction in the TB
model. This is a widely used
approach.~\cite{RaAl2003,ZiJa2005,ScSc2006}

\section{Site-controlled (111)-{InGaAs/GaAs} QDs: Experimental data, QD geometry and supercell}
\label{sec:Exp_SC}
Having presented the theoretical framework we describe here
the available experimental data on the structural properties of
site-controlled (111)-oriented InGaAs/GaAs QDs. We discuss also how their
characteristic features are included in our theoretical framework.

Site-controlled (111)-oriented InGaAs/GaAs
QDs, grown by metal organic vapor phase epitaxy (MOVPE) in inverted
pyramidal recesses etched in a (111)-B-oriented GaAs
substrate,~\cite{MeDi2009} exhibit a triangular
shape.~\cite{PeWa2004,DiPe2012,JuDi2014} The base length of the
triangle is of order 50-80 nm, while the height of the nanostructure
is only 1-2 nm.~\cite{HeYo2010} The experimentally reported InAs
contents in the QDs range from 15\% to 65\%.~\cite{JuDi2014}

\begin{figure}
\includegraphics[width=0.49\columnwidth]{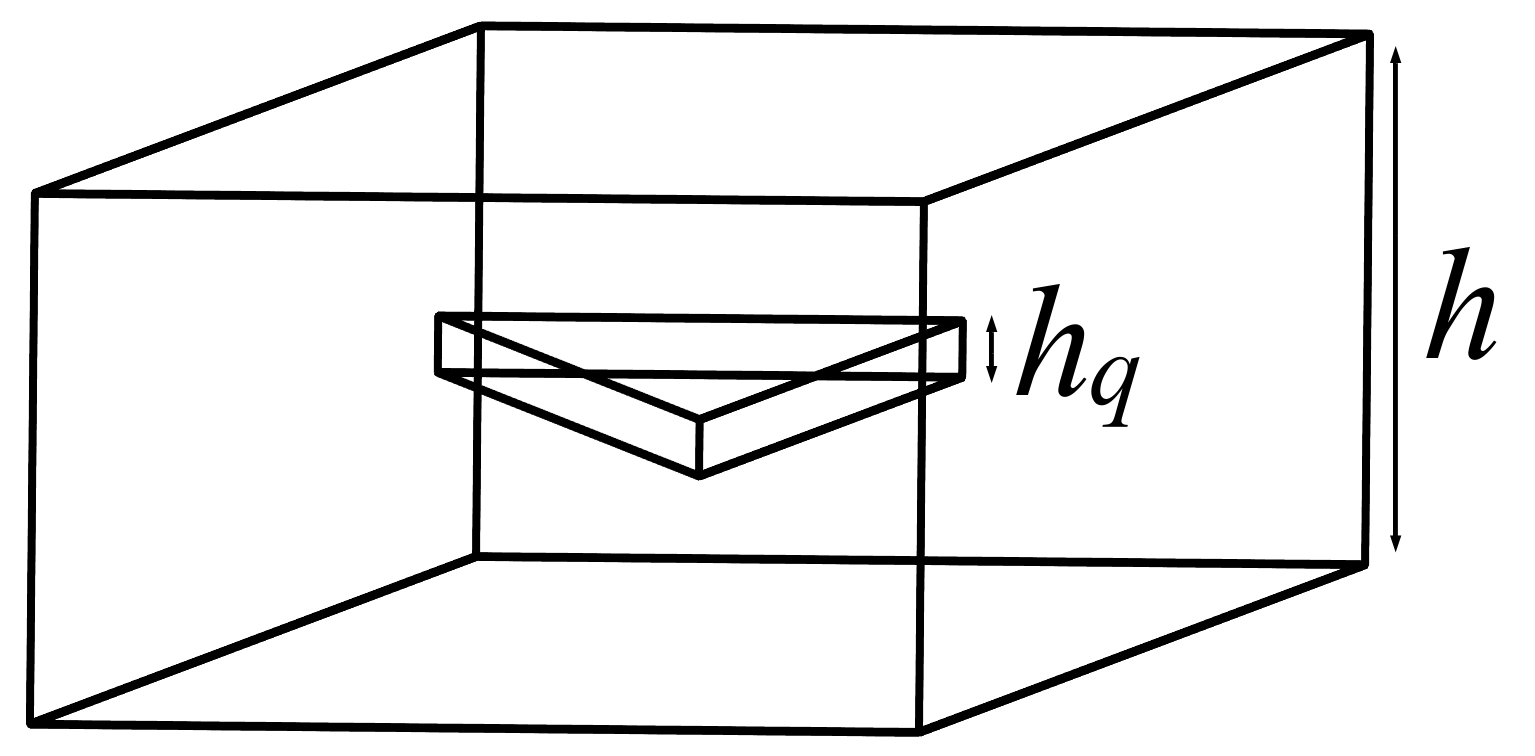}
\includegraphics[width=0.49\columnwidth]{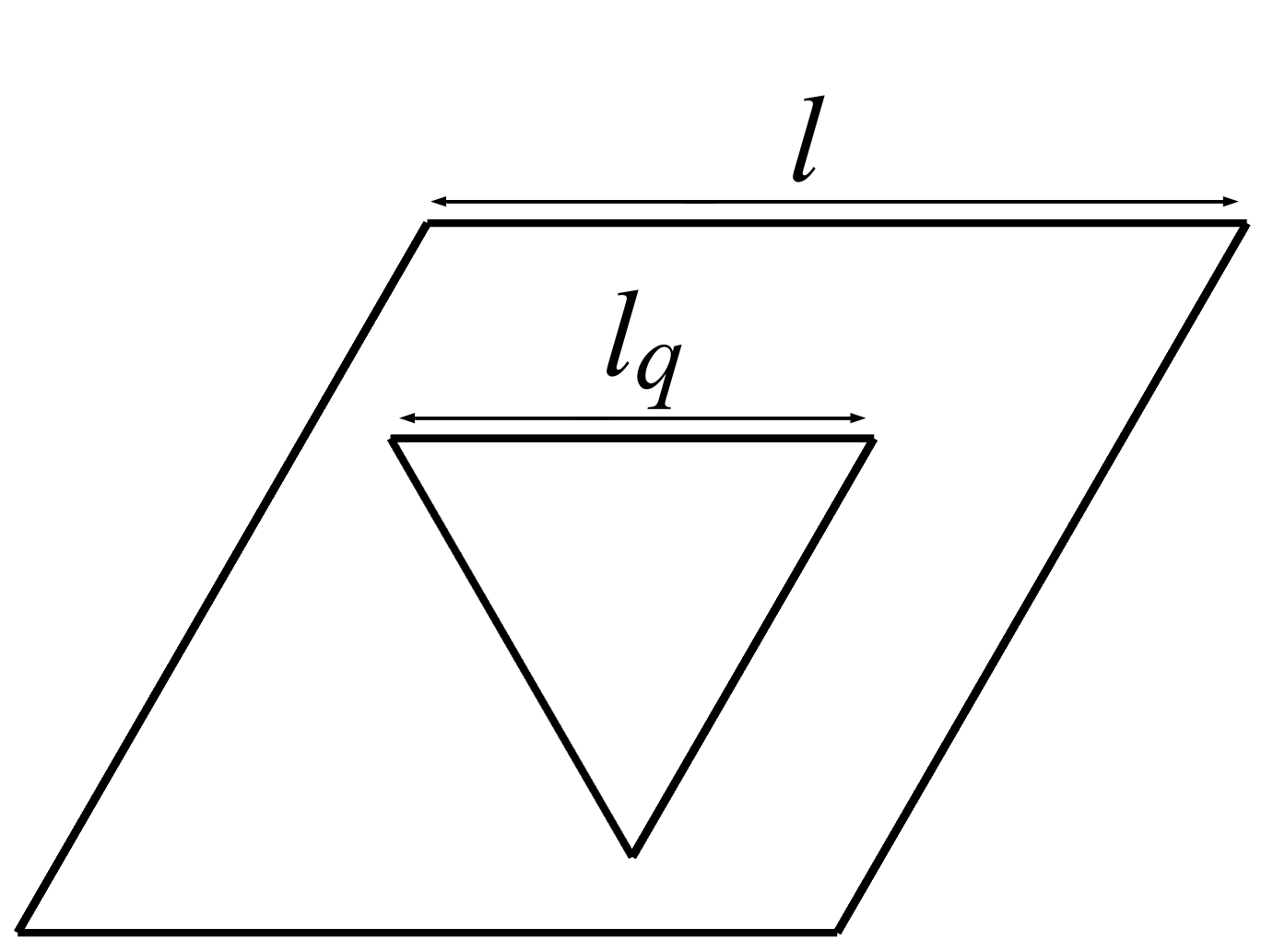}
\caption{Schematic perspective view (left) and top view (right) of
the supercell. The base length and the height of the supercell are denoted by $l$ and $h$, respectively. The QD base length is denoted by $l_q$ while its height is  given by $h_q$.}\label{fig:supercell}
\end{figure}

Equipped with this knowledge about the experimental data we model
the QD on our atomistic grid, introduced in
Sec.~\ref{sec:Theory_SC}, in the following way. The QD is placed
inside the GaAs matrix by substituting Ga atoms located in the QD
volume by In atoms. Following the experimental findings, we assume a
triangular-shaped QD. The edges of the triangular basis are along
the $[\bar{1},1,0]$, $[\bar{1},0,1]$ and $[0,\bar{1},1]$ directions.
A schematic illustration of the QD geometry inside the SC is shown
in Fig.~\ref{fig:supercell}. The base length of the triangle is
denoted by $l_q $ and the height by $h_q $. Taking into account
periodic boundary conditions, the generated SC preserves the
$C_{3v}$ symmetry of the underlying atomic grid.

For our investigation we have constructed a SC characterized by
$l=86$~nm and $h=18$~nm (see Fig.~\ref{fig:supercell}). This SC
contains more than 5 million atoms. We assume in the following a QD
with the dimensions $h_q=2$~nm (height) and $l_q=55$~nm (base
length), with some results also presented for QDs with a smaller
base length, $l_q=15$~nm.

\section{Results}
\label{sec:Results}

In this section we present the results of our calculations. To
analyze the impact of spin-orbit coupling, strain and built-in
fields on the electronic structure of the QD system in question we
study in a first step a pure InAs/GaAs QD. Then we introduce random
alloy fluctuations and compare the electronic structures of three
different random microscopic configurations of an alloyed
In$_{0.25}$Ga$_{0.75}$As/GaAs QD with the results from a
corresponding dot described in the virtual crystal approximation
(VCA). We also investigate the impact of dot base length by
comparing the optical properties of a 25\% InAs QDs with $l_q=15$
and 55~nm.

\subsection{Pure InAs/GaAs QD}
\label{sec:Results_small_QD}
\begin{figure}
\includegraphics[width=0.95\columnwidth]{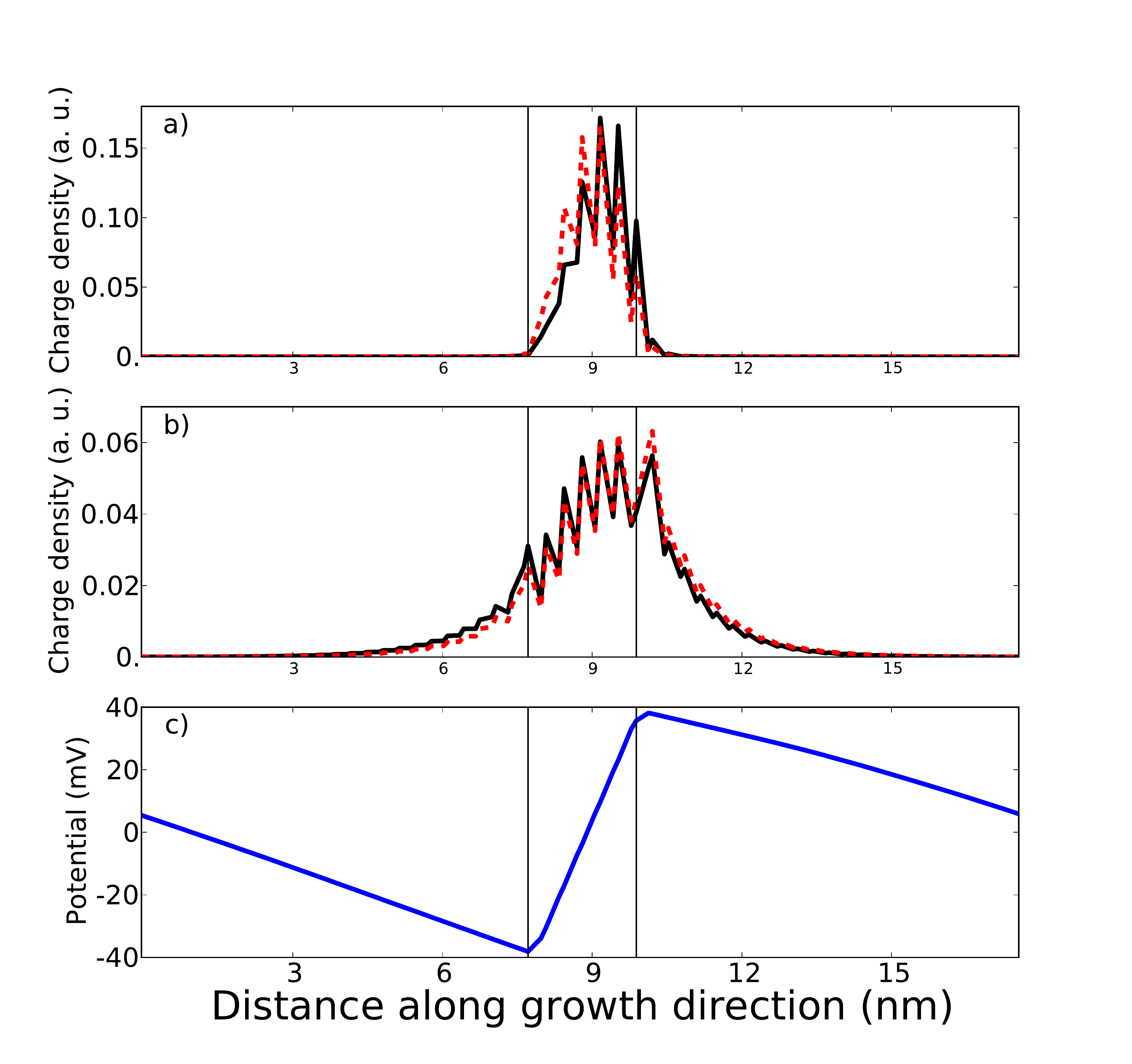}
\caption{a) Hole and b) electron charge densities projected on the growth direction, in the presence (solid line) and in the absence (dotted line) of the built-in potential. The profile of the piezoelectric potential along the growth direction and across the centre of a pure InAs/GaAs QD is shown in c).}\label{fig:pot}
\end{figure}

To gain detailed understanding of the electronic structure of
site-controlled InGaAs/GaAs QDs, we start with a pure system and
study spin-orbit coupling, strain and built-in fields separately.
The QD dimensions are $l_q=55$ nm and $h_q=2$ nm. Since the
constructed SC should preserve the $C_{3v}$ symmetry of the system
(QD geometry plus underlying ZB lattice), we expect that for example
the electron $p$-states should remain degenerate when including all
effects other than spin-orbit coupling. The results of our analysis
are summarized in Table~\ref{tab:no_so}.

In a first step we neglect strain, built-in field and spin-orbit
coupling and focus purely on quantum confinement (QC) effects. Our
results confirm the expected degeneracy of the electron and hole
$p$-states. In a next step, we introduce strain effects by relaxing
the atomic positions using the VFF model described in
Sec.~\ref{sec:Theory_Strain}. We still neglect spin-orbit coupling
and built-in field effects. Our calculations show that when
including strain effects the hole and electron ground states are
shifted to higher energies due to the presence of hydrostatic and
biaxial compressive strains. We note also that in this particular
case where spin-orbit coupling is neglected, including strain
switches the order of the $s$- and $p$-shell in the valence band.
The electron and hole $p$-states are still degenerate. Next, we
include the total (first- plus second-order) piezoelectric potential
in the calculations. The piezoelectric potential increases the
calculated energy gap between the highest hole and the electron
state by 1.1 meV. This is initially surprising, as one normally
expects for a symmetric dot that a potential variation such as that
in Fig.~\ref{fig:pot}(c) will reduce both the electron and hole
confinement energies (quantum confined Stark effect). The anomalous
behaviour observed in this case can however be understood from
Figs.~\ref{fig:pot}(a) and (b), which show that both the electron
and hole wavefunctions are shifted towards the top of the dot. This
asymmetry in the electron and hole wavefunctions would not be
predicted using a continuum model to describe the electronic
structure, but arises here as a natural consequence of the
underlying atomic structure of the (111)-oriented dot.

Table~\ref{tab:no_so} summarizes also the energy gap and the
electron and hole $s$- and $p$-shell energies without spin-orbit
coupling. Please note that the $p$-states remain degenerate when
including the piezoelectric potential. All this shows that our SC
and QD geometry accurately reflects and preserves the $C_{3v}$
symmetry. No artificial symmetry breaking is introduced.
\begin{table}
\caption{\label{tab:no_so} Energy gap ($ E_g$), the hole $s$- and
$p$-shell state energies ($ E_h^s$ and $E_h^p$ ) and the electron
$s$- and $p$-shell state energies ($ E_e^s$ and $E_e^p$)  for a pure
InAs/GaAs QD ($l_q=55$ nm, $h_q=2$ nm) calculated without spin-orbit
coupling, considering only the quantum confinement (QC), the quantum
confinement and the strain (QC+$\epsilon$), and the quantum
confinement, the strain and the built-in potential
(QC+$\epsilon$+PZ). The energies (in meV) are given with respect to
the bulk GaAs valence band maximum.}
\begin{tabular*}{\columnwidth}{@{\extracolsep{\fill}}lccc}
\hline\hline
       & QC    & QC+$\epsilon$ &  QC+$\epsilon$+PZ \\
\hline
$ E_g$ & 992.3  &   1089.6 &  1090.2 \\
$ E_h^s$ &   171.1  &    407.1 &   392.        \\
$ E_h^p$ &   170.6  &  408.6   &  392.6    \\

$ E_e^s$ &   1163.4 &   1497.7   &  1482.8      \\
$ E_e^p$ &   1175.8 &   1510.0   &   1492.0  \\
\hline\hline
\end{tabular*}
\end{table}

\begin{figure}
\includegraphics[width=\columnwidth]{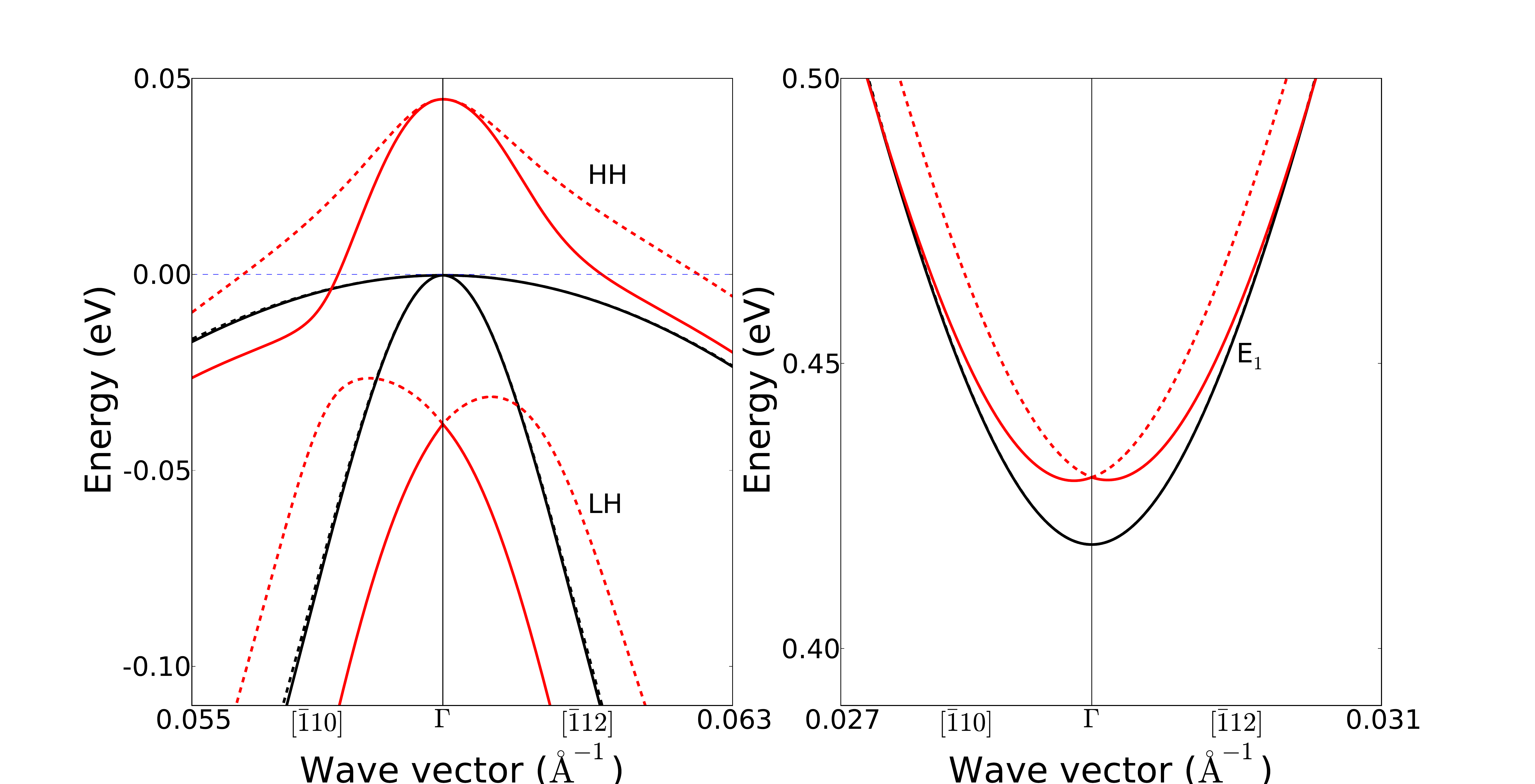}
\caption{Heavy and light hole (left) and lowest electron states band
structure around Brillouin zone centre of bulk InAs, before (black
lines) and after (red lines) applying a (111) biaxial strain of 1\%.
}\label{fig:band_str}
\end{figure}

Without the spin-orbit coupling the symmetry properties of the
system were determined by geometrical symmetries. In terms of group
theory, in the absence of spin-orbit coupling effects the
irreducible representations of the \textit{single} group $C_{3v}$
are important. When including spin-orbit coupling effects one has to
deal with \textit{double} groups, in this case $\bar{C}_{3v}$. This
group contains only two dimensional irreducible
representations.~\cite{joshi1997elements} Consequently, each state
can only be two-fold (Kramers) degenerate. Table~\ref{tab:so}
summarizes the energy gap ($ E_g$), the electron $p$-state splitting
($\Delta E^{e}_{p}$) between the two-fold Kramers degenerate
states and the electron ($ E_e^0$) and hole ($ E_h^0$) ground state
 energies. For the unstrained QD, including the
spin-orbit coupling, the $p$-state splittings are tiny (0.06~meV for
holes and 0.04~meV for electrons). When including strain effects the
$p$-state splittings are one order of magnitude larger (1.31~meV for
holes and 0.36~meV for electrons). The effect of the piezoelectric
potential on the $p$-state splittings is negligible (cf.
Table~\ref{tab:so}). In combination with our calculations without
spin-orbit coupling, we have shown that spin-orbit effects are the
origin of the here observed tiny $p$-state splitting in a
triangular-shaped (111)-oriented InAs/GaAs QD  and that the lattice
constant mismatch between InAs and GaAs amplifies this splitting.
This is a consequence of the increase of electron and hole spin
splitting induced by a (111)-oriented biaxial strain. In order to
confirm this statement, we show in Fig.~\ref{fig:band_str} the band
structure of bulk InAs before and after applying a (111)-oriented
biaxial strain of 1\%. Our model gives a large strain-induced
spin-splitting for this particular strain type, associated with the
$C_5$ shear-strain linear-k term in the Pikus-Bir
Hamiltonian.\cite{Bir-Pikus,Trebin} We note also that the light hole
state spin splitting is more sensitive to strain, this is in line
with previous investigation of strain-induced spin-splitting
associated with the $C_4$ axial-strain linear-k term.~\cite{EOR}

\begin{table}
\caption{\label{tab:so} Energy gap ($ E_g$) , the two highest
valence states energies ($ E_h^0$ and $ E_h^1$), the two lowest
electron energies ($ E_e^0$ and $ E_e^1$) and electron and hole
$p$-states splittings ($\Delta E_{p}^e$ and $\Delta E_{p}^h$)
for a pure InAs/GaAs QD calculated including spin-orbit coupling.
The energies (in meV) are given with respect to the bulk GaAs
valence band maximum.  }
\begin{tabular*}{\columnwidth}{@{\extracolsep{\fill}}lccc}
\hline\hline
       & QC    & QC+$\epsilon$ &  QC+$\epsilon$+PZ \\
\hline
$ E_g$ &  872.7     &  980  &  982.4 \\
$ E_h^0$ &  182.0   &  407.3   &  397.4        \\
$ E_h^1$ & 176.6    &  398.9   &  386.9    \\
$\Delta E_{p}^h$ & 0.06 & 1.31 & 1.4 \\

$ E_e^0$ & 1054.7   &  1387.3    &  1379.8     \\
$ E_e^1$ & 1067.4   &   1399.5   &   1.388.7  \\
$\Delta E_{p}^e$ & 0.04 & 0.36 &  0.3 \\
\hline\hline
\end{tabular*}
\end{table}

It should be noted that the $p$-state splitting introduced by the
spin-orbit coupling does not lead to a FSS. This is discussed in
detail by Karlsson {\it et al.} (Ref.~\citenum{Karlsson}) in terms
of group theory. Thus, in contrast to (001)-oriented InAs/GaAs QDs,
where the $p$-state splitting is introduced mainly by strain field
anisotropy or piezoelectric built-in potentials, a non-vanishing
$p$-state splitting in (111)-oriented InAs/GaAs QDs is not directly
indicative of the presence of a FSS.
\begin{figure*}
\begin{tabular*}{\textwidth}{@{\extracolsep{\fill}}ccccc}
\hline \hline
 $\mathbf{h_0}$ & $\mathbf{h_1}$   &  $\mathbf{h_2}$   &  $\mathbf{h_3}$       \\
\hline
  &   &     &         \\

\includegraphics[width=0.47\columnwidth]{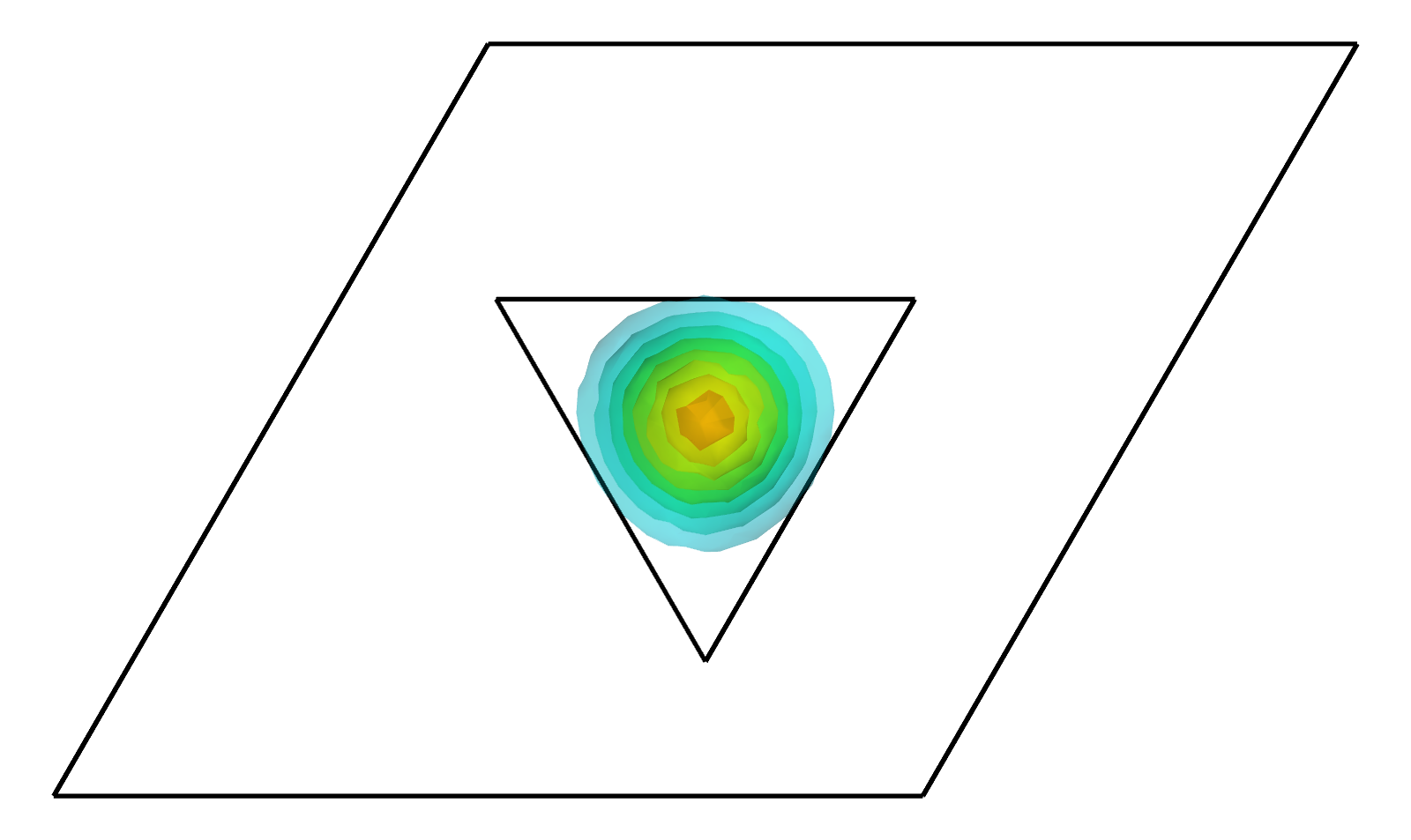}
 &\includegraphics[width=0.47\columnwidth]{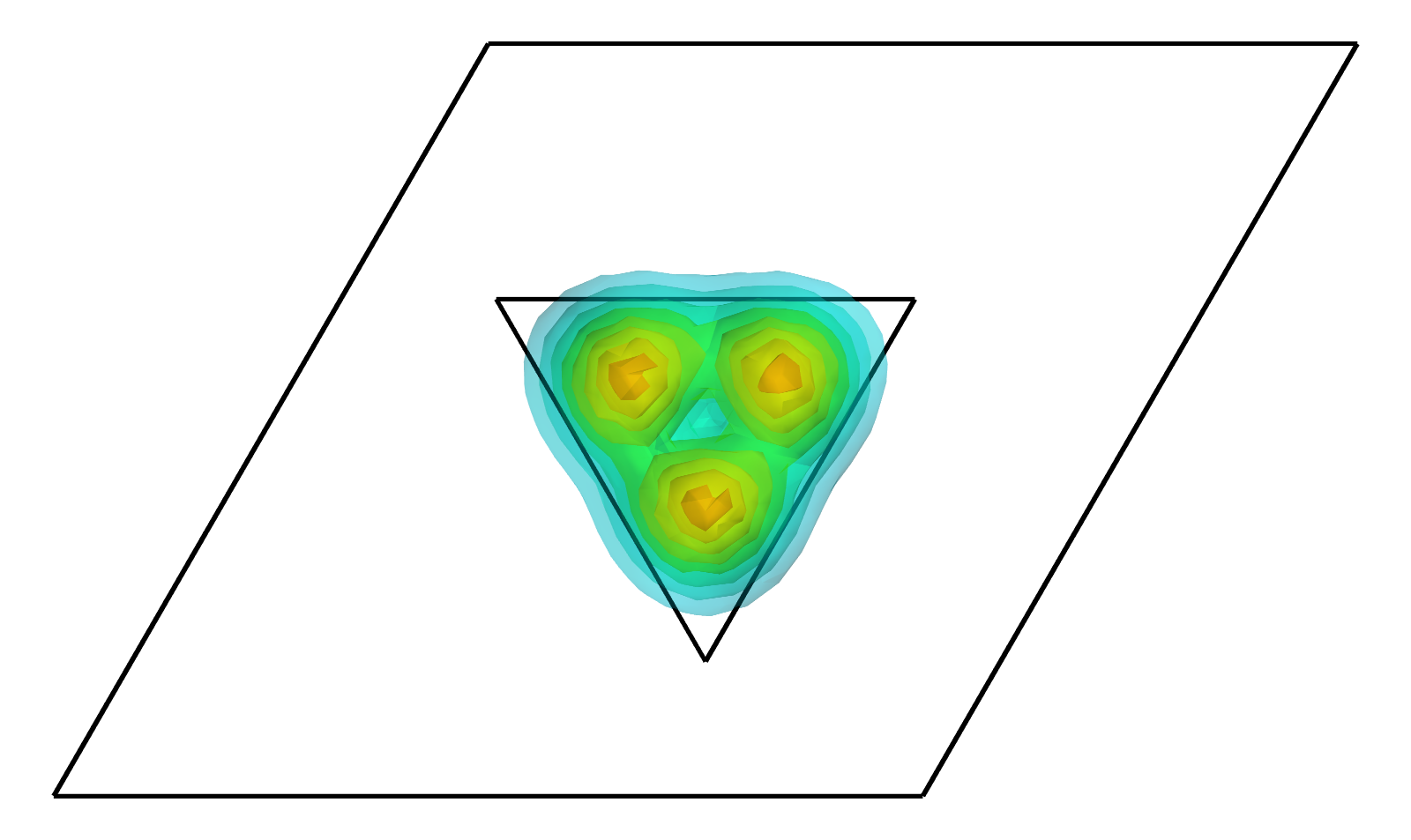}
 &\includegraphics[width=0.47\columnwidth]{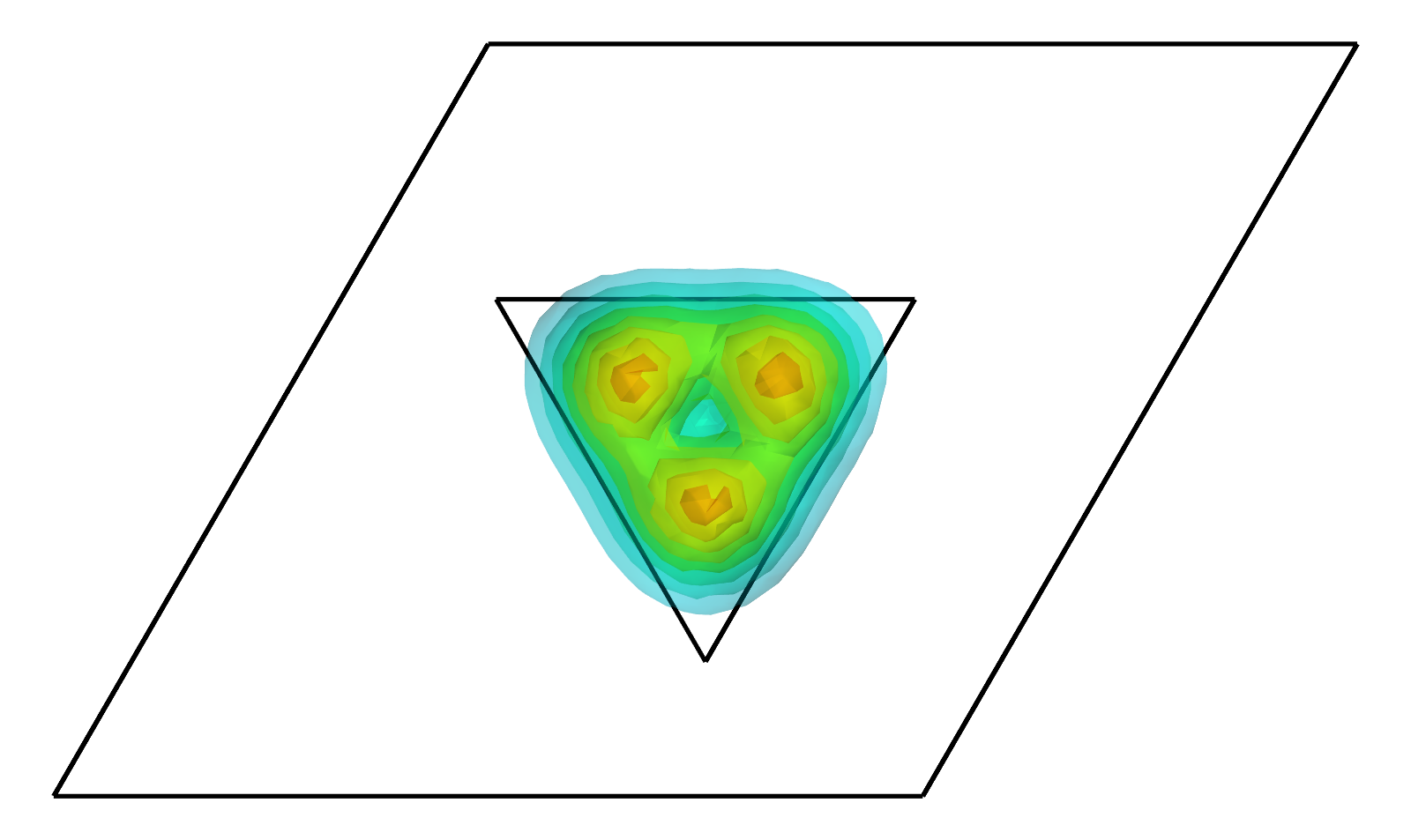}
 &\includegraphics[width=0.47\columnwidth]{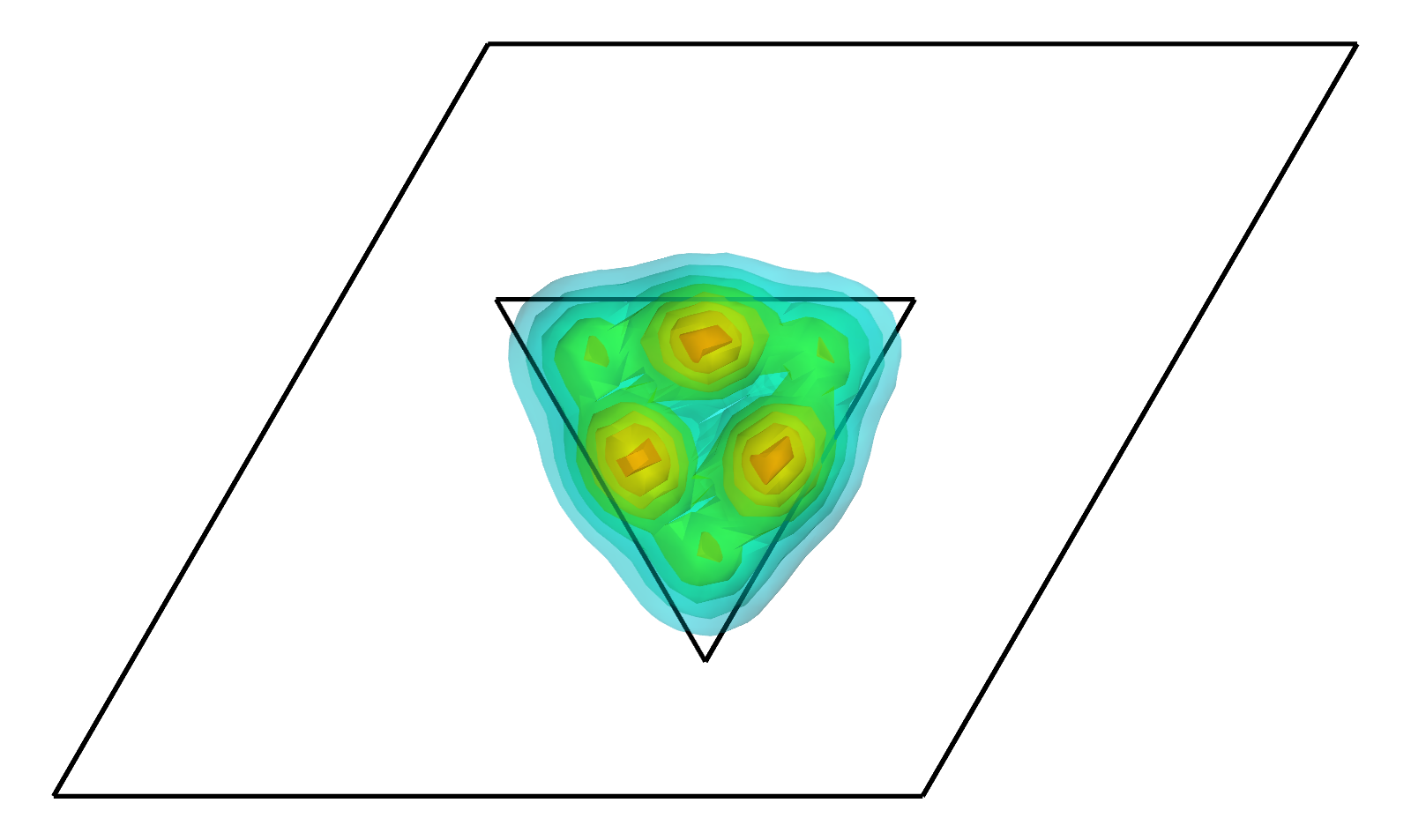}\\
  &   &     &         \\

\hline

 $\mathbf{e_0}$ & $\mathbf{e_1}$   &  $\mathbf{e_2}$   &  $\mathbf{e_3}$       \\
\hline
  &   &     &         \\

\includegraphics[width=0.47\columnwidth]{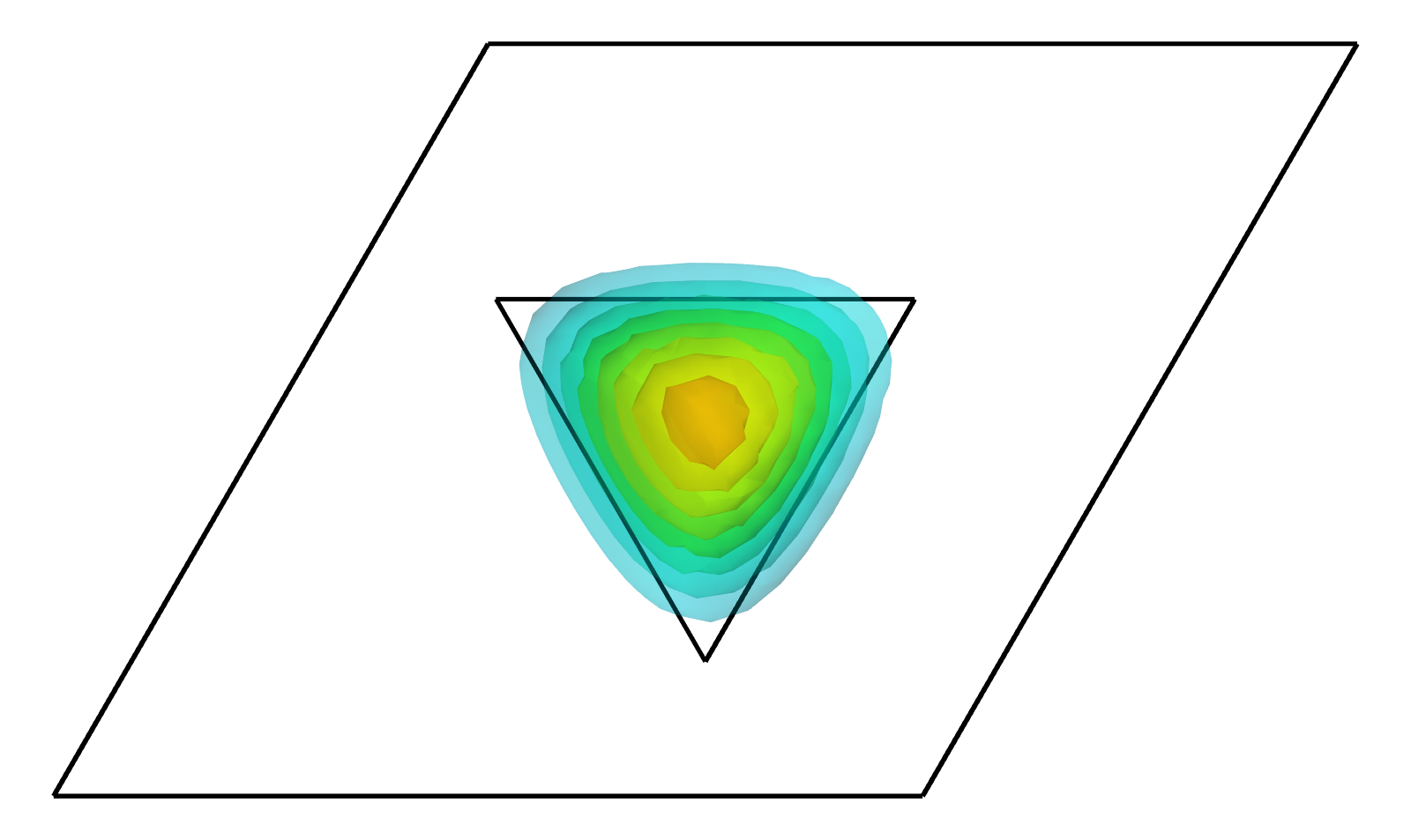}
 &\includegraphics[width=0.47\columnwidth]{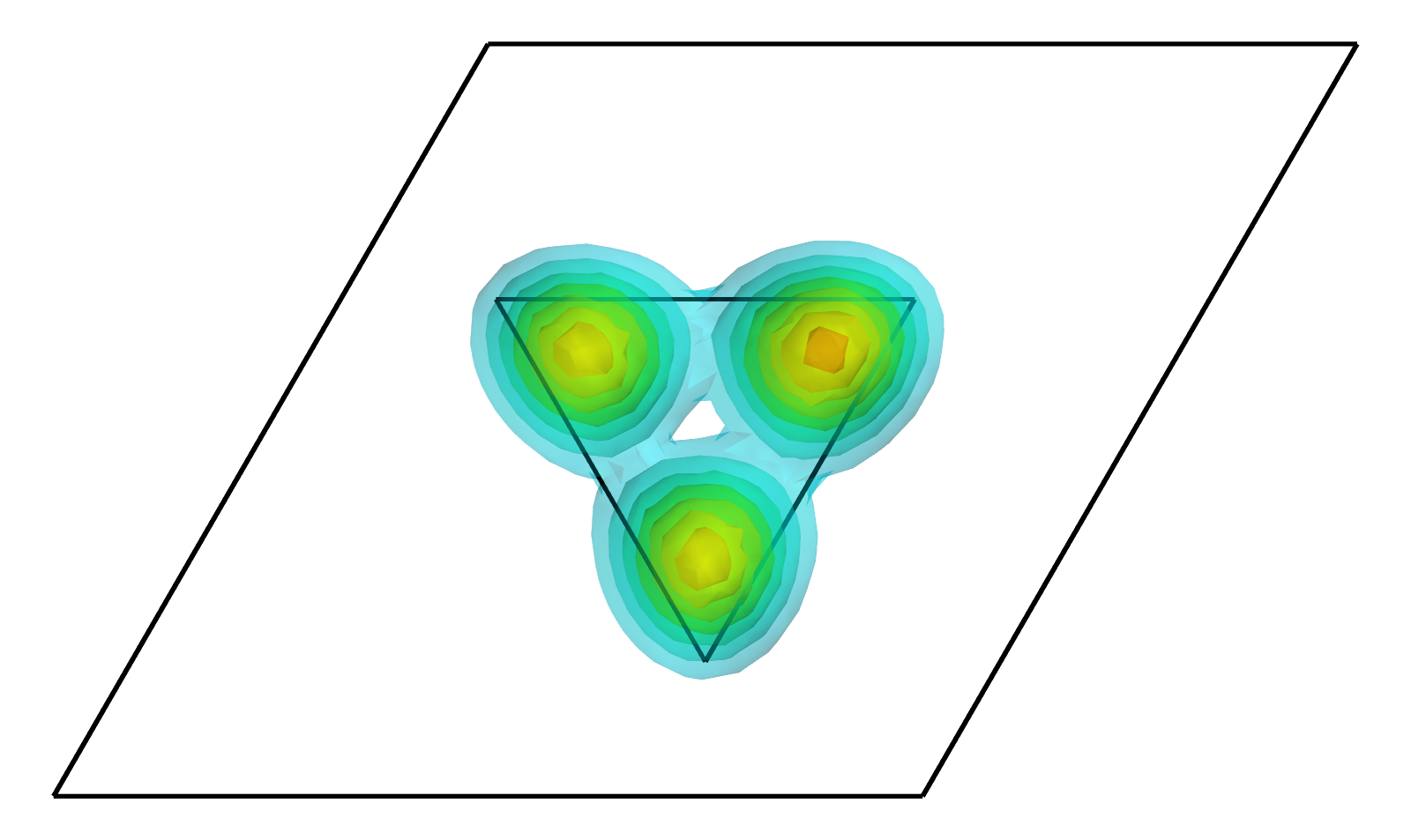}
 &\includegraphics[width=0.47\columnwidth]{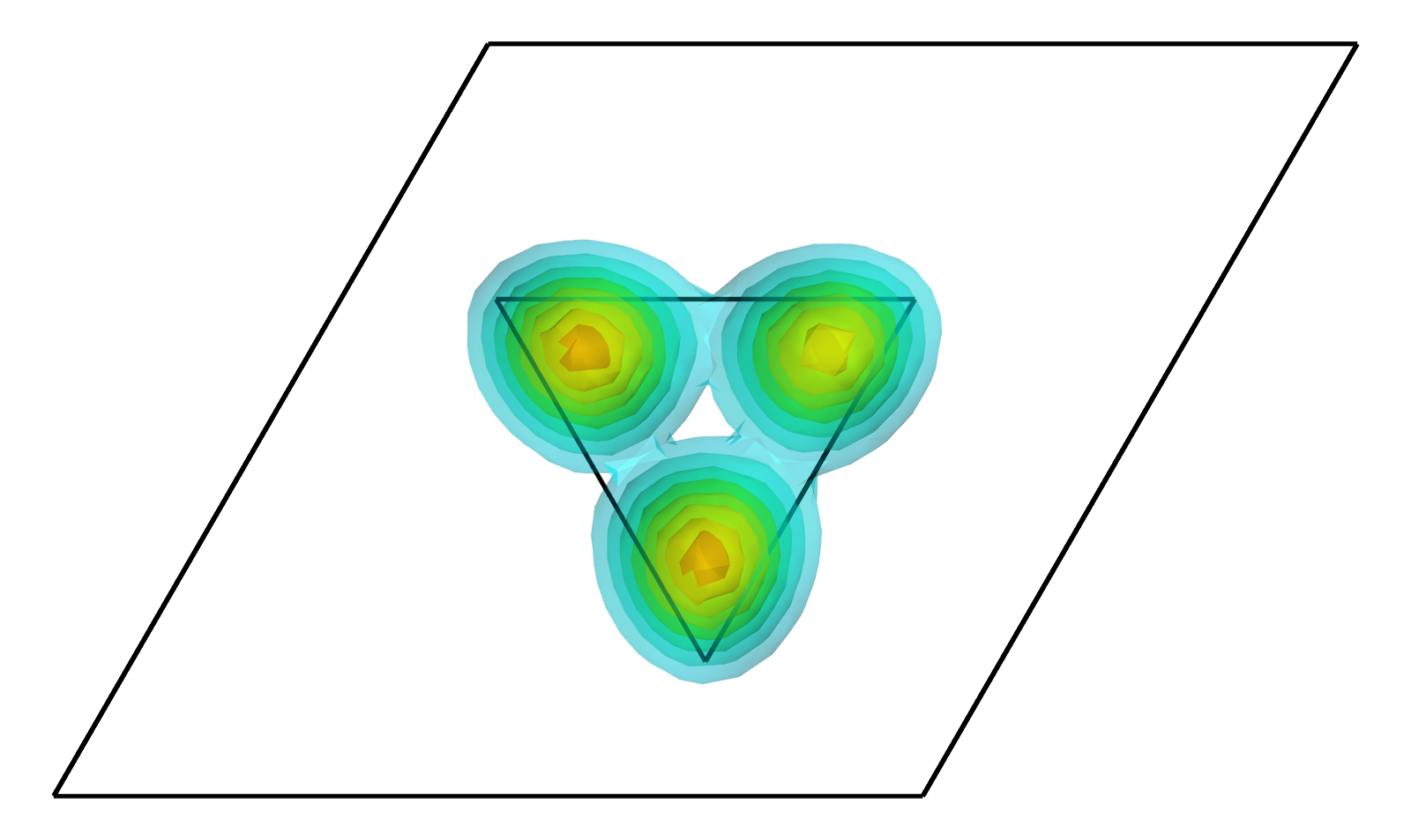}
 &\includegraphics[width=0.47\columnwidth]{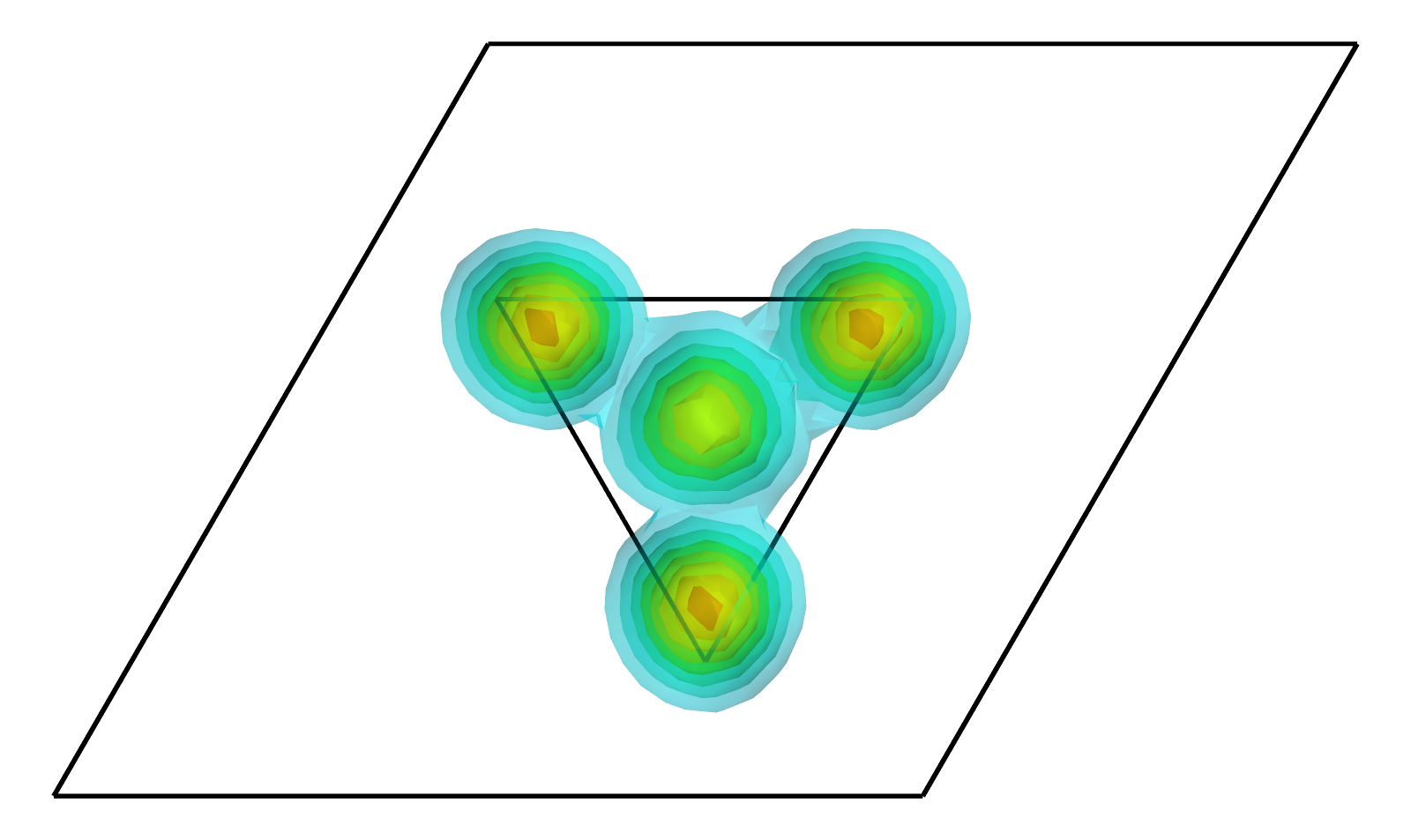} \\
  &   &     &         \\

\hline \hline
\end{tabular*}
\caption{Isosurfaces plot of charge densities of the four
energetically highest valence states ($h_0$, $h_1$, $h_2$, $h_3$)
and the four energetically lowest conduction states ($e_0$, $e_1$,
$e_2$, $e_3$) in a triangular-shaped (111)-oriented InAs/GaAs QD.
}\label{pure_InAs_WF}
\end{figure*}

To shed more light on the electronic structure of site-controlled
(111)-oriented InAs/GaAs QDs,  Fig.~\ref{pure_InAs_WF} displays the
charge densities of the four energetically highest valence states
and the four energetically lowest conduction states when taking all
the different contributions (strain, built-in fields) into account.
The charge densities of Fig.~\ref{pure_InAs_WF} reflect the
triangular-shaped QD confinement. It can also be seen from
Figs.~\ref{fig:pot} and \ref{pure_InAs_WF} that the holes are more
localized than electrons, both along the growth direction and in the
growth plane, primarily because of the larger effective masses of
holes than electrons.

\subsection{Alloyed QD}

\begin{figure}
\includegraphics[width=0.95\columnwidth]{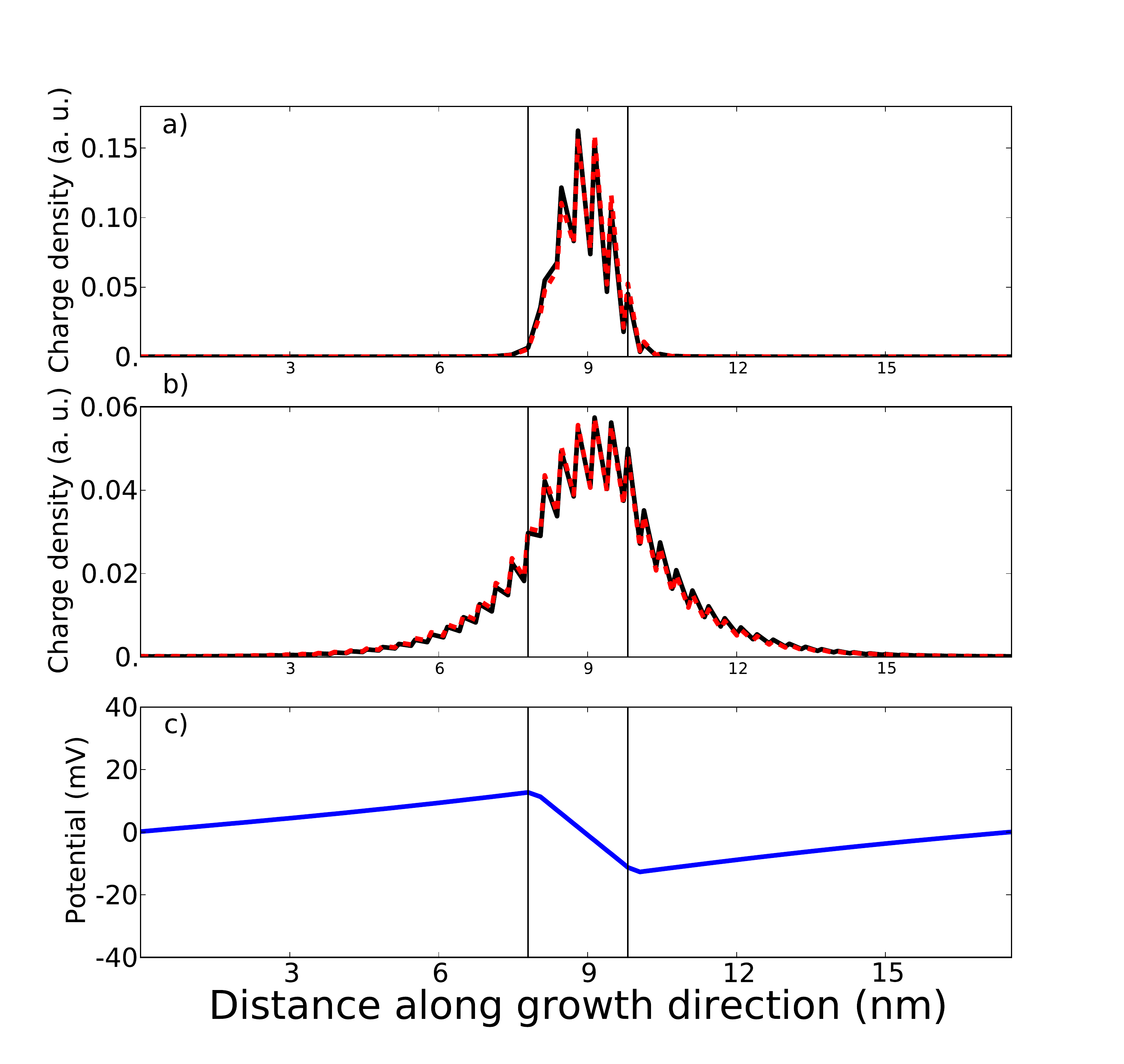}
\caption{Same as Fig.~\ref{fig:pot} for a In$_{0.25}$Ga$_{0.75}$As/GaAs QD within VCA.}\label{fig:pot1}
\end{figure}

Having discussed the influence of spin-orbit coupling, strain and
built-in fields on the electronic properties of site-controlled
(111)-oriented InAs/GaAs QDs, we turn now and study the impact of
random alloy fluctuations on their electronic and optical
properties. We consider here an InAs content of 25\%, which is well
within the experimentally relevant range (cf.
Sec.~\ref{sec:Exp_SC}). As a reference point for our random alloy
calculations we perform in a first step a VCA calculation. In this
calculation the QD region is filled with a virtual crystal for which
the TB parameters, the elastic and piezoelectric constants are
simply a concentration weighted average of the corresponding bulk
InAs and GaAs parameters. Since we are mainly interested in
investigating the robustness of the electronic structure of the QD
system against alloy fluctuations, the adjustment of the VCA results
by introducing bowing parameters for the involved TB parameters is
beyond the scope of the present work. Obviously, using the VCA for
electronic structure calculations of site-controlled
In$_{0.25}$Ga$_{0.75}$As/GaAs QD does not change the symmetry of the
system. Therefore, in terms of the $p$-state splitting, similar
results as for the pure InAs/GaAs QD are expected. Quantitatively,
the energy gap for the 25\% InAs VCA calculation is larger than in
the pure InAs case, since GaAs has a larger band gap and we have
less InAs content in the QD. The smaller InAs content results also
in a reduced carrier confinement. Figure~\ref{fig:pot1} show the
charge densities of hole and electron ground states projected on the
growth direction, in the presence and in the absence of the built-in
potential. The profile of the piezoelectric potential along the
growth direction and across the centre of the QD is also displayed
in Fig.~\ref{fig:pot1}. As in the case of a pure InAs/GaAs QD, the
wavefunctions are asymmetric with respect to the (111) mirror plane;
however this asymmetry is less pronounced in the case of a virtual
In$_{0.25}$Ga$_{0.75}$As/GaAs QD when compared with the pure system
(cf. Fig.~\ref{fig:pot}). Also, we note that the piezoelectric
potential profile is inverted compared to the case of a pure InAs QD
(Fig.~\ref{fig:pot}), where the large lattice mismatch leads to the
second-order piezoelectric potential contribution being larger than
the first-order one.

\begin{figure*}
\begin{tabular*}{\textwidth}{@{\extracolsep{\fill}}ccccc}
\hline \hline
 & \textbf{VCA} & \textbf{Config. 1}   &  \textbf{Config. 2}   &  \textbf{Config. 3}\\
\hline
 & &    &   &  \\
 \raisebox{1.25cm}{\rotatebox{90}{$\mathbf{h_0}$}} &
\includegraphics[width=0.44\columnwidth]{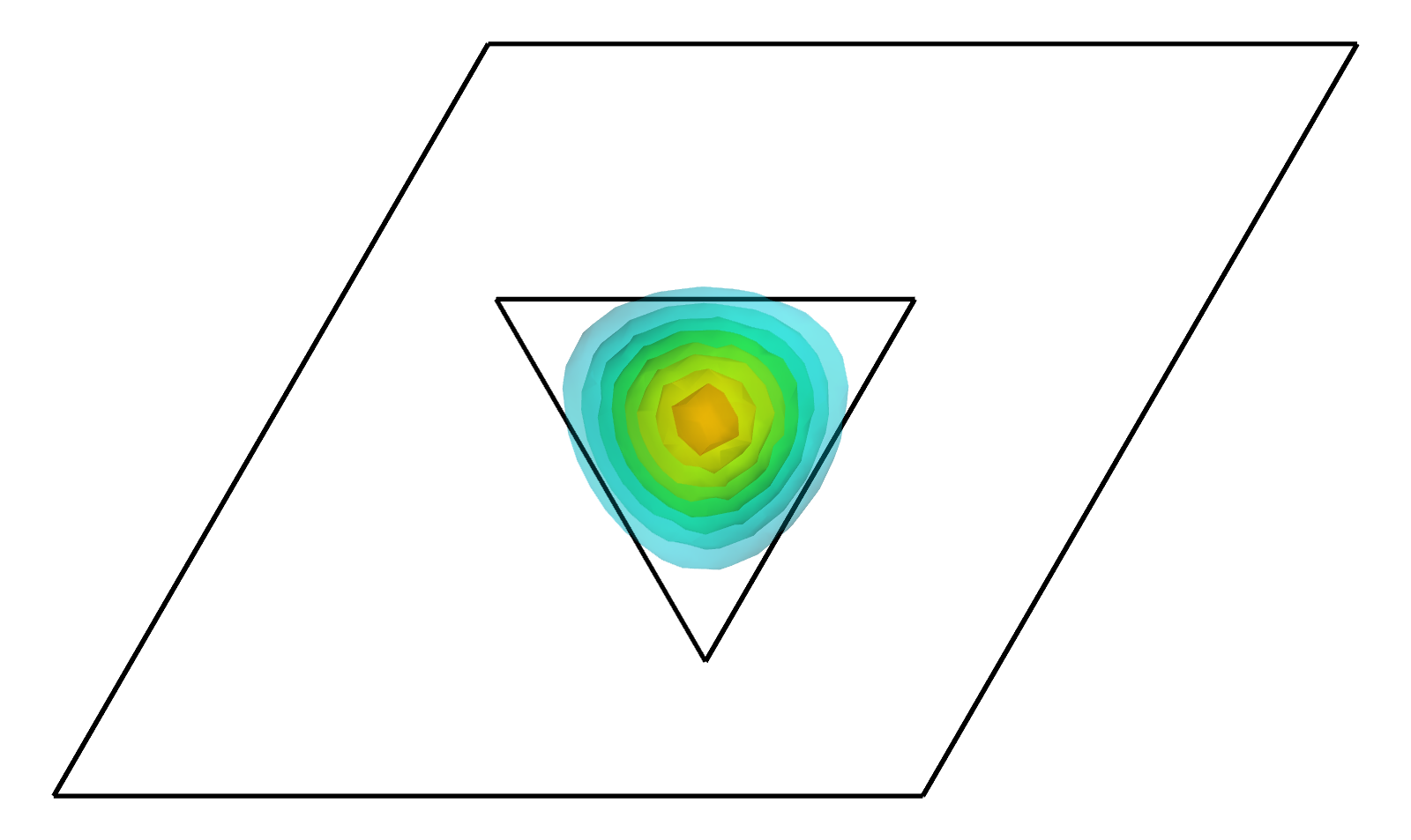}
 &\includegraphics[width=0.44\columnwidth]{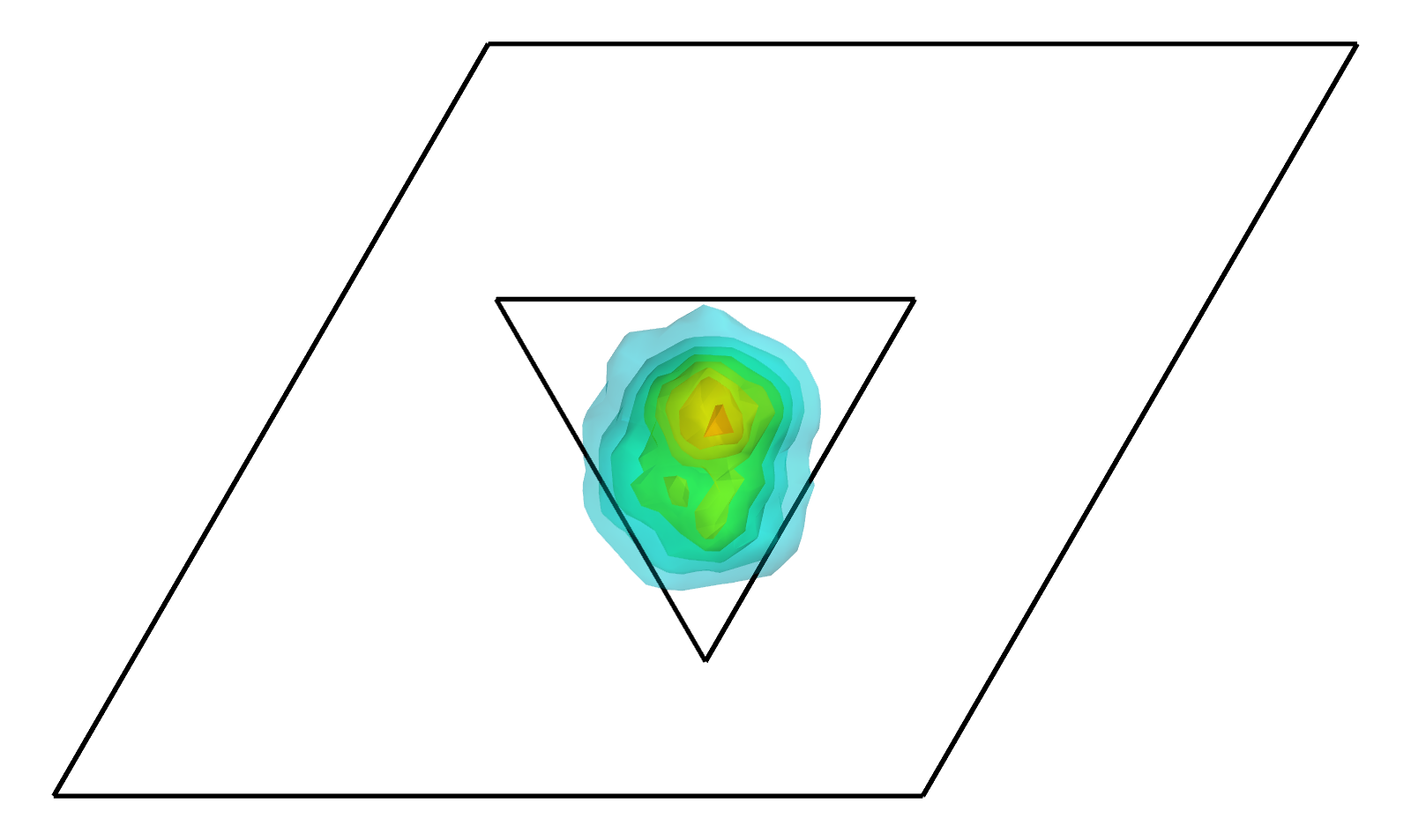}
 &\includegraphics[width=0.44\columnwidth]{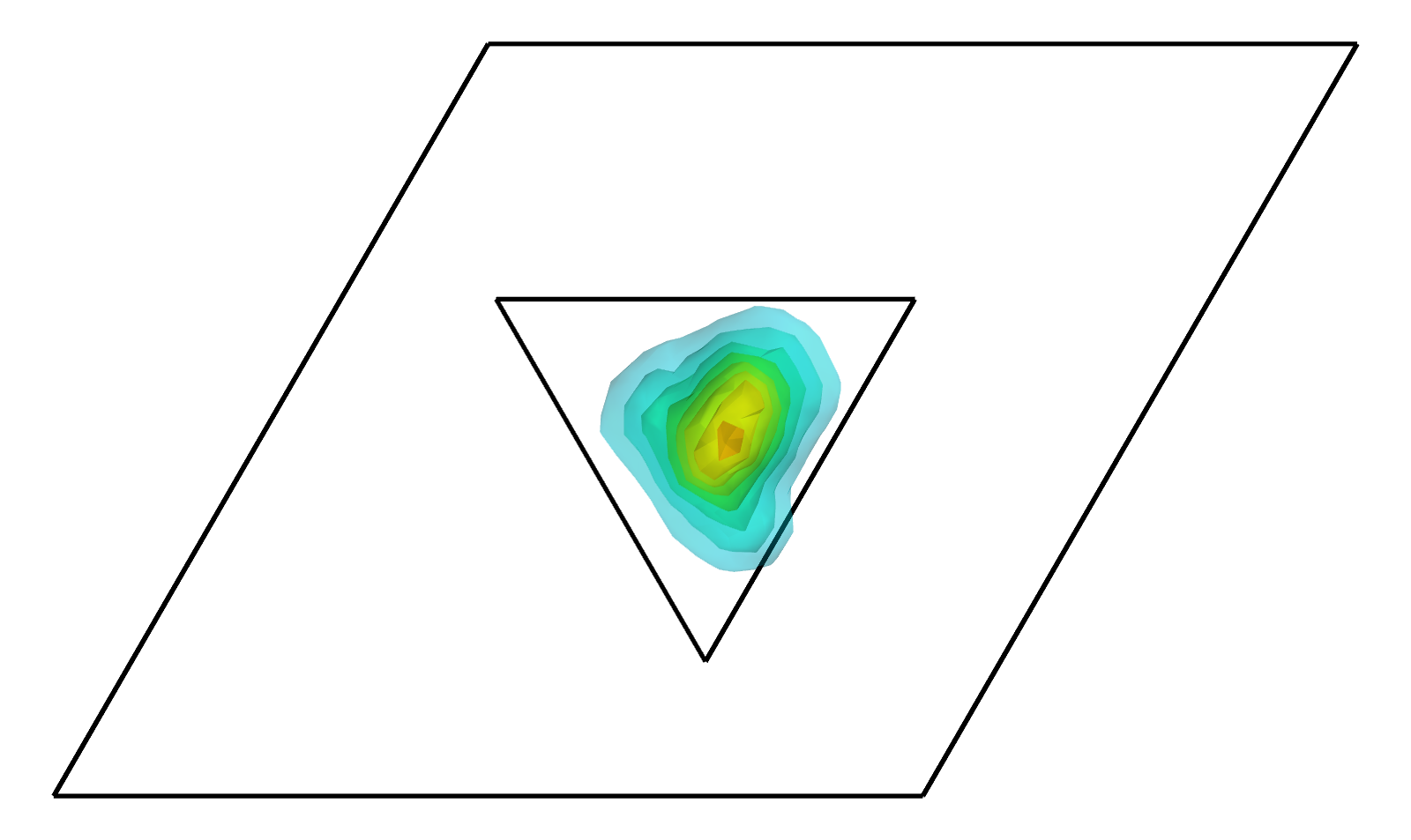}
 &\includegraphics[width=0.44\columnwidth]{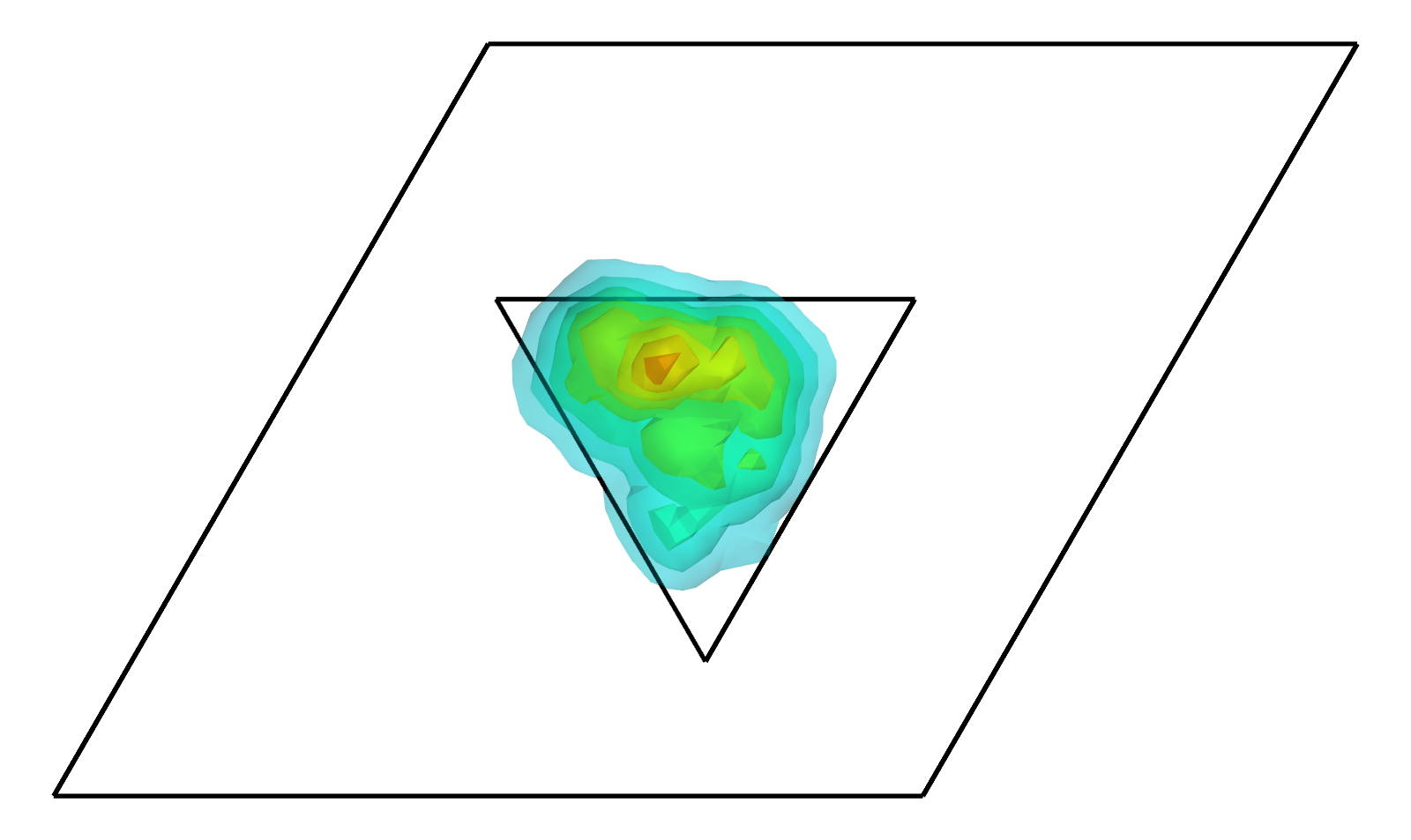}\\
 & &    &   &  \\

  & \begin{tabular}{lcr} $|s\rangle$ & $=$ & $0.0008 $ \\
  $|s^*\rangle$ & $=$ & $0.0001 $ \\
  $|p_x\rangle$ & $=$ & $0.2605$ \\
  $|p_y\rangle$ & $=$ & $0.2605 $ \\
  $|p_z\rangle$ & $=$ & $0.2605 $ \\
  $|d_{xy}\rangle$ & $=$ & $0.0724 $ \\
  $|d_{yz}\rangle$ & $=$ & $0.0724 $\\
  $|d_{zx}\rangle$ & $=$ & $0.0724 $\\
  $|d_{x^2-y^2}\rangle$ & $=$ & $0.0000 $\\
  $|d_{3z^2-r^2}\rangle$ & $=$ & $0.0000 $ \end{tabular} &

   \begin{tabular}{lcr} $|s\rangle$ & $=$ & $0.0014 $\\     
    $|s^*\rangle$ & $=$ & $0.0002 $\\
    $|p_x\rangle$ & $=$ & $0.2588$ \\
    $|p_y\rangle$ & $=$ & $0.2598 $ \\
    $|p_z\rangle$ & $=$ & $0.2612 $ \\
    $|d_{xy}\rangle$ & $=$ & $0.0728 $\\
    $|d_{yz}\rangle$ & $=$ & $0.0721 $ \\
    $|d_{zx}\rangle$ & $=$ & $0.0724 $ \\
    $|d_{x^2-y^2}\rangle$ & $=$ & $0.0005 $ \\
    $|d_{3z^2-r^2}\rangle$ & $=$ & $0.0005 $ \end{tabular}  &

   \begin{tabular}{lcr} $|s\rangle$ & $=$ & $0.0015 $\\
    $|s^*\rangle$ & $=$ & $0.0003 $\\
    $|p_x\rangle$ & $=$ & $0.2618$ \\
    $|p_y\rangle$ & $=$ & $0.2535 $ \\
    $|p_z\rangle$ & $=$ & $0.2644 $ \\
    $|d_{xy}\rangle$ & $=$ & $0.0737 $\\
    $|d_{yz}\rangle$ & $=$ & $0.0729 $ \\
    $|d_{zx}\rangle$ & $=$ & $0.0706 $ \\
    $|d_{x^2-y^2}\rangle$ & $=$ & $0.0006 $ \\
    $|d_{3z^2-r^2}\rangle$ & $=$ & $0.0006 $ \end{tabular} &

   \begin{tabular}{lcr} $|s\rangle$ & $=$ & $0.0014 $\\
    $|s^*\rangle$ & $=$ & $0.0002 $\\
    $|p_x\rangle$ & $=$ & $0.2570$ \\
    $|p_y\rangle$ & $=$ & $0.2645 $ \\
    $|p_z\rangle$ & $=$ & $0.2583 $ \\
    $|d_{xy}\rangle$ & $=$ & $0.0720 $\\
    $|d_{yz}\rangle$ & $=$ & $0.0716 $ \\
    $|d_{zx}\rangle$ & $=$ & $0.0737 $ \\
    $|d_{x^2-y^2}\rangle$ & $=$ & $0.0006 $ \\
    $|d_{3z^2-r^2}\rangle$ & $=$ & $0.0006 $ \end{tabular} \\

\hline
 & &    &   &  \\
 \raisebox{1.25cm}{\rotatebox{90}{$\mathbf{e_0}$}} &
\includegraphics[width=0.44\columnwidth]{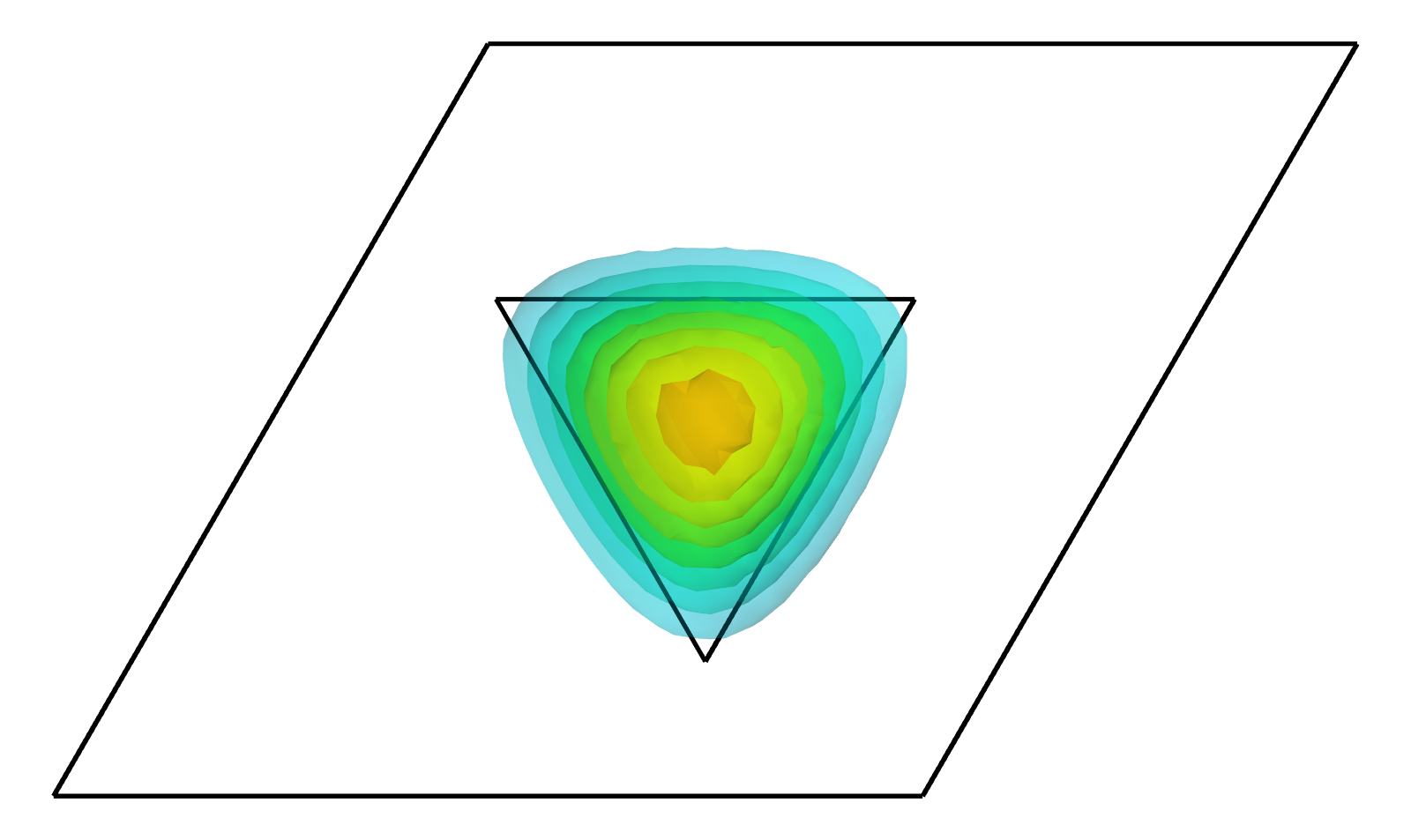}
 &\includegraphics[width=0.44\columnwidth]{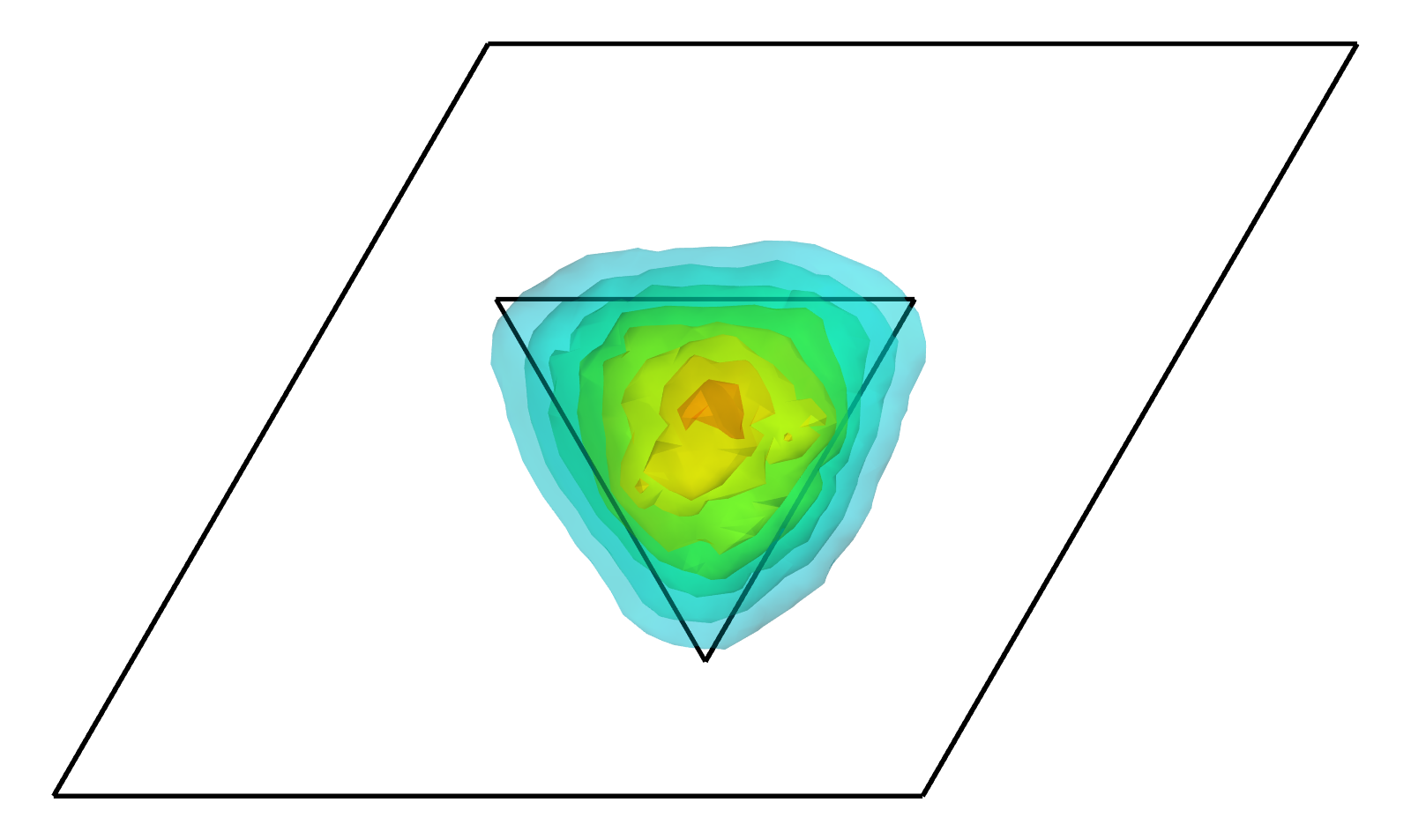}
 &\includegraphics[width=0.44\columnwidth]{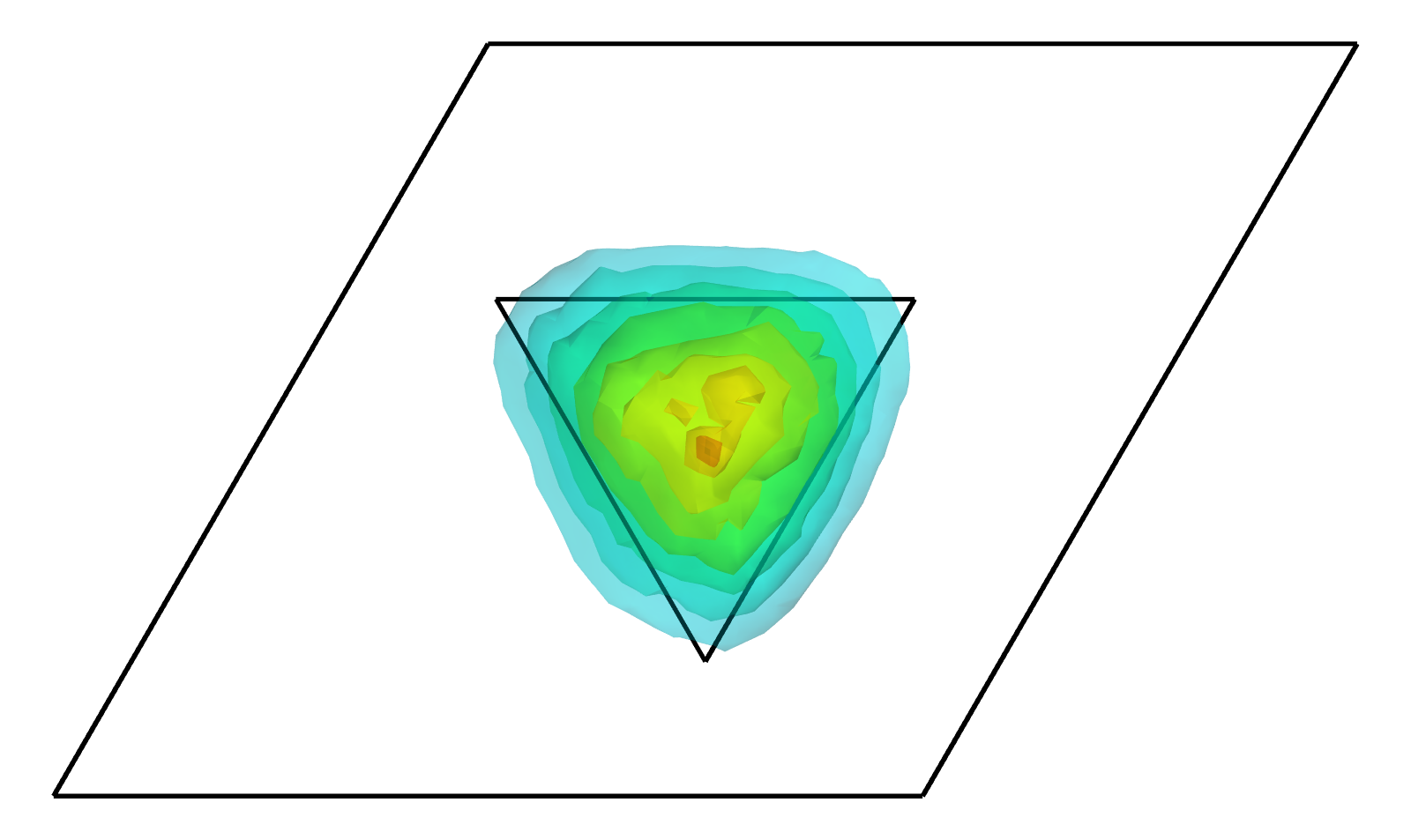}
 &\includegraphics[width=0.44\columnwidth]{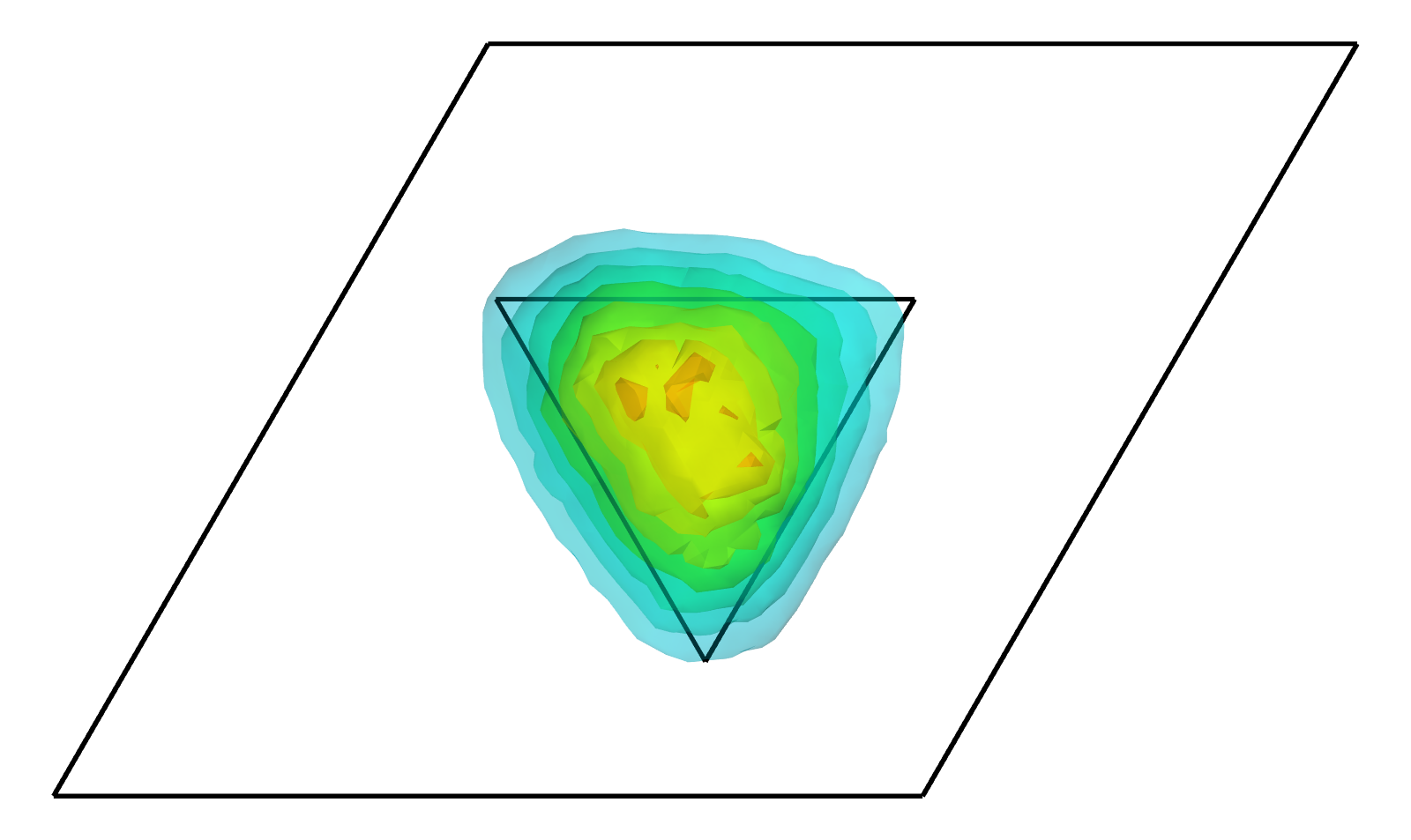}\\
 & &    &   &  \\
 & \begin{tabular}{lcr} $|s\rangle$ & $=$ & $0.8652 $\\
    $|s^*\rangle$ & $=$ & $0.1329 $\\
    $|p_x\rangle$ & $=$ & $0.0005$ \\
    $|p_y\rangle$ & $=$ & $0.0005 $ \\
    $|p_z\rangle$ & $=$ & $0.0005 $ \\
    $|d_{xy}\rangle$ & $=$ & $0.0001 $\\
    $|d_{yz}\rangle$ & $=$ & $0.0001 $ \\
    $|d_{zx}\rangle$ & $=$ & $0.0001 $ \\
    $|d_{x^2-y^2}\rangle$ & $=$ & $0.0000 $ \\
    $|d_{3z^2-r^2}\rangle$ & $=$ & $0.0000 $ \end{tabular} &

   \begin{tabular}{lcr} $|s\rangle$ & $=$ & $0.8554 $\\     
    $|s^*\rangle$ & $=$ & $0.1357 $\\
    $|p_x\rangle$ & $=$ & $0.0021$ \\
    $|p_y\rangle$ & $=$ & $0.0021 $ \\
    $|p_z\rangle$ & $=$ & $0.0021 $ \\
    $|d_{xy}\rangle$ & $=$ & $0.0006 $\\
    $|d_{yz}\rangle$ & $=$ & $0.0006 $ \\
    $|d_{zx}\rangle$ & $=$ & $0.0006 $ \\
    $|d_{x^2-y^2}\rangle$ & $=$ & $0.0003 $ \\
    $|d_{3z^2-r^2}\rangle$ & $=$ & $0.0003 $ \end{tabular}  &

   \begin{tabular}{lcr} $|s\rangle$ & $=$ & $0.8558 $\\     
    $|s^*\rangle$ & $=$ & $0.1356 $\\
    $|p_x\rangle$ & $=$ & $0.0020$ \\
    $|p_y\rangle$ & $=$ & $0.0020 $ \\
    $|p_z\rangle$ & $=$ & $0.0020 $ \\
    $|d_{xy}\rangle$ & $=$ & $0.0006 $\\
    $|d_{yz}\rangle$ & $=$ & $0.0006 $ \\
    $|d_{zx}\rangle$ & $=$ & $0.0006 $ \\
    $|d_{x^2-y^2}\rangle$ & $=$ & $0.0003 $ \\
    $|d_{3z^2-r^2}\rangle$ & $=$ & $0.0003 $ \end{tabular} &

   \begin{tabular}{lcr} $|s\rangle$ & $=$ & $0.8556 $\\     
    $|s^*\rangle$ & $=$ & $0.1356 $\\
    $|p_x\rangle$ & $=$ & $0.0021$ \\
    $|p_y\rangle$ & $=$ & $0.0020 $ \\
    $|p_z\rangle$ & $=$ & $0.0020 $ \\
    $|d_{xy}\rangle$ & $=$ & $0.0006 $\\
    $|d_{yz}\rangle$ & $=$ & $0.0006 $ \\
    $|d_{zx}\rangle$ & $=$ & $0.0006 $ \\
    $|d_{x^2-y^2}\rangle$ & $=$ & $0.0003 $ \\
    $|d_{3z^2-r^2}\rangle$ & $=$ & $0.0003 $ \end{tabular} \\
\hline \hline
\end{tabular*}
\caption{ Isosurfaces plot of hole (top) and electron (bottom)
ground state charge densities for a triangular-shaped (111)-oriented
In$_{0.25}$Ga$_{0.75}$As/GaAs QD obtained within the VCA and for the
three different random alloy configurations considered
(Configs.~1-3). Also we show for each case what fraction of the
state projects onto each type of atomic orbitals used in the
$sp^3d^5s^*$ TB model.}\label{small_InGaAs_WF}
\end{figure*}

To analyze the impact of random alloy fluctuations at a microscopic
level, we randomly replace Ga atoms in the QD region by In atoms so
that the (111)-oriented QD contains 25\% InAs. To study the
influence of the alloy microstructure on the electronic and optical
properties, we have constructed three different microscopic random
configurations. Figure~\ref{small_InGaAs_WF} shows the electron
($e_0$) and hole ($h_0$) ground state charge densities for the VCA
case together with the results from the three different random
configurations (Configs. 1 - 3). The charge densities obtained from
the VCA calculation are symmetric with respect to rotations of
120$^\circ$ around the central QD axis. This is in contrast to the
results of the alloyed calculations. Here the charge densities
appear deformed due to the random distribution of the In and Ga
atoms inside the QD. The deformation is more pronounced for the hole
states due to the larger hole effective mass compared with the
electron mass, and also due to the stronger hole confinement in the
QD.

The contribution of the different orbital types to the electronic
wavefunctions are also presented in Fig.~\ref{small_InGaAs_WF}. With
the Hamiltonian that we are using, the zone centre bulk valence band
maximum states are made up of $p$ and $d$ orbitals, with the lowest
conduction states having $s$ and $s^*$ character.  As expected, the
highest valence states in the QD are mainly made of $p$ and $d$
orbitals, with $< 0.2\%$ $s$ and $s^*$ conduction character.
Likewise, the electron states are made of $s$ and $s^*$ orbitals,
with $< 1\%$ valence character. This is to be compared with (001)
InGaAs/GaAs QDs, which generally have a much lower base length to
height ratio, and for which the results of typical calculations give
$1\%$ conduction character in the highest valence states and
$\sim10\%$ valence character in the lowest conduction
state.~\cite{Bimberg} Mixing of valence and conduction states is one
of the factors that can contribute to a finite FSS
value.~\cite{krapek} Looking at the results by
Krapek \etal\cite{krapek} the electron-hole exchange interaction
matrix elements $EX$, which mainly determine the magnitude of the
FSS,  can be expressed as a sum of three contributions
$EX=EX_0+EX_1+EX_2$. In Ref.~\onlinecite{krapek}, the authors show
that $EX_0$ and $EX_1$ depend on the conduction and valence band
mixing contributions. Consequently, if conduction and valence bands
are decoupled, for instance in a two plus six band
$\mathbf{k}\cdot\mathbf{p}$ model, ~\cite{Singh2000,Marquardt2008}
$EX_0=EX_1=0$ and only $EX_2$ contributes. Additionally, Krapek and
co-workers show that the contribution from these three terms to the
FSS scale as $EX_0\sim 1/L$, $EX_1\sim 1/L^2$ and $EX_2\sim 1/L^3$,
where $L$ is the extension of the wavefunction, which is then
to a good approximation proportional to the dot size. Thus the
larger the QD, the smaller the contribution of the electron-hole
exchange interaction and consequently the FSS. Therefore, the
large aspect ratio ($\frac{l_q}{h_q}\sim25$) of the here considered
realistic site-controlled (111)-oriented InGaAs/GaAs QDs has to two
consequences. Firstly, the conduction-valence mixing is significantly
reduced. Secondly, the magnitude of the term $EX_2$ is strongly
reduced. All in all, this highlights again that the geometrical
features of realistic site-controlled (111)-oriented InGaAs/GaAs
dots lead to electronic and optical properties that are very
different from standard (001)-oriented InGaAs/GaAs dot
systems. Obviously all this is of benefit for achieving a high
proportion of dots with minimal FSS, as reported in the
literature.~\cite{Juska} We will return to the calculation of
many-body effects, such as the FSS, in more detail below. We note
also that the $p_x$, $p_y$ and $p_z$ states contribute equally to
the hole ground state wave function in the case of the VCA
calculation, while their contributions are slightly different when
including alloy fluctuations. The impact of these fluctuations on
the optical properties of the QDs will be discussed further below.

Table~\ref{tab:alloy} summarizes the energy gap ($E_g$) and the
electron ($\Delta E_{p}^e$) and hole ($\Delta E_{p}^h$)
$p$-state splitting for the different configurations. One can infer
from this table, that in terms of the variation in the energy gap
($E_g$), there is very little difference between the different
random configurations. Thus, with a small number of random
configurations one obtains already reliable insight into the physics
of site-controlled (111)-oriented InGaAs/GaAs QDs. This is in stark
contrast to wurtzite nitride-based alloys containing InN (such as
AlInN or InGaN), where the electronic and optical properties are
strongly influenced by the microscopic structure of the
alloy.~\cite{SSC2015a,SSC2014,SSC2015b} When looking at the
$p$-state splitting, cf. Table~\ref{tab:alloy}, we find that all
three random configurations give a splitting below 1 meV. This
supports that the symmetry of realistically shaped site-controlled
InGaAs/GaAs QDs is only weakly affected by random alloy
fluctuations, consistent with the recent demonstration of their
potential for on demand entangled photon
emitters.~\cite{JuDi2013,JuMu2015}
\begin{table}
\caption{\label{tab:alloy}Energy gap ($ E_g$) , the two highest hole
states energies ($ E_h^0$ and $ E_h^1$), the two lowest electron
energies ($ E_e^0$ and $ E_e^1$) and electron ($\Delta E_{p}^e$) and
hole ($\Delta E_{p}^h$) $p$-states splittings for a
triangular-shaped (111)-oriented In$_{0.25}$Ga$_{0.75}$As/GaAs QD
calculated within (VCA) and for three different random alloy
configurations (Config. 1, 2 and 3).
 The energies (in meV) are given with respect to the bulk GaAs valence band maximum.
}
\begin{tabular*}{\columnwidth}{@{\extracolsep{\fill}}lcccc}
\hline\hline
                & VCA   & Config. 1  &  Config. 2  &  Config. 3   \\
\hline
$E_g$     & 1484.5 & 1427.2 & 1427.8 & 1427.6 \\

$ E_h^0$ &  58.4  & 72.4 &  71.0    &  71.2 \\
$ E_h^1$ & 52.4  & 67.8  &  65.3    &  67.5 \\

$\Delta E_{p}^h$  & 0.48 & 1.3  & 1.0  &  1.9 \\

$ E_e^0$ &  1511.0   &  1499.6   & 1498.8    & 1498.8\\
$ E_e^1$ &   1516.5  &  1504.7   &  1503.9   & 1503.6\\

$\Delta E_{p}^e$  & 0.01 & 0.15  &  0.03  & 0.41 \\
\hline\hline
\end{tabular*}
\end{table}

It is beyond the scope of the present study to carry
out the full many-body calculations required to determine the FSS
accurately; we can however give an overview here what is required
and highlight the challenges of such a calculation for realistically
sized and shaped (111)-oriented site-controlled InGaAs/GaAs QDs.
Given the particularities of the system under consideration,
approaches used for (001)-oriented systems are not directly
applicable here for the following reasons. Usually, to determine the
FSS, configuration interaction schemes are applied.~\cite{BeNa2003,Schliwa2009}
 This involves the calculation of Coulomb matrix
elements describing the attractive electron-hole interaction as well
as the electron-hole exchange interaction.~\cite{Franceschetti,Franceschetti98,Rohlfing} These matrix elements are calculated
from the bound electron and hole single-particle wavefunctions. For
an accurate CI calculation two main factors are important. First, to
construct the many-body Hamiltonian in the basis of anti-symmetrized
products of bound single-particle electron and hole states, a
sufficiently large number of these bound single-particle states is
required. This has been highlighted and discussed in detail by
Wimmer \etal\cite{Wimmer} and Schliwa \etal\cite{Schliwa2009}. For
instance, the data presented in Ref.~\onlinecite{Schliwa2009} shows
for a (001)-oriented InGaAs/GaAs QD that at least 6 bound electron
and 6 bound hole states have to be included to describe correlation
effects accurately. The pseudo-potential calculations by Bester
\etal\cite{BeNa2003} included even more basis states, namely 12
bound electron and 12 bound hole states, to achieve an accurate
value for the FSS in (001)-oriented InAs/GaAs QDs. However, this
approach presents a problem for realistic site-controlled
(111)-oriented InGaAs/GaAs QDs, since we have here a strong
asymmetry between the bound electron and hole states. In fact we
find only three weakly bound electron states. Such a small number of
basis states should make it difficult to apply the standard CI
approache used in (001)-oriented system in a straightforward way.
For other QD systems, such as InGaN/GaN QDs similar arguments and
discussions have been made.~\cite{Bimberg06} One approach to circumvent
problems arising from the small number of bound electron and hole
states is to perform first a self-consistent Hartree-Fock
calculation and use the resulting wave functions as input for CI
calculations.~\cite{Kindel} Since the supercells that have to be considered
for realistic site-controlled (111)-oriented InGaAs/GaAs QDs are
huge ($>$ 5 million atoms), self-consistent calculations are
computationally extremely demanding.

However, leaving the issue of the small number of
bound electron states aside, a further challenge that one encounters here is
the accurate calculation of the Coulomb matrix elements
required for the CI. For an accurate calculation of the
FSS the electron-hole exchange matrix elements need in particular to be calculated
accurately. These matrix elements are usually one order of magnitude
smaller than the direct electron-hole Coulomb matrix
elements.~\cite{Schulz06} Small changes in the values of the
exchange matrix elements might lead to large changes in the FSS,
which is only of the order of $\mu$eV.~\cite{BeNa2003} In the framework
of an empirical tight-binding model, calculations of the Coulomb
matrix elements present a challenge since usually the underlying
atomic-like basis states are not explicitly
known.~\cite{Benchamekh2015} The standard approach is to make
assumptions about the atomic-like basis states (Most often Slater
orbitals are used), and calculate with those states the on-site
Coulomb contributions. For nearest neighbor and more distant contributions
the on-site Coulomb matrix elements are scaled by $1/r$ and weighted
by the tight-binding expansion coefficients.~\cite{Schulz06} Thus
the precise structure of the underlying basis states is neglected
and the Coulomb interaction is treated on the length scale of
lattice vectors. For the direct attractive Coulomb interaction this
approach has been shown to be a good approximation.~\cite{Schulz06}
However, since the electron-hole exchange terms are dominated by
short range contributions, more precise knowledge about the
underlying wavefunctions is required. This has been discussed in
detail by Korkusinski~\etal\cite{Korkusinski}, for instance. For an
accurate treatment of the electron-hole exchange terms more advanced
approaches are required, which fit not only the bulk band structures
but also the wave functions to DFT wave
functions.~\cite{Benchamekh2015} But, using such an approach in
conjunction with a self-consistent Hartree-Fock calculation as input
for CI calculations is beyond the scope of the present work.

\begin{figure}
\includegraphics[width=\columnwidth]{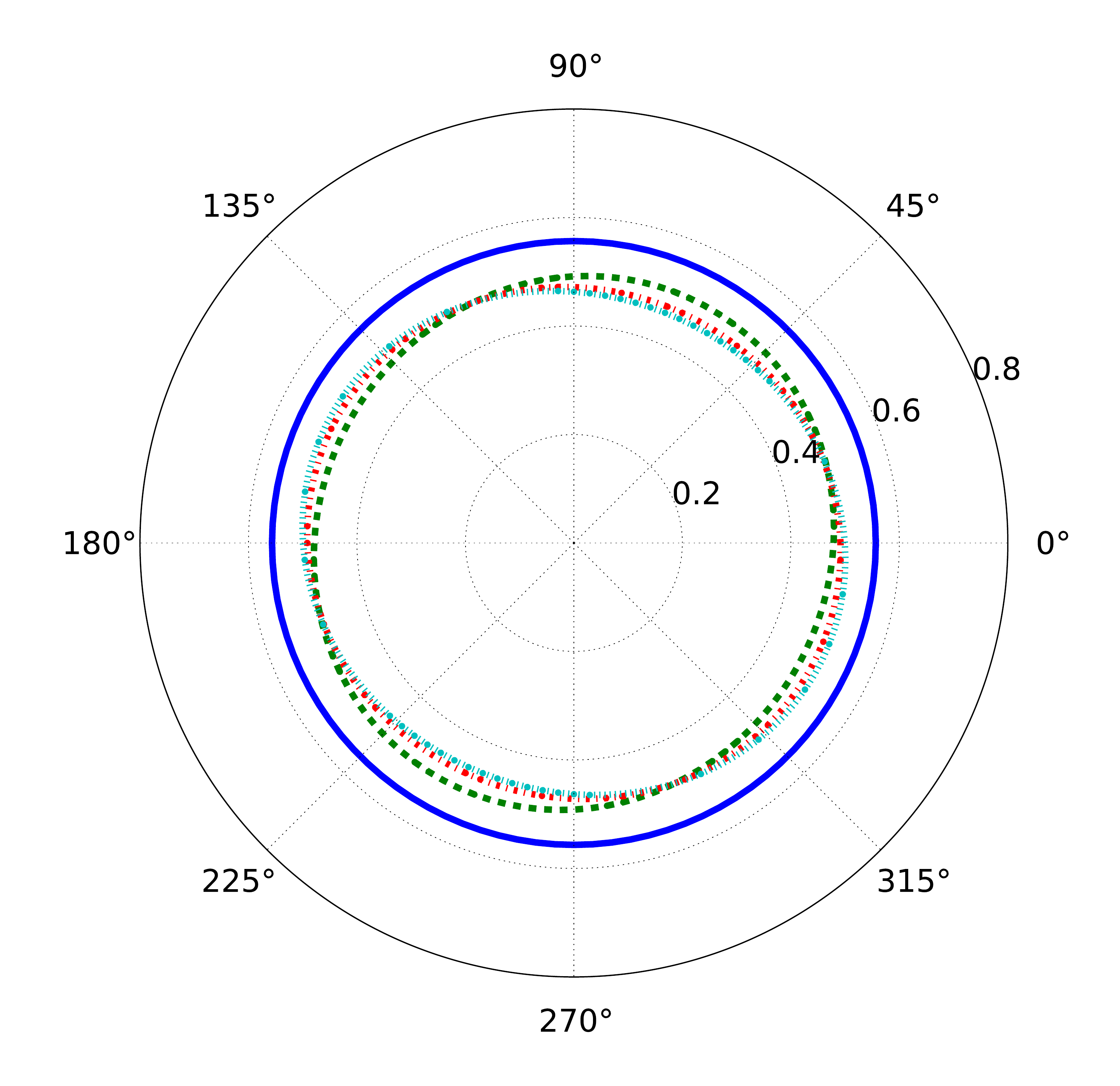}
\caption{Angular dependence of the in-plane optical matrix element (in eV) of a small ($l_q=15$~nm)
In$_{0.25}$Ga$_{0.75}$As/GaAs QD calculated within the VCA (solid blue line) and three different alloyed
configurations (Config.1: dashed green line, Config.2: dashed-dotted red line, Config.3: blue
dotted line)}\label{fig:dipole_alloy}
\end{figure}

Nevertheless, even without the explicit calculation
of the FSS, we can gain insight into key parameters which are
reliable indicators for the symmetry breaking due to random alloy
effects. One of these indicators is the $p$-state splitting
discussed above. As a further measure, we can use
the angular dependent interband transition optical
matrix element, $E_p^{\alpha}$:~\cite{Fishman}
\begin{equation}
E_p^{\alpha}=\frac{2}{m_0} p_{\alpha}^2\,\, .
\end{equation}
Here $p_{\alpha}$ is  the optical momentum component in the
direction $u_{\alpha}$ in the (111) plane and $\alpha $ denotes the
angle between $u_{\alpha}$ and the [$\bar{1}10$]-direction. We have
used the method in Ref.~\citenum{Lew_Yan_Voon} to calculate
$p_\alpha$ between electron and hole ground states for two dot
sizes, firstly with base length $l_q=15$ nm, and then with $l_q=55$
nm. For both systems the height of the dot is kept the same.
Figure~\ref{fig:dipole_alloy} shows $E_p^{\alpha}$ as a function of
the angle $\alpha$ for the VCA results (solid blue line) and the
three different random alloy configurations for the dot with base
length $l_q=15$ nm (Config.1: dashed green line, Config.2:
dashed-dotted red line, Config.3: blue dotted line). Without alloy
fluctuations $E_p^{\alpha}$ is independent of $\alpha$. For the
random alloy configurations we find that $E^\alpha_p$ depends on
$\alpha$. To quantify the impact of the random alloy fluctuations on
the ideally circular symmetric $E^\alpha_p$ we introduce the
following measure:
\begin{equation}
\beta=\left[\text{Max}(E_p^{\alpha})-\text{Min}(E_p^{\alpha})\right]/\text{Min}(E_p^{\alpha})\,
. \label{eq:beta}
\end{equation}
The ratio $\beta$ is equal to 0 for a symmetric system, i.e. a pure
InAs/GaAs QD or an alloyed system treated within VCA. $\beta$
measures exclusively the effect of alloy fluctuations on the QD
symmetry. For the three considered configurations, $\beta $ is equal
to 0.060, 0.055 and 0.109, respectively. These numbers reflect and
confirm the data shown in Fig.~\ref{fig:dipole_alloy}. This angular
asymmetry in the single particle recombination properties will
contribute to an increase in the FSS, reflecting that orthogonal
axes in the QD now have distinct properties.

\begin{figure}
\includegraphics[width=\columnwidth]{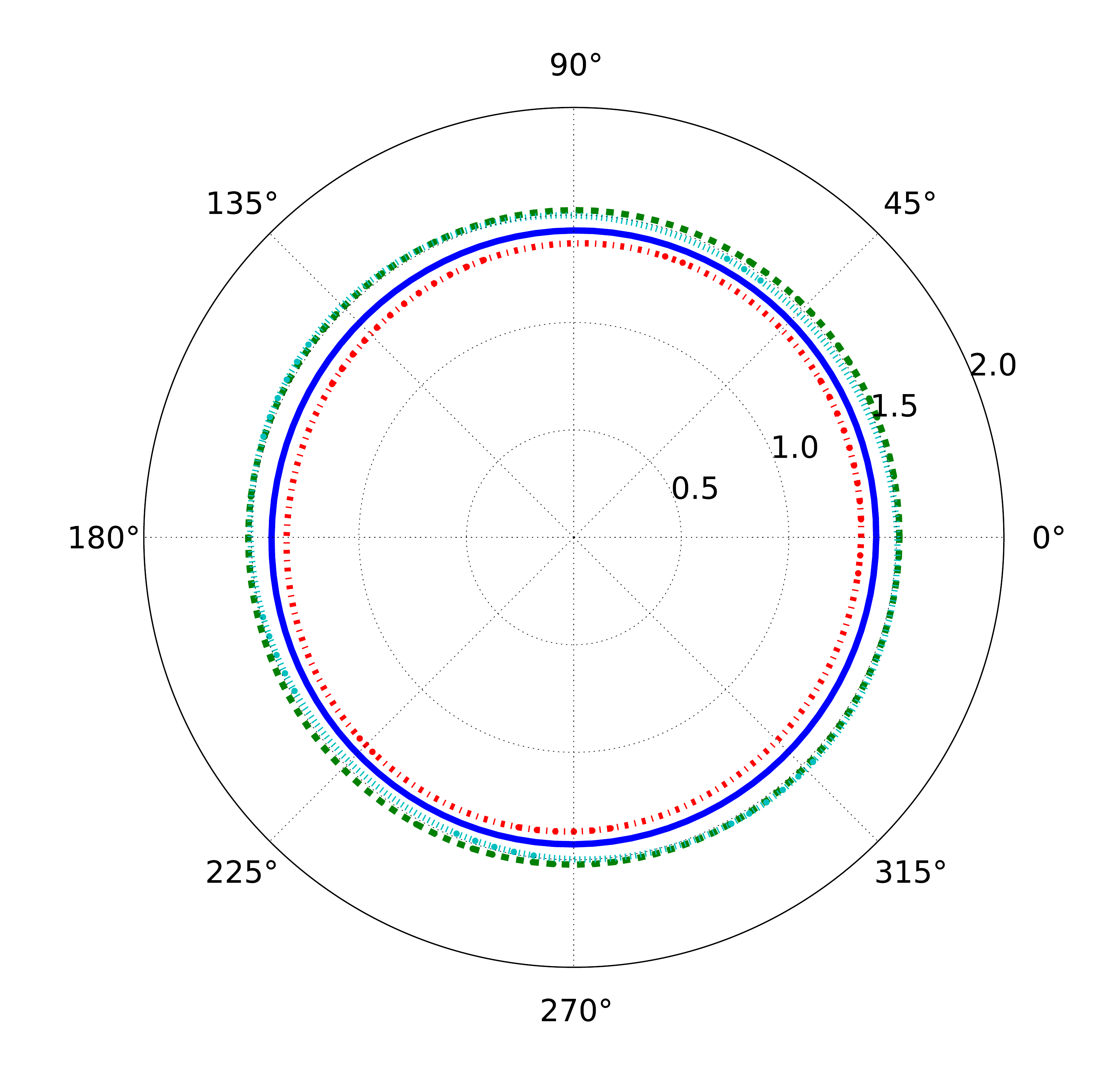}
\caption{Same as in Fig.~\ref{fig:dipole_alloy} for a larger ($l_q=55$~nm) In$_{0.25}$Ga$_{0.75}$As/GaAs QD.
}\label{fig:dipole_alloy1}
\end{figure}

Figure~\ref{fig:dipole_alloy1} shows the angular dependence of
$E_p^{\alpha}$  between electron and hole ground states calculated
within the VCA and for three different random alloy configurations
for an In$_{0.25}$Ga$_{0.75}$As/GaAs QD with the dimensions
$l_q=55$~nm and keeping unchanged the height $h_q=2$~nm. Compared to
Fig.~ \ref{fig:dipole_alloy}, we notice a smaller anisotropy of the
optical momentum matrix element, with $\beta $ respectively equal to
0.006, 0.033 and 0.033, for the three different configurations
considered. This is indicative that the effects of random alloy
fluctuations reduce with increasing QD base size.

Overall, based on both the $p$-state splitting and the optical
momentum matrix elements, our calculations reveal that random alloy
fluctuations affect the electronic and optical properties of
realistic site-controlled InGaAs/GaAs QDs only slightly. Keeping in
mind that the size of the considered QDs is at the lower limit of
the experimental size range (50-80~nm) and that the anisotropy in
$E_p^{\alpha}$ should further reduce with increasing size, our
analysis confirms that site-controlled (111)-oriented InGaAs/GaAs
QDs are promising candidates as future entangled photon sources.
Additionally, since the electron-hole exchange
matrix elements are inversely proportional to the dot size, this
further supports that for even larger dots, the FSS values should
tend to much lower values than in standard (001)-oriented
InGaAs/GaAs dots. This finding is in line with recent experimental
results where an area with up to 15\% of polarization-entangled
photon emitters was obtained.~\cite{JuDi2013}

\section{Conclusion}
\label{sec:Conclusion}

In conclusion, we investigated the optoelectronic properties of
triangular-shaped (111)-oriented InGaAs/GaAs QDs using an
$sp^3d^5s^*$ TB model including local strain field and piezoelectric
potential effects. Overall, we find that the
electronic and optical properties of these systems are vastly
different from standard (001)-oriented InGaAs/GaAs systems.

In contrast with continuum models previously used to described
realistic site-controlled (111)-oriented InGaAs/GaAs QDs, all the
symmetry operations of these structures are well described by our
atomistic model including the absence of the horizontal reflection
plane. This is reflected in the asymmetry of the obtained charge
densities. Furthermore, our atomistic description clarifies the
importance of random alloy fluctuations on the electronic and
optical properties of site-controlled (111)-oriented QDs which
cannot be addressed by continuum-based models. We considered
electron and hole $p$-state splittings as well as the anisotropy in
the interband transition optical matrix element as  measures of the
$C_{3v}$ symmetry reduction. We show that the $p$-state degeneracy
is lifted  either by including spin-orbit coupling or by random
alloy fluctuations. The value of the $p$-state splitting is mainly
governed by the amplitude of the strain due to the induced spin
splitting in the case of realistically sized (111)-oriented
InGaAs/GaAs QD structures. The electron $p$-state splitting remains
smaller than 0.2~meV for all the here considered configurations and
the interband transition anisotropy becomes very small with
increasing dot base size. Also we find that conduction and valence
band mixing effects are strongly reduced in comparison with standard
(001)-oriented InGaAs/GaAs QD systems. This reduced mixing is also
indicative of a reduced FSS in site-controlled (111)-oriented
InGaAs/GaAs QDs.~\cite{krapek} Furthermore, we show that the
anisotropy in $E_p^{\alpha}$ and the $p$-state splitting  reduce with increasing the QD
lateral size. Since the size of the here considered QDs is at the
lower limit of the experimentally realized structures, our results
indicate that the larger QDs are even more promising for achieving
entangled photon generation, and support thus the potential of
site-controlled (111)-oriented InGaAs/GaAs QDs as polarization
entangled photon emitters in future quantum information
applications.

\section*{Acknowledgement}
This work was carried out with the financial support of Science
Foundation Ireland (project number 10/IN.1/I2994). The authors would
like to acknowledge the SFI/HEA Irish Center for High-End Computing
for computational resources.

\bibliography{InGaAs_QD}

\begin{thebibliography}{76}%
\makeatletter
\providecommand \@ifxundefined [1]{%
 \@ifx{#1\undefined}
}%
\providecommand \@ifnum [1]{%
 \ifnum #1\expandafter \@firstoftwo
 \else \expandafter \@secondoftwo
 \fi
}%
\providecommand \@ifx [1]{%
 \ifx #1\expandafter \@firstoftwo
 \else \expandafter \@secondoftwo
 \fi
}%
\providecommand \natexlab [1]{#1}%
\providecommand \enquote  [1]{``#1''}%
\providecommand \bibnamefont  [1]{#1}%
\providecommand \bibfnamefont [1]{#1}%
\providecommand \citenamefont [1]{#1}%
\providecommand \href@noop [0]{\@secondoftwo}%
\providecommand \href [0]{\begingroup \@sanitize@url \@href}%
\providecommand \@href[1]{\@@startlink{#1}\@@href}%
\providecommand \@@href[1]{\endgroup#1\@@endlink}%
\providecommand \@sanitize@url [0]{\catcode `\\12\catcode `\$12\catcode
  `\&12\catcode `\#12\catcode `\^12\catcode `\_12\catcode `\%12\relax}%
\providecommand \@@startlink[1]{}%
\providecommand \@@endlink[0]{}%
\providecommand \url  [0]{\begingroup\@sanitize@url \@url }%
\providecommand \@url [1]{\endgroup\@href {#1}{\urlprefix }}%
\providecommand \urlprefix  [0]{URL }%
\providecommand \Eprint [0]{\href }%
\providecommand \doibase [0]{http://dx.doi.org/}%
\providecommand \selectlanguage [0]{\@gobble}%
\providecommand \bibinfo  [0]{\@secondoftwo}%
\providecommand \bibfield  [0]{\@secondoftwo}%
\providecommand \translation [1]{[#1]}%
\providecommand \BibitemOpen [0]{}%
\providecommand \bibitemStop [0]{}%
\providecommand \bibitemNoStop [0]{.\EOS\space}%
\providecommand \EOS [0]{\spacefactor3000\relax}%
\providecommand \BibitemShut  [1]{\csname bibitem#1\endcsname}%
\let\auto@bib@innerbib\@empty
\bibitem [{\citenamefont {Bouwmeester}\ \emph {et~al.}(2000)\citenamefont
  {Bouwmeester}, \citenamefont {Ekert},\ and\ \citenamefont
  {Zeilinger}}]{BoEk2000}%
  \BibitemOpen
  \bibfield  {author} {\bibinfo {author} {\bibfnamefont {D.}~\bibnamefont
  {Bouwmeester}}, \bibinfo {author} {\bibfnamefont {A.}~\bibnamefont {Ekert}},
  \ and\ \bibinfo {author} {\bibfnamefont {A.}~\bibnamefont {Zeilinger}},\
  }\href@noop {} {\emph {\bibinfo {title} {The Physics of Quantum
  Information}}}\ (\bibinfo  {publisher} {Springer},\ \bibinfo {address}
  {Berlin},\ \bibinfo {year} {2000})\BibitemShut {NoStop}%
\bibitem [{\citenamefont {M\"uller}\ \emph {et~al.}(2014)\citenamefont
  {M\"uller}, \citenamefont {Bounouar}, \citenamefont {J\"ons}, \citenamefont
  {Gl\"assl},\ and\ \citenamefont {Michler}}]{MuBo2014}%
  \BibitemOpen
  \bibfield  {author} {\bibinfo {author} {\bibfnamefont {M.}~\bibnamefont
  {M\"uller}}, \bibinfo {author} {\bibfnamefont {S.}~\bibnamefont {Bounouar}},
  \bibinfo {author} {\bibfnamefont {K.~D.}\ \bibnamefont {J\"ons}}, \bibinfo
  {author} {\bibfnamefont {M.}~\bibnamefont {Gl\"assl}}, \ and\ \bibinfo
  {author} {\bibfnamefont {P.}~\bibnamefont {Michler}},\ }\href@noop {}
  {\bibfield  {journal} {\bibinfo  {journal} {Nat. Photonics}\ }\textbf
  {\bibinfo {volume} {8}},\ \bibinfo {pages} {224} (\bibinfo {year}
  {2014})}\BibitemShut {NoStop}%
\bibitem [{\citenamefont {Benson}\ \emph {et~al.}(2000)\citenamefont {Benson},
  \citenamefont {Santori}, \citenamefont {Pelton},\ and\ \citenamefont
  {Yamamoto}}]{Benson}%
  \BibitemOpen
  \bibfield  {author} {\bibinfo {author} {\bibfnamefont {O.}~\bibnamefont
  {Benson}}, \bibinfo {author} {\bibfnamefont {C.}~\bibnamefont {Santori}},
  \bibinfo {author} {\bibfnamefont {M.}~\bibnamefont {Pelton}}, \ and\ \bibinfo
  {author} {\bibfnamefont {Y.}~\bibnamefont {Yamamoto}},\ }\href {\doibase
  10.1103/PhysRevLett.84.2513} {\bibfield  {journal} {\bibinfo  {journal}
  {Phys. Rev. Lett.}\ }\textbf {\bibinfo {volume} {84}},\ \bibinfo {pages}
  {2513} (\bibinfo {year} {2000})}\BibitemShut {NoStop}%
\bibitem [{\citenamefont {Akopian}\ \emph {et~al.}(2006)\citenamefont
  {Akopian}, \citenamefont {Lindner}, \citenamefont {Poem}, \citenamefont
  {Berlatzky}, \citenamefont {Avron}, \citenamefont {Gershoni}, \citenamefont
  {Gerardot},\ and\ \citenamefont {Petroff}}]{Akopian}%
  \BibitemOpen
  \bibfield  {author} {\bibinfo {author} {\bibfnamefont {N.}~\bibnamefont
  {Akopian}}, \bibinfo {author} {\bibfnamefont {N.~H.}\ \bibnamefont
  {Lindner}}, \bibinfo {author} {\bibfnamefont {E.}~\bibnamefont {Poem}},
  \bibinfo {author} {\bibfnamefont {Y.}~\bibnamefont {Berlatzky}}, \bibinfo
  {author} {\bibfnamefont {J.}~\bibnamefont {Avron}}, \bibinfo {author}
  {\bibfnamefont {D.}~\bibnamefont {Gershoni}}, \bibinfo {author}
  {\bibfnamefont {B.~D.}\ \bibnamefont {Gerardot}}, \ and\ \bibinfo {author}
  {\bibfnamefont {P.~M.}\ \bibnamefont {Petroff}},\ }\href {\doibase
  10.1103/PhysRevLett.96.130501} {\bibfield  {journal} {\bibinfo  {journal}
  {Phys. Rev. Lett.}\ }\textbf {\bibinfo {volume} {96}},\ \bibinfo {pages}
  {130501} (\bibinfo {year} {2006})}\BibitemShut {NoStop}%
\bibitem [{\citenamefont {Plumhof}\ \emph {et~al.}(2012)\citenamefont
  {Plumhof}, \citenamefont {Trotta}, \citenamefont {Rastelli},\ and\
  \citenamefont {Schmidt}}]{22726724}%
  \BibitemOpen
  \bibfield  {author} {\bibinfo {author} {\bibfnamefont {J.}~\bibnamefont
  {Plumhof}}, \bibinfo {author} {\bibfnamefont {R.}~\bibnamefont {Trotta}},
  \bibinfo {author} {\bibfnamefont {A.}~\bibnamefont {Rastelli}}, \ and\
  \bibinfo {author} {\bibfnamefont {O.}~\bibnamefont {Schmidt}},\ }\href
  {\doibase 10.1186/1556-276X-7-336} {\bibfield  {journal} {\bibinfo  {journal}
  {Nanoscale Research Letters}\ }\textbf {\bibinfo {volume} {7}},\ \bibinfo
  {pages} {336} (\bibinfo {year} {2012})}\BibitemShut {NoStop}%
\bibitem [{\citenamefont {Seguin}\ \emph {et~al.}(2005)\citenamefont {Seguin},
  \citenamefont {Schliwa}, \citenamefont {Rodt}, \citenamefont {Potschke},
  \citenamefont {Pohl},\ and\ \citenamefont {Bimberg}}]{SeSc2005}%
  \BibitemOpen
  \bibfield  {author} {\bibinfo {author} {\bibfnamefont {R.}~\bibnamefont
  {Seguin}}, \bibinfo {author} {\bibfnamefont {A.}~\bibnamefont {Schliwa}},
  \bibinfo {author} {\bibfnamefont {S.}~\bibnamefont {Rodt}}, \bibinfo {author}
  {\bibfnamefont {K.}~\bibnamefont {Potschke}}, \bibinfo {author}
  {\bibfnamefont {U.~W.}\ \bibnamefont {Pohl}}, \ and\ \bibinfo {author}
  {\bibfnamefont {D.}~\bibnamefont {Bimberg}},\ }\href@noop {} {\bibfield
  {journal} {\bibinfo  {journal} {Phys. Rev. Lett.}\ }\textbf {\bibinfo
  {volume} {95}},\ \bibinfo {pages} {257402} (\bibinfo {year}
  {2005})}\BibitemShut {NoStop}%
\bibitem [{\citenamefont {Bester}\ \emph {et~al.}(2003)\citenamefont {Bester},
  \citenamefont {Nair},\ and\ \citenamefont {Zunger}}]{BeNa2003}%
  \BibitemOpen
  \bibfield  {author} {\bibinfo {author} {\bibfnamefont {G.}~\bibnamefont
  {Bester}}, \bibinfo {author} {\bibfnamefont {S.}~\bibnamefont {Nair}}, \ and\
  \bibinfo {author} {\bibfnamefont {A.}~\bibnamefont {Zunger}},\ }\href@noop {}
  {\bibfield  {journal} {\bibinfo  {journal} {Phys. Rev. B}\ }\textbf {\bibinfo
  {volume} {67}},\ \bibinfo {pages} {161306(R)} (\bibinfo {year}
  {2003})}\BibitemShut {NoStop}%
\bibitem [{\citenamefont {Luo}\ \emph {et~al.}(2012)\citenamefont {Luo},
  \citenamefont {Singh}, \citenamefont {Zunger},\ and\ \citenamefont
  {Bester}}]{LuSi2012}%
  \BibitemOpen
  \bibfield  {author} {\bibinfo {author} {\bibfnamefont {J.-W.}\ \bibnamefont
  {Luo}}, \bibinfo {author} {\bibfnamefont {R.}~\bibnamefont {Singh}}, \bibinfo
  {author} {\bibfnamefont {A.}~\bibnamefont {Zunger}}, \ and\ \bibinfo {author}
  {\bibfnamefont {G.}~\bibnamefont {Bester}},\ }\href {\doibase
  10.1103/PhysRevB.86.161302} {\bibfield  {journal} {\bibinfo  {journal} {Phys.
  Rev. B}\ }\textbf {\bibinfo {volume} {86}},\ \bibinfo {pages} {161302}
  (\bibinfo {year} {2012})}\BibitemShut {NoStop}%
\bibitem [{\citenamefont {Rastelli}\ \emph {et~al.}(2012)\citenamefont
  {Rastelli}, \citenamefont {Ding}, \citenamefont {Plumhof}, \citenamefont
  {Kumar}, \citenamefont {Trotta}, \citenamefont {Deneke}, \citenamefont
  {Malachias}, \citenamefont {Atkinson}, \citenamefont {Zallo}, \citenamefont
  {Zander}, \citenamefont {Herklotz}, \citenamefont {Singh}, \citenamefont
  {K\v{r}\'{a}pek}, \citenamefont {Schr\"oter}, \citenamefont {Kiravittaya},
  \citenamefont {Benyoucef}, \citenamefont {Hafenbrak}, \citenamefont {J\"ons},
  \citenamefont {Thurmer}, \citenamefont {Grimm}, \citenamefont {Bester},
  \citenamefont {D\"orr}, \citenamefont {Michler},\ and\ \citenamefont
  {Schmidt}}]{RaDi2012}%
  \BibitemOpen
  \bibfield  {author} {\bibinfo {author} {\bibfnamefont {A.}~\bibnamefont
  {Rastelli}}, \bibinfo {author} {\bibfnamefont {F.}~\bibnamefont {Ding}},
  \bibinfo {author} {\bibfnamefont {J.~D.}\ \bibnamefont {Plumhof}}, \bibinfo
  {author} {\bibfnamefont {S.}~\bibnamefont {Kumar}}, \bibinfo {author}
  {\bibfnamefont {R.}~\bibnamefont {Trotta}}, \bibinfo {author} {\bibfnamefont
  {C.}~\bibnamefont {Deneke}}, \bibinfo {author} {\bibfnamefont
  {A.}~\bibnamefont {Malachias}}, \bibinfo {author} {\bibfnamefont
  {P.}~\bibnamefont {Atkinson}}, \bibinfo {author} {\bibfnamefont
  {E.}~\bibnamefont {Zallo}}, \bibinfo {author} {\bibfnamefont
  {T.}~\bibnamefont {Zander}}, \bibinfo {author} {\bibfnamefont
  {A.}~\bibnamefont {Herklotz}}, \bibinfo {author} {\bibfnamefont
  {R.}~\bibnamefont {Singh}}, \bibinfo {author} {\bibfnamefont
  {V.}~\bibnamefont {K\v{r}\'{a}pek}}, \bibinfo {author} {\bibfnamefont
  {J.~R.}\ \bibnamefont {Schr\"oter}}, \bibinfo {author} {\bibfnamefont
  {S.}~\bibnamefont {Kiravittaya}}, \bibinfo {author} {\bibfnamefont
  {M.}~\bibnamefont {Benyoucef}}, \bibinfo {author} {\bibfnamefont
  {R.}~\bibnamefont {Hafenbrak}}, \bibinfo {author} {\bibfnamefont {K.~D.}\
  \bibnamefont {J\"ons}}, \bibinfo {author} {\bibfnamefont {D.~J.}\
  \bibnamefont {Thurmer}}, \bibinfo {author} {\bibfnamefont {D.}~\bibnamefont
  {Grimm}}, \bibinfo {author} {\bibfnamefont {G.}~\bibnamefont {Bester}},
  \bibinfo {author} {\bibfnamefont {K.}~\bibnamefont {D\"orr}}, \bibinfo
  {author} {\bibfnamefont {P.}~\bibnamefont {Michler}}, \ and\ \bibinfo
  {author} {\bibfnamefont {O.~G.}\ \bibnamefont {Schmidt}},\ }\href {\doibase
  10.1002/pssb.201100775} {\bibfield  {journal} {\bibinfo  {journal} {physica
  status solidi (b)}\ }\textbf {\bibinfo {volume} {249}},\ \bibinfo {pages}
  {687} (\bibinfo {year} {2012})}\BibitemShut {NoStop}%
\bibitem [{\citenamefont {Pooley}\ \emph {et~al.}(2014)\citenamefont {Pooley},
  \citenamefont {Bennett}, \citenamefont {Stevenson}, \citenamefont {Shields},
  \citenamefont {Farrer},\ and\ \citenamefont {Ritchie}}]{PoBe2014}%
  \BibitemOpen
  \bibfield  {author} {\bibinfo {author} {\bibfnamefont {M.~A.}\ \bibnamefont
  {Pooley}}, \bibinfo {author} {\bibfnamefont {A.~J.}\ \bibnamefont {Bennett}},
  \bibinfo {author} {\bibfnamefont {R.~M.}\ \bibnamefont {Stevenson}}, \bibinfo
  {author} {\bibfnamefont {A.~J.}\ \bibnamefont {Shields}}, \bibinfo {author}
  {\bibfnamefont {I.}~\bibnamefont {Farrer}}, \ and\ \bibinfo {author}
  {\bibfnamefont {D.~A.}\ \bibnamefont {Ritchie}},\ }\href {\doibase
  10.1103/PhysRevApplied.1.024002} {\bibfield  {journal} {\bibinfo  {journal}
  {Phys. Rev. Applied}\ }\textbf {\bibinfo {volume} {1}},\ \bibinfo {pages}
  {024002} (\bibinfo {year} {2014})}\BibitemShut {NoStop}%
\bibitem [{\citenamefont {Huo}\ \emph {et~al.}(2014)\citenamefont {Huo},
  \citenamefont {Witek}, \citenamefont {Kumar}, \citenamefont {Cardenas},
  \citenamefont {Zhang}, \citenamefont {Akopian}, \citenamefont {Singh},
  \citenamefont {Zallo}, \citenamefont {Grifone}, \citenamefont {Kriegner},
  \citenamefont {Trotta}, \citenamefont {Ding}, \citenamefont {Stangl},
  \citenamefont {Zwiller}, \citenamefont {Bester}, \citenamefont {Rastelli},\
  and\ \citenamefont {Schmidt}}]{HuWi2014}%
  \BibitemOpen
  \bibfield  {author} {\bibinfo {author} {\bibfnamefont {Y.~H.}\ \bibnamefont
  {Huo}}, \bibinfo {author} {\bibfnamefont {B.~J.}\ \bibnamefont {Witek}},
  \bibinfo {author} {\bibfnamefont {S.}~\bibnamefont {Kumar}}, \bibinfo
  {author} {\bibfnamefont {J.~R.}\ \bibnamefont {Cardenas}}, \bibinfo {author}
  {\bibfnamefont {J.~X.}\ \bibnamefont {Zhang}}, \bibinfo {author}
  {\bibfnamefont {N.}~\bibnamefont {Akopian}}, \bibinfo {author} {\bibfnamefont
  {R.}~\bibnamefont {Singh}}, \bibinfo {author} {\bibfnamefont
  {E.}~\bibnamefont {Zallo}}, \bibinfo {author} {\bibfnamefont
  {R.}~\bibnamefont {Grifone}}, \bibinfo {author} {\bibfnamefont
  {D.}~\bibnamefont {Kriegner}}, \bibinfo {author} {\bibfnamefont
  {R.}~\bibnamefont {Trotta}}, \bibinfo {author} {\bibfnamefont
  {F.}~\bibnamefont {Ding}}, \bibinfo {author} {\bibfnamefont {J.}~\bibnamefont
  {Stangl}}, \bibinfo {author} {\bibfnamefont {V.}~\bibnamefont {Zwiller}},
  \bibinfo {author} {\bibfnamefont {G.}~\bibnamefont {Bester}}, \bibinfo
  {author} {\bibfnamefont {A.}~\bibnamefont {Rastelli}}, \ and\ \bibinfo
  {author} {\bibfnamefont {O.~G.}\ \bibnamefont {Schmidt}},\ }\href@noop {}
  {\bibfield  {journal} {\bibinfo  {journal} {Nat. Physics}\ }\textbf {\bibinfo
  {volume} {10}},\ \bibinfo {pages} {46} (\bibinfo {year} {2014})}\BibitemShut
  {NoStop}%
\bibitem [{\citenamefont {Schliwa}\ \emph
  {et~al.}(2009{\natexlab{a}})\citenamefont {Schliwa}, \citenamefont
  {Winkelnkemper}, \citenamefont {Lochmann}, \citenamefont {Stock},\ and\
  \citenamefont {Bimberg}}]{Schliwa}%
  \BibitemOpen
  \bibfield  {author} {\bibinfo {author} {\bibfnamefont {A.}~\bibnamefont
  {Schliwa}}, \bibinfo {author} {\bibfnamefont {M.}~\bibnamefont
  {Winkelnkemper}}, \bibinfo {author} {\bibfnamefont {A.}~\bibnamefont
  {Lochmann}}, \bibinfo {author} {\bibfnamefont {E.}~\bibnamefont {Stock}}, \
  and\ \bibinfo {author} {\bibfnamefont {D.}~\bibnamefont {Bimberg}},\ }\href
  {\doibase 10.1103/PhysRevB.80.161307} {\bibfield  {journal} {\bibinfo
  {journal} {Phys. Rev. B}\ }\textbf {\bibinfo {volume} {80}},\ \bibinfo
  {pages} {161307} (\bibinfo {year} {2009}{\natexlab{a}})}\BibitemShut
  {NoStop}%
\bibitem [{\citenamefont {Singh}\ and\ \citenamefont {Bester}(2009)}]{Singh}%
  \BibitemOpen
  \bibfield  {author} {\bibinfo {author} {\bibfnamefont {R.}~\bibnamefont
  {Singh}}\ and\ \bibinfo {author} {\bibfnamefont {G.}~\bibnamefont {Bester}},\
  }\href {\doibase 10.1103/PhysRevLett.103.063601} {\bibfield  {journal}
  {\bibinfo  {journal} {Phys. Rev. Lett.}\ }\textbf {\bibinfo {volume} {103}},\
  \bibinfo {pages} {063601} (\bibinfo {year} {2009})}\BibitemShut {NoStop}%
\bibitem [{\citenamefont {Juska}\ \emph {et~al.}(2013)\citenamefont {Juska},
  \citenamefont {Dimastrodonato}, \citenamefont {Mereni}, \citenamefont
  {Gocalinska},\ and\ \citenamefont {Pelucchi}}]{JuDi2013}%
  \BibitemOpen
  \bibfield  {author} {\bibinfo {author} {\bibfnamefont {G.}~\bibnamefont
  {Juska}}, \bibinfo {author} {\bibfnamefont {V.}~\bibnamefont
  {Dimastrodonato}}, \bibinfo {author} {\bibfnamefont {L.~O.}\ \bibnamefont
  {Mereni}}, \bibinfo {author} {\bibfnamefont {A.}~\bibnamefont {Gocalinska}},
  \ and\ \bibinfo {author} {\bibfnamefont {E.}~\bibnamefont {Pelucchi}},\
  }\href@noop {} {\bibfield  {journal} {\bibinfo  {journal} {Nat. Photonics}\
  }\textbf {\bibinfo {volume} {7}},\ \bibinfo {pages} {527} (\bibinfo {year}
  {2013})}\BibitemShut {NoStop}%
\bibitem [{\citenamefont {Karlsson}\ \emph {et~al.}(2010)\citenamefont
  {Karlsson}, \citenamefont {Dupertuis}, \citenamefont {Oberli}, \citenamefont
  {Pelucchi}, \citenamefont {Rudra}, \citenamefont {Holtz},\ and\ \citenamefont
  {Kapon}}]{Karlsson}%
  \BibitemOpen
  \bibfield  {author} {\bibinfo {author} {\bibfnamefont {K.~F.}\ \bibnamefont
  {Karlsson}}, \bibinfo {author} {\bibfnamefont {M.~A.}\ \bibnamefont
  {Dupertuis}}, \bibinfo {author} {\bibfnamefont {D.~Y.}\ \bibnamefont
  {Oberli}}, \bibinfo {author} {\bibfnamefont {E.}~\bibnamefont {Pelucchi}},
  \bibinfo {author} {\bibfnamefont {A.}~\bibnamefont {Rudra}}, \bibinfo
  {author} {\bibfnamefont {P.~O.}\ \bibnamefont {Holtz}}, \ and\ \bibinfo
  {author} {\bibfnamefont {E.}~\bibnamefont {Kapon}},\ }\href {\doibase
  10.1103/PhysRevB.81.161307} {\bibfield  {journal} {\bibinfo  {journal} {Phys.
  Rev. B}\ }\textbf {\bibinfo {volume} {81}},\ \bibinfo {pages} {161307}
  (\bibinfo {year} {2010})}\BibitemShut {NoStop}%
\bibitem [{\citenamefont {Juska}\ \emph {et~al.}(2015)\citenamefont {Juska},
  \citenamefont {Murray}, \citenamefont {Dimastrodonato}, \citenamefont
  {Chung}, \citenamefont {Moroni}, \citenamefont {Gocalinska},\ and\
  \citenamefont {Pelucchi}}]{JuMu2015}%
  \BibitemOpen
  \bibfield  {author} {\bibinfo {author} {\bibfnamefont {G.}~\bibnamefont
  {Juska}}, \bibinfo {author} {\bibfnamefont {E.}~\bibnamefont {Murray}},
  \bibinfo {author} {\bibfnamefont {V.}~\bibnamefont {Dimastrodonato}},
  \bibinfo {author} {\bibfnamefont {T.~H.}\ \bibnamefont {Chung}}, \bibinfo
  {author} {\bibfnamefont {S.~T.}\ \bibnamefont {Moroni}}, \bibinfo {author}
  {\bibfnamefont {A.}~\bibnamefont {Gocalinska}}, \ and\ \bibinfo {author}
  {\bibfnamefont {E.}~\bibnamefont {Pelucchi}},\ }\href@noop {} {\bibfield
  {journal} {\bibinfo  {journal} {J. Appl. Phys}\ }\textbf {\bibinfo {volume}
  {117}},\ \bibinfo {pages} {134302} (\bibinfo {year} {2015})}\BibitemShut
  {NoStop}%
\bibitem [{\citenamefont {Wang}\ \emph {et~al.}(2015)\citenamefont {Wang},
  \citenamefont {Gong}, \citenamefont {Guo},\ and\ \citenamefont
  {He}}]{Wang2015}%
  \BibitemOpen
  \bibfield  {author} {\bibinfo {author} {\bibfnamefont {J.}~\bibnamefont
  {Wang}}, \bibinfo {author} {\bibfnamefont {M.}~\bibnamefont {Gong}}, \bibinfo
  {author} {\bibfnamefont {G.-C.}\ \bibnamefont {Guo}}, \ and\ \bibinfo
  {author} {\bibfnamefont {L.}~\bibnamefont {He}},\ }\href {\doibase
  10.1103/PhysRevLett.115.067401} {\bibfield  {journal} {\bibinfo  {journal}
  {Phys. Rev. Lett.}\ }\textbf {\bibinfo {volume} {115}},\ \bibinfo {pages}
  {067401} (\bibinfo {year} {2015})}\BibitemShut {NoStop}%
\bibitem [{\citenamefont {Healy}\ \emph {et~al.}(2010)\citenamefont {Healy},
  \citenamefont {Young}, \citenamefont {Mereni}, \citenamefont
  {Dimastrodonato}, \citenamefont {Pelucchi},\ and\ \citenamefont
  {O'Reilly}}]{HeYo2010}%
  \BibitemOpen
  \bibfield  {author} {\bibinfo {author} {\bibfnamefont {S.~B.}\ \bibnamefont
  {Healy}}, \bibinfo {author} {\bibfnamefont {R.~J.}\ \bibnamefont {Young}},
  \bibinfo {author} {\bibfnamefont {L.~O.}\ \bibnamefont {Mereni}}, \bibinfo
  {author} {\bibfnamefont {V.}~\bibnamefont {Dimastrodonato}}, \bibinfo
  {author} {\bibfnamefont {E.}~\bibnamefont {Pelucchi}}, \ and\ \bibinfo
  {author} {\bibfnamefont {E.~P.}\ \bibnamefont {O'Reilly}},\ }\href@noop {}
  {\bibfield  {journal} {\bibinfo  {journal} {Physica E (Amsterdam)}\ }\textbf
  {\bibinfo {volume} {42}},\ \bibinfo {pages} {2761} (\bibinfo {year}
  {2010})}\BibitemShut {NoStop}%
\bibitem [{\citenamefont {Mereni}\ \emph {et~al.}(2009)\citenamefont {Mereni},
  \citenamefont {Dimastrodonato}, \citenamefont {Young},\ and\ \citenamefont
  {Pelucchi}}]{MeDi2009}%
  \BibitemOpen
  \bibfield  {author} {\bibinfo {author} {\bibfnamefont {L.~O.}\ \bibnamefont
  {Mereni}}, \bibinfo {author} {\bibfnamefont {V.}~\bibnamefont
  {Dimastrodonato}}, \bibinfo {author} {\bibfnamefont {R.~J.}\ \bibnamefont
  {Young}}, \ and\ \bibinfo {author} {\bibfnamefont {E.}~\bibnamefont
  {Pelucchi}},\ }\href@noop {} {\bibfield  {journal} {\bibinfo  {journal}
  {Appl. Phys. Lett.}\ }\textbf {\bibinfo {volume} {94}},\ \bibinfo {pages}
  {223121} (\bibinfo {year} {2009})}\BibitemShut {NoStop}%
\bibitem [{\citenamefont {Marquardt}\ \emph {et~al.}(2014)\citenamefont
  {Marquardt}, \citenamefont {O'Reilly},\ and\ \citenamefont
  {Schulz}}]{MaORe2014}%
  \BibitemOpen
  \bibfield  {author} {\bibinfo {author} {\bibfnamefont {O.}~\bibnamefont
  {Marquardt}}, \bibinfo {author} {\bibfnamefont {E.~P.}\ \bibnamefont
  {O'Reilly}}, \ and\ \bibinfo {author} {\bibfnamefont {S.}~\bibnamefont
  {Schulz}},\ }\href@noop {} {\bibfield  {journal} {\bibinfo  {journal} {J.
  Phys.: Condens. Matter}\ }\textbf {\bibinfo {volume} {26}},\ \bibinfo {pages}
  {035303} (\bibinfo {year} {2014})}\BibitemShut {NoStop}%
\bibitem [{\citenamefont {K\ifmmode~\check{r}\else \v{r}\fi{}\'apek}\ \emph
  {et~al.}(2015)\citenamefont {K\ifmmode~\check{r}\else \v{r}\fi{}\'apek},
  \citenamefont {Klenovsk\'y},\ and\ \citenamefont {\ifmmode~\check{S}\else
  \v{S}\fi{}ikola}}]{krapek}%
  \BibitemOpen
  \bibfield  {author} {\bibinfo {author} {\bibfnamefont {V.}~\bibnamefont
  {K\ifmmode~\check{r}\else \v{r}\fi{}\'apek}}, \bibinfo {author}
  {\bibfnamefont {P.}~\bibnamefont {Klenovsk\'y}}, \ and\ \bibinfo {author}
  {\bibfnamefont {T.}~\bibnamefont {\ifmmode~\check{S}\else \v{S}\fi{}ikola}},\
  }\href {\doibase 10.1103/PhysRevB.92.195430} {\bibfield  {journal} {\bibinfo
  {journal} {Phys. Rev. B}\ }\textbf {\bibinfo {volume} {92}},\ \bibinfo
  {pages} {195430} (\bibinfo {year} {2015})}\BibitemShut {NoStop}%
\bibitem [{\citenamefont {Bester}\ and\ \citenamefont
  {Zunger}(2005)}]{Bester2005}%
  \BibitemOpen
  \bibfield  {author} {\bibinfo {author} {\bibfnamefont {G.}~\bibnamefont
  {Bester}}\ and\ \bibinfo {author} {\bibfnamefont {A.}~\bibnamefont
  {Zunger}},\ }\href {\doibase 10.1103/PhysRevB.71.045318} {\bibfield
  {journal} {\bibinfo  {journal} {Phys. Rev. B}\ }\textbf {\bibinfo {volume}
  {71}},\ \bibinfo {pages} {045318} (\bibinfo {year} {2005})}\BibitemShut
  {NoStop}%
\bibitem [{\citenamefont {Keating}(1966)}]{Keating}%
  \BibitemOpen
  \bibfield  {author} {\bibinfo {author} {\bibfnamefont {P.~N.}\ \bibnamefont
  {Keating}},\ }\href {\doibase 10.1103/PhysRev.145.637} {\bibfield  {journal}
  {\bibinfo  {journal} {Phys. Rev.}\ }\textbf {\bibinfo {volume} {145}},\
  \bibinfo {pages} {637} (\bibinfo {year} {1966})}\BibitemShut {NoStop}%
\bibitem [{\citenamefont {Vurgaftman}\ \emph {et~al.}(2001)\citenamefont
  {Vurgaftman}, \citenamefont {Meyer},\ and\ \citenamefont
  {Ram-Mohan}}]{Ram-mohan}%
  \BibitemOpen
  \bibfield  {author} {\bibinfo {author} {\bibfnamefont {I.}~\bibnamefont
  {Vurgaftman}}, \bibinfo {author} {\bibfnamefont {J.~R.}\ \bibnamefont
  {Meyer}}, \ and\ \bibinfo {author} {\bibfnamefont {L.~R.}\ \bibnamefont
  {Ram-Mohan}},\ }\href@noop {} {\bibfield  {journal} {\bibinfo  {journal}
  {Journal of Applied Physics}\ }\textbf {\bibinfo {volume} {89}} (\bibinfo
  {year} {2001})}\BibitemShut {NoStop}%
\bibitem [{\citenamefont {Nye}(1985)}]{Nye85}%
  \BibitemOpen
  \bibfield  {author} {\bibinfo {author} {\bibfnamefont {J.~F.}\ \bibnamefont
  {Nye}},\ }\href@noop {} {\emph {\bibinfo {title} {Physical Properties of
  Crystals: Their Representation by Tensors and Matrices}}}\ (\bibinfo
  {publisher} {Oxford University Press},\ \bibinfo {address} {New York},\
  \bibinfo {year} {1985})\BibitemShut {NoStop}%
\bibitem [{\citenamefont {Grimmer}(2007)}]{Grim2007}%
  \BibitemOpen
  \bibfield  {author} {\bibinfo {author} {\bibfnamefont {H.}~\bibnamefont
  {Grimmer}},\ }\href@noop {} {\bibfield  {journal} {\bibinfo  {journal} {Acta
  Cryst.}\ }\textbf {\bibinfo {volume} {A63}},\ \bibinfo {pages} {441}
  (\bibinfo {year} {2007})}\BibitemShut {NoStop}%
\bibitem [{\citenamefont {Smith}\ and\ \citenamefont
  {Mailhiot}(1990)}]{Smith90}%
  \BibitemOpen
  \bibfield  {author} {\bibinfo {author} {\bibfnamefont {D.~L.}\ \bibnamefont
  {Smith}}\ and\ \bibinfo {author} {\bibfnamefont {C.}~\bibnamefont
  {Mailhiot}},\ }\href {\doibase 10.1103/RevModPhys.62.173} {\bibfield
  {journal} {\bibinfo  {journal} {Rev. Mod. Phys.}\ }\textbf {\bibinfo {volume}
  {62}},\ \bibinfo {pages} {173} (\bibinfo {year} {1990})}\BibitemShut
  {NoStop}%
\bibitem [{\citenamefont {Niquet}(2006)}]{Niquet06}%
  \BibitemOpen
  \bibfield  {author} {\bibinfo {author} {\bibfnamefont {Y.~M.}\ \bibnamefont
  {Niquet}},\ }\href {\doibase 10.1103/PhysRevB.74.155304} {\bibfield
  {journal} {\bibinfo  {journal} {Phys. Rev. B}\ }\textbf {\bibinfo {volume}
  {74}},\ \bibinfo {pages} {155304} (\bibinfo {year} {2006})}\BibitemShut
  {NoStop}%
\bibitem [{\citenamefont {Beya-Wakata}\ \emph {et~al.}(2011)\citenamefont
  {Beya-Wakata}, \citenamefont {Prodhomme},\ and\ \citenamefont
  {Bester}}]{Bester2011}%
  \BibitemOpen
  \bibfield  {author} {\bibinfo {author} {\bibfnamefont {A.}~\bibnamefont
  {Beya-Wakata}}, \bibinfo {author} {\bibfnamefont {P.-Y.}\ \bibnamefont
  {Prodhomme}}, \ and\ \bibinfo {author} {\bibfnamefont {G.}~\bibnamefont
  {Bester}},\ }\href {\doibase 10.1103/PhysRevB.84.195207} {\bibfield
  {journal} {\bibinfo  {journal} {Phys. Rev. B}\ }\textbf {\bibinfo {volume}
  {84}},\ \bibinfo {pages} {195207} (\bibinfo {year} {2011})}\BibitemShut
  {NoStop}%
\bibitem [{\citenamefont {Smith}(1986)}]{Smith86}%
  \BibitemOpen
  \bibfield  {author} {\bibinfo {author} {\bibfnamefont {D.}~\bibnamefont
  {Smith}},\ }\href {\doibase http://dx.doi.org/10.1016/0038-1098(86)90924-5}
  {\bibfield  {journal} {\bibinfo  {journal} {Solid State Communications}\
  }\textbf {\bibinfo {volume} {57}},\ \bibinfo {pages} {919 } (\bibinfo {year}
  {1986})}\BibitemShut {NoStop}%
\bibitem [{\citenamefont {Smith}\ and\ \citenamefont
  {Mailhiot}(1987{\natexlab{a}})}]{SmithJVST87}%
  \BibitemOpen
  \bibfield  {author} {\bibinfo {author} {\bibfnamefont {D.~L.}\ \bibnamefont
  {Smith}}\ and\ \bibinfo {author} {\bibfnamefont {C.}~\bibnamefont
  {Mailhiot}},\ }\href {\doibase http://dx.doi.org/10.1116/1.574919} {\bibfield
   {journal} {\bibinfo  {journal} {Journal of Vacuum Science \& Technology A}\
  }\textbf {\bibinfo {volume} {5}},\ \bibinfo {pages} {2060} (\bibinfo {year}
  {1987}{\natexlab{a}})}\BibitemShut {NoStop}%
\bibitem [{\citenamefont {Mailhiot}\ and\ \citenamefont
  {Smith}(1987)}]{SmithPRB87}%
  \BibitemOpen
  \bibfield  {author} {\bibinfo {author} {\bibfnamefont {C.}~\bibnamefont
  {Mailhiot}}\ and\ \bibinfo {author} {\bibfnamefont {D.~L.}\ \bibnamefont
  {Smith}},\ }\href {\doibase 10.1103/PhysRevB.35.1242} {\bibfield  {journal}
  {\bibinfo  {journal} {Phys. Rev. B}\ }\textbf {\bibinfo {volume} {35}},\
  \bibinfo {pages} {1242} (\bibinfo {year} {1987})}\BibitemShut {NoStop}%
\bibitem [{\citenamefont {Smith}\ and\ \citenamefont
  {Mailhiot}(1987{\natexlab{b}})}]{SmithPRL87}%
  \BibitemOpen
  \bibfield  {author} {\bibinfo {author} {\bibfnamefont {D.~L.}\ \bibnamefont
  {Smith}}\ and\ \bibinfo {author} {\bibfnamefont {C.}~\bibnamefont
  {Mailhiot}},\ }\href {\doibase 10.1103/PhysRevLett.58.1264} {\bibfield
  {journal} {\bibinfo  {journal} {Phys. Rev. Lett.}\ }\textbf {\bibinfo
  {volume} {58}},\ \bibinfo {pages} {1264} (\bibinfo {year}
  {1987}{\natexlab{b}})}\BibitemShut {NoStop}%
\bibitem [{\citenamefont {Mailhiot}\ and\ \citenamefont
  {Smith}(1989)}]{Smith89}%
  \BibitemOpen
  \bibfield  {author} {\bibinfo {author} {\bibfnamefont {C.}~\bibnamefont
  {Mailhiot}}\ and\ \bibinfo {author} {\bibfnamefont {D.~L.}\ \bibnamefont
  {Smith}},\ }\href {\doibase http://dx.doi.org/10.1116/1.575897} {\bibfield
  {journal} {\bibinfo  {journal} {Journal of Vacuum Science \& Technology A}\
  }\textbf {\bibinfo {volume} {7}},\ \bibinfo {pages} {609} (\bibinfo {year}
  {1989})}\BibitemShut {NoStop}%
\bibitem [{\citenamefont {Smith}\ and\ \citenamefont
  {Mailhiot}(1988)}]{Smith88}%
  \BibitemOpen
  \bibfield  {author} {\bibinfo {author} {\bibfnamefont {D.~L.}\ \bibnamefont
  {Smith}}\ and\ \bibinfo {author} {\bibfnamefont {C.}~\bibnamefont
  {Mailhiot}},\ }\href {\doibase http://dx.doi.org/10.1063/1.340965} {\bibfield
   {journal} {\bibinfo  {journal} {Journal of Applied Physics}\ }\textbf
  {\bibinfo {volume} {63}},\ \bibinfo {pages} {2717} (\bibinfo {year}
  {1988})}\BibitemShut {NoStop}%
\bibitem [{\citenamefont {Bester}\ \emph {et~al.}(2006)\citenamefont {Bester},
  \citenamefont {Zunger}, \citenamefont {Wu},\ and\ \citenamefont
  {Vanderbilt}}]{Bester06}%
  \BibitemOpen
  \bibfield  {author} {\bibinfo {author} {\bibfnamefont {G.}~\bibnamefont
  {Bester}}, \bibinfo {author} {\bibfnamefont {A.}~\bibnamefont {Zunger}},
  \bibinfo {author} {\bibfnamefont {X.}~\bibnamefont {Wu}}, \ and\ \bibinfo
  {author} {\bibfnamefont {D.}~\bibnamefont {Vanderbilt}},\ }\href {\doibase
  10.1103/PhysRevB.74.081305} {\bibfield  {journal} {\bibinfo  {journal} {Phys.
  Rev. B}\ }\textbf {\bibinfo {volume} {74}},\ \bibinfo {pages} {081305}
  (\bibinfo {year} {2006})}\BibitemShut {NoStop}%
\bibitem [{\citenamefont {Schliwa}\ \emph {et~al.}(2007)\citenamefont
  {Schliwa}, \citenamefont {Winkelnkemper},\ and\ \citenamefont
  {Bimberg}}]{ScWi2007}%
  \BibitemOpen
  \bibfield  {author} {\bibinfo {author} {\bibfnamefont {A.}~\bibnamefont
  {Schliwa}}, \bibinfo {author} {\bibfnamefont {M.}~\bibnamefont
  {Winkelnkemper}}, \ and\ \bibinfo {author} {\bibfnamefont {D.}~\bibnamefont
  {Bimberg}},\ }\href@noop {} {\bibfield  {journal} {\bibinfo  {journal} {Phys.
  Rev. B}\ }\textbf {\bibinfo {volume} {76}},\ \bibinfo {pages} {205324}
  (\bibinfo {year} {2007})}\BibitemShut {NoStop}%
\bibitem [{\citenamefont {Schulz}\ \emph {et~al.}(2011)\citenamefont {Schulz},
  \citenamefont {Caro}, \citenamefont {O'Reilly},\ and\ \citenamefont
  {Marquardt}}]{Schulz11}%
  \BibitemOpen
  \bibfield  {author} {\bibinfo {author} {\bibfnamefont {S.}~\bibnamefont
  {Schulz}}, \bibinfo {author} {\bibfnamefont {M.~A.}\ \bibnamefont {Caro}},
  \bibinfo {author} {\bibfnamefont {E.~P.}\ \bibnamefont {O'Reilly}}, \ and\
  \bibinfo {author} {\bibfnamefont {O.}~\bibnamefont {Marquardt}},\ }\href
  {\doibase 10.1103/PhysRevB.84.125312} {\bibfield  {journal} {\bibinfo
  {journal} {Phys. Rev. B}\ }\textbf {\bibinfo {volume} {84}},\ \bibinfo
  {pages} {125312} (\bibinfo {year} {2011})}\BibitemShut {NoStop}%
\bibitem [{\citenamefont {Caro}\ \emph {et~al.}(2015)\citenamefont {Caro},
  \citenamefont {Schulz},\ and\ \citenamefont {O'Reilly}}]{CaSc2015}%
  \BibitemOpen
  \bibfield  {author} {\bibinfo {author} {\bibfnamefont {M.~A.}\ \bibnamefont
  {Caro}}, \bibinfo {author} {\bibfnamefont {S.}~\bibnamefont {Schulz}}, \ and\
  \bibinfo {author} {\bibfnamefont {E.~P.}\ \bibnamefont {O'Reilly}},\
  }\href@noop {} {\bibfield  {journal} {\bibinfo  {journal} {Phys. Rev. B}\
  }\textbf {\bibinfo {volume} {91}},\ \bibinfo {pages} {075203} (\bibinfo
  {year} {2015})}\BibitemShut {NoStop}%
\bibitem [{\citenamefont {Jancu}\ \emph {et~al.}(1998)\citenamefont {Jancu},
  \citenamefont {Scholz}, \citenamefont {Beltram},\ and\ \citenamefont
  {Bassani}}]{Jancu98}%
  \BibitemOpen
  \bibfield  {author} {\bibinfo {author} {\bibfnamefont {J.-M.}\ \bibnamefont
  {Jancu}}, \bibinfo {author} {\bibfnamefont {R.}~\bibnamefont {Scholz}},
  \bibinfo {author} {\bibfnamefont {F.}~\bibnamefont {Beltram}}, \ and\
  \bibinfo {author} {\bibfnamefont {F.}~\bibnamefont {Bassani}},\ }\href
  {\doibase 10.1103/PhysRevB.57.6493} {\bibfield  {journal} {\bibinfo
  {journal} {Phys. Rev. B}\ }\textbf {\bibinfo {volume} {57}},\ \bibinfo
  {pages} {6493} (\bibinfo {year} {1998})}\BibitemShut {NoStop}%
\bibitem [{\citenamefont {Raouafi}\ \emph {et~al.}(2016)\citenamefont
  {Raouafi}, \citenamefont {Benchamekh}, \citenamefont {Nestoklon},
  \citenamefont {Jancu},\ and\ \citenamefont {Voisin}}]{Raouafi}%
  \BibitemOpen
  \bibfield  {author} {\bibinfo {author} {\bibfnamefont {F.}~\bibnamefont
  {Raouafi}}, \bibinfo {author} {\bibfnamefont {R.}~\bibnamefont {Benchamekh}},
  \bibinfo {author} {\bibfnamefont {M.~O.}\ \bibnamefont {Nestoklon}}, \bibinfo
  {author} {\bibfnamefont {J.-M.}\ \bibnamefont {Jancu}}, \ and\ \bibinfo
  {author} {\bibfnamefont {P.}~\bibnamefont {Voisin}},\ }\href
  {http://stacks.iop.org/0953-8984/28/i=4/a=045001} {\bibfield  {journal}
  {\bibinfo  {journal} {Journal of Physics: Condensed Matter}\ }\textbf
  {\bibinfo {volume} {28}},\ \bibinfo {pages} {045001} (\bibinfo {year}
  {2016})}\BibitemShut {NoStop}%
\bibitem [{\citenamefont {Soucail}\ \emph {et~al.}(1990)\citenamefont
  {Soucail}, \citenamefont {Dupuis}, \citenamefont {Ferreira}, \citenamefont
  {Voisin}, \citenamefont {Roth}, \citenamefont {Morris}, \citenamefont
  {Gibb},\ and\ \citenamefont {Lacelle}}]{Soucail}%
  \BibitemOpen
  \bibfield  {author} {\bibinfo {author} {\bibfnamefont {B.}~\bibnamefont
  {Soucail}}, \bibinfo {author} {\bibfnamefont {N.}~\bibnamefont {Dupuis}},
  \bibinfo {author} {\bibfnamefont {R.}~\bibnamefont {Ferreira}}, \bibinfo
  {author} {\bibfnamefont {P.}~\bibnamefont {Voisin}}, \bibinfo {author}
  {\bibfnamefont {A.~P.}\ \bibnamefont {Roth}}, \bibinfo {author}
  {\bibfnamefont {D.}~\bibnamefont {Morris}}, \bibinfo {author} {\bibfnamefont
  {K.}~\bibnamefont {Gibb}}, \ and\ \bibinfo {author} {\bibfnamefont
  {C.}~\bibnamefont {Lacelle}},\ }\href {\doibase 10.1103/PhysRevB.41.8568}
  {\bibfield  {journal} {\bibinfo  {journal} {Phys. Rev. B}\ }\textbf {\bibinfo
  {volume} {41}},\ \bibinfo {pages} {8568} (\bibinfo {year}
  {1990})}\BibitemShut {NoStop}%
\bibitem [{\citenamefont {Jancu}\ and\ \citenamefont {Voisin}(2007)}]{Jancu07}%
  \BibitemOpen
  \bibfield  {author} {\bibinfo {author} {\bibfnamefont {J.-M.}\ \bibnamefont
  {Jancu}}\ and\ \bibinfo {author} {\bibfnamefont {P.}~\bibnamefont {Voisin}},\
  }\href {\doibase 10.1103/PhysRevB.76.115202} {\bibfield  {journal} {\bibinfo
  {journal} {Phys. Rev. B}\ }\textbf {\bibinfo {volume} {76}},\ \bibinfo
  {pages} {115202} (\bibinfo {year} {2007})}\BibitemShut {NoStop}%
\bibitem [{\citenamefont {Niquet}\ \emph {et~al.}(2009)\citenamefont {Niquet},
  \citenamefont {Rideau}, \citenamefont {Tavernier}, \citenamefont {Jaouen},\
  and\ \citenamefont {Blase}}]{Niquet}%
  \BibitemOpen
  \bibfield  {author} {\bibinfo {author} {\bibfnamefont {Y.~M.}\ \bibnamefont
  {Niquet}}, \bibinfo {author} {\bibfnamefont {D.}~\bibnamefont {Rideau}},
  \bibinfo {author} {\bibfnamefont {C.}~\bibnamefont {Tavernier}}, \bibinfo
  {author} {\bibfnamefont {H.}~\bibnamefont {Jaouen}}, \ and\ \bibinfo {author}
  {\bibfnamefont {X.}~\bibnamefont {Blase}},\ }\href {\doibase
  10.1103/PhysRevB.79.245201} {\bibfield  {journal} {\bibinfo  {journal} {Phys.
  Rev. B}\ }\textbf {\bibinfo {volume} {79}},\ \bibinfo {pages} {245201}
  (\bibinfo {year} {2009})}\BibitemShut {NoStop}%
\bibitem [{\citenamefont {Boykin}\ and\ \citenamefont {Vogl}(2001)}]{Boykin}%
  \BibitemOpen
  \bibfield  {author} {\bibinfo {author} {\bibfnamefont {T.~B.}\ \bibnamefont
  {Boykin}}\ and\ \bibinfo {author} {\bibfnamefont {P.}~\bibnamefont {Vogl}},\
  }\href@noop {} {\bibfield  {journal} {\bibinfo  {journal} {Phys.\ Rev.\ B}\
  }\textbf {\bibinfo {volume} {65}},\ \bibinfo {pages} {035202} (\bibinfo
  {year} {2001})}\BibitemShut {NoStop}%
\bibitem [{\citenamefont {Zieli\ifmmode~\acute{n}\else
  \'{n}\fi{}ski}(2012)}]{Zielinski}%
  \BibitemOpen
  \bibfield  {author} {\bibinfo {author} {\bibfnamefont {M.}~\bibnamefont
  {Zieli\ifmmode~\acute{n}\else \'{n}\fi{}ski}},\ }\href {\doibase
  10.1103/PhysRevB.86.115424} {\bibfield  {journal} {\bibinfo  {journal} {Phys.
  Rev. B}\ }\textbf {\bibinfo {volume} {86}},\ \bibinfo {pages} {115424}
  (\bibinfo {year} {2012})}\BibitemShut {NoStop}%
\bibitem [{\citenamefont {Ren}\ \emph {et~al.}(1982)\citenamefont {Ren},
  \citenamefont {Dow},\ and\ \citenamefont {Wolford}}]{Ren}%
  \BibitemOpen
  \bibfield  {author} {\bibinfo {author} {\bibfnamefont {S.~Y.}\ \bibnamefont
  {Ren}}, \bibinfo {author} {\bibfnamefont {J.~D.}\ \bibnamefont {Dow}}, \ and\
  \bibinfo {author} {\bibfnamefont {D.~J.}\ \bibnamefont {Wolford}},\ }\href
  {\doibase 10.1103/PhysRevB.25.7661} {\bibfield  {journal} {\bibinfo
  {journal} {Phys. Rev. B}\ }\textbf {\bibinfo {volume} {25}},\ \bibinfo
  {pages} {7661} (\bibinfo {year} {1982})}\BibitemShut {NoStop}%
\bibitem [{\citenamefont {Ranjan}\ \emph {et~al.}(2003)\citenamefont {Ranjan},
  \citenamefont {Allan}, \citenamefont {Priester},\ and\ \citenamefont
  {Delerue}}]{RaAl2003}%
  \BibitemOpen
  \bibfield  {author} {\bibinfo {author} {\bibfnamefont {V.}~\bibnamefont
  {Ranjan}}, \bibinfo {author} {\bibfnamefont {G.}~\bibnamefont {Allan}},
  \bibinfo {author} {\bibfnamefont {C.}~\bibnamefont {Priester}}, \ and\
  \bibinfo {author} {\bibfnamefont {C.}~\bibnamefont {Delerue}},\ }\href@noop
  {} {\bibfield  {journal} {\bibinfo  {journal} {Physical Review B}\ }\textbf
  {\bibinfo {volume} {68}},\ \bibinfo {pages} {115305} (\bibinfo {year}
  {2003})}\BibitemShut {NoStop}%
\bibitem [{\citenamefont {Zielinski}\ \emph {et~al.}(2005)\citenamefont
  {Zielinski}, \citenamefont {Jaskolski}, \citenamefont {Aizpurua},\ and\
  \citenamefont {Bryant}}]{ZiJa2005}%
  \BibitemOpen
  \bibfield  {author} {\bibinfo {author} {\bibfnamefont {M.}~\bibnamefont
  {Zielinski}}, \bibinfo {author} {\bibfnamefont {W.}~\bibnamefont
  {Jaskolski}}, \bibinfo {author} {\bibfnamefont {J.}~\bibnamefont {Aizpurua}},
  \ and\ \bibinfo {author} {\bibfnamefont {G.~W.}\ \bibnamefont {Bryant}},\
  }\href@noop {} {\bibfield  {journal} {\bibinfo  {journal} {Acta Physica
  Polonica A}\ }\textbf {\bibinfo {volume} {108}},\ \bibinfo {pages} {929}
  (\bibinfo {year} {2005})}\BibitemShut {NoStop}%
\bibitem [{\citenamefont {Schulz}\ \emph
  {et~al.}(2006{\natexlab{a}})\citenamefont {Schulz}, \citenamefont
  {Schumacher},\ and\ \citenamefont {Czycholl}}]{ScSc2006}%
  \BibitemOpen
  \bibfield  {author} {\bibinfo {author} {\bibfnamefont {S.}~\bibnamefont
  {Schulz}}, \bibinfo {author} {\bibfnamefont {S.}~\bibnamefont {Schumacher}},
  \ and\ \bibinfo {author} {\bibfnamefont {G.}~\bibnamefont {Czycholl}},\
  }\href@noop {} {\bibfield  {journal} {\bibinfo  {journal} {Phys. Rev. B}\
  }\textbf {\bibinfo {volume} {73}},\ \bibinfo {pages} {245327} (\bibinfo
  {year} {2006}{\natexlab{a}})}\BibitemShut {NoStop}%
\bibitem [{\citenamefont {Pelucchi}\ \emph {et~al.}(2004)\citenamefont
  {Pelucchi}, \citenamefont {Watanabe}, \citenamefont {Leifer}, \citenamefont
  {Dwir},\ and\ \citenamefont {Kapon}}]{PeWa2004}%
  \BibitemOpen
  \bibfield  {author} {\bibinfo {author} {\bibfnamefont {E.}~\bibnamefont
  {Pelucchi}}, \bibinfo {author} {\bibfnamefont {S.}~\bibnamefont {Watanabe}},
  \bibinfo {author} {\bibfnamefont {K.}~\bibnamefont {Leifer}}, \bibinfo
  {author} {\bibfnamefont {B.}~\bibnamefont {Dwir}}, \ and\ \bibinfo {author}
  {\bibfnamefont {E.}~\bibnamefont {Kapon}},\ }\href@noop {} {\bibfield
  {journal} {\bibinfo  {journal} {Physica E}\ }\textbf {\bibinfo {volume}
  {23}},\ \bibinfo {pages} {476} (\bibinfo {year} {2004})}\BibitemShut
  {NoStop}%
\bibitem [{\citenamefont {Dimastrodonato}\ \emph {et~al.}(2012)\citenamefont
  {Dimastrodonato}, \citenamefont {Pelucchi},\ and\ \citenamefont
  {Vvedensky}}]{DiPe2012}%
  \BibitemOpen
  \bibfield  {author} {\bibinfo {author} {\bibfnamefont {V.}~\bibnamefont
  {Dimastrodonato}}, \bibinfo {author} {\bibfnamefont {E.}~\bibnamefont
  {Pelucchi}}, \ and\ \bibinfo {author} {\bibfnamefont {D.~D.}\ \bibnamefont
  {Vvedensky}},\ }\href@noop {} {\bibfield  {journal} {\bibinfo  {journal}
  {Phys. Rev. Lett.}\ }\textbf {\bibinfo {volume} {108}},\ \bibinfo {pages}
  {256102} (\bibinfo {year} {2012})}\BibitemShut {NoStop}%
\bibitem [{\citenamefont {Juska}\ \emph {et~al.}(2014)\citenamefont {Juska},
  \citenamefont {Dimastrodonato}, \citenamefont {Mereni}, \citenamefont
  {Chung}, \citenamefont {Gocalinska}, \citenamefont {Pelucchi}, \citenamefont
  {Hattem}, \citenamefont {Ediger},\ and\ \citenamefont {Corfdir}}]{JuDi2014}%
  \BibitemOpen
  \bibfield  {author} {\bibinfo {author} {\bibfnamefont {G.}~\bibnamefont
  {Juska}}, \bibinfo {author} {\bibfnamefont {V.}~\bibnamefont
  {Dimastrodonato}}, \bibinfo {author} {\bibfnamefont {L.~O.}\ \bibnamefont
  {Mereni}}, \bibinfo {author} {\bibfnamefont {T.~H.}\ \bibnamefont {Chung}},
  \bibinfo {author} {\bibfnamefont {A.}~\bibnamefont {Gocalinska}}, \bibinfo
  {author} {\bibfnamefont {E.}~\bibnamefont {Pelucchi}}, \bibinfo {author}
  {\bibfnamefont {B.~V.}\ \bibnamefont {Hattem}}, \bibinfo {author}
  {\bibfnamefont {M.}~\bibnamefont {Ediger}}, \ and\ \bibinfo {author}
  {\bibfnamefont {P.}~\bibnamefont {Corfdir}},\ }\href@noop {} {\bibfield
  {journal} {\bibinfo  {journal} {Phys. Rev. B}\ }\textbf {\bibinfo {volume}
  {89}},\ \bibinfo {pages} {205430} (\bibinfo {year} {2014})}\BibitemShut
  {NoStop}%
\bibitem [{\citenamefont {Joshi}(1997)}]{joshi1997elements}%
  \BibitemOpen
  \bibfield  {author} {\bibinfo {author} {\bibfnamefont {A.}~\bibnamefont
  {Joshi}},\ }\href {https://books.google.ie/books?id=f4f2wNPv5hYC} {\emph
  {\bibinfo {title} {Elements of Group Theory for Physicists}}}\ (\bibinfo
  {publisher} {New Age International},\ \bibinfo {year} {1997})\BibitemShut
  {NoStop}%
\bibitem [{\citenamefont {Bir}\ and\ \citenamefont {Pikus}(1962)}]{Bir-Pikus}%
  \BibitemOpen
  \bibfield  {author} {\bibinfo {author} {\bibfnamefont {G.~L.}\ \bibnamefont
  {Bir}}\ and\ \bibinfo {author} {\bibfnamefont {G.~E.}\ \bibnamefont
  {Pikus}},\ }\href@noop {} {\bibfield  {journal} {\bibinfo  {journal} {Sov.
  Phys. Solid State}\ }\textbf {\bibinfo {volume} {3}},\ \bibinfo {pages}
  {2221. [Fiz. Trevd. Tela (Leningrad), 1961, {\bf 3}, 3050]} (\bibinfo {year}
  {1962})},\ \bibinfo {note} {[Fiz. Trevd. Tela (Leningrad), 1961, {\bf 3},
  3050.]}\BibitemShut {NoStop}%
\bibitem [{\citenamefont {Trebin}\ \emph {et~al.}(1979)\citenamefont {Trebin},
  \citenamefont {R\"ossler},\ and\ \citenamefont {Ranvaud}}]{Trebin}%
  \BibitemOpen
  \bibfield  {author} {\bibinfo {author} {\bibfnamefont {H.~R.}\ \bibnamefont
  {Trebin}}, \bibinfo {author} {\bibfnamefont {U.}~\bibnamefont {R\"ossler}}, \
  and\ \bibinfo {author} {\bibfnamefont {R.}~\bibnamefont {Ranvaud}},\ }\href
  {\doibase 10.1103/PhysRevB.20.686} {\bibfield  {journal} {\bibinfo  {journal}
  {Phys. Rev. B}\ }\textbf {\bibinfo {volume} {20}},\ \bibinfo {pages} {686}
  (\bibinfo {year} {1979})}\BibitemShut {NoStop}%
\bibitem [{\citenamefont {Silver}\ \emph {et~al.}(1992)\citenamefont {Silver},
  \citenamefont {Batty}, \citenamefont {Ghiti},\ and\ \citenamefont
  {O'Reilly}}]{EOR}%
  \BibitemOpen
  \bibfield  {author} {\bibinfo {author} {\bibfnamefont {M.}~\bibnamefont
  {Silver}}, \bibinfo {author} {\bibfnamefont {W.}~\bibnamefont {Batty}},
  \bibinfo {author} {\bibfnamefont {A.}~\bibnamefont {Ghiti}}, \ and\ \bibinfo
  {author} {\bibfnamefont {E.~P.}\ \bibnamefont {O'Reilly}},\ }\href {\doibase
  10.1103/PhysRevB.46.6781} {\bibfield  {journal} {\bibinfo  {journal} {Phys.
  Rev. B}\ }\textbf {\bibinfo {volume} {46}},\ \bibinfo {pages} {6781}
  (\bibinfo {year} {1992})}\BibitemShut {NoStop}%
\bibitem [{\citenamefont {Stier}\ \emph {et~al.}(1999)\citenamefont {Stier},
  \citenamefont {Grundmann},\ and\ \citenamefont {Bimberg}}]{Bimberg}%
  \BibitemOpen
  \bibfield  {author} {\bibinfo {author} {\bibfnamefont {O.}~\bibnamefont
  {Stier}}, \bibinfo {author} {\bibfnamefont {M.}~\bibnamefont {Grundmann}}, \
  and\ \bibinfo {author} {\bibfnamefont {D.}~\bibnamefont {Bimberg}},\ }\href
  {\doibase 10.1103/PhysRevB.59.5688} {\bibfield  {journal} {\bibinfo
  {journal} {Phys. Rev. B}\ }\textbf {\bibinfo {volume} {59}},\ \bibinfo
  {pages} {5688} (\bibinfo {year} {1999})}\BibitemShut {NoStop}%
\bibitem [{\citenamefont {Wang}\ \emph {et~al.}(2000)\citenamefont {Wang},
  \citenamefont {Williamson}, \citenamefont {Zunger}, \citenamefont {Jiang},\
  and\ \citenamefont {Singh}}]{Singh2000}%
  \BibitemOpen
  \bibfield  {author} {\bibinfo {author} {\bibfnamefont {L.~W.}\ \bibnamefont
  {Wang}}, \bibinfo {author} {\bibfnamefont {A.~J.}\ \bibnamefont
  {Williamson}}, \bibinfo {author} {\bibfnamefont {A.}~\bibnamefont {Zunger}},
  \bibinfo {author} {\bibfnamefont {H.}~\bibnamefont {Jiang}}, \ and\ \bibinfo
  {author} {\bibfnamefont {J.}~\bibnamefont {Singh}},\ }\href@noop {}
  {\bibfield  {journal} {\bibinfo  {journal} {Applied Physics Letters}\
  }\textbf {\bibinfo {volume} {76}} (\bibinfo {year} {2000})}\BibitemShut
  {NoStop}%
\bibitem [{\citenamefont {Marquardt}\ \emph {et~al.}(2008)\citenamefont
  {Marquardt}, \citenamefont {Mourad}, \citenamefont {Schulz}, \citenamefont
  {Hickel}, \citenamefont {Czycholl},\ and\ \citenamefont
  {Neugebauer}}]{Marquardt2008}%
  \BibitemOpen
  \bibfield  {author} {\bibinfo {author} {\bibfnamefont {O.}~\bibnamefont
  {Marquardt}}, \bibinfo {author} {\bibfnamefont {D.}~\bibnamefont {Mourad}},
  \bibinfo {author} {\bibfnamefont {S.}~\bibnamefont {Schulz}}, \bibinfo
  {author} {\bibfnamefont {T.}~\bibnamefont {Hickel}}, \bibinfo {author}
  {\bibfnamefont {G.}~\bibnamefont {Czycholl}}, \ and\ \bibinfo {author}
  {\bibfnamefont {J.}~\bibnamefont {Neugebauer}},\ }\href {\doibase
  10.1103/PhysRevB.78.235302} {\bibfield  {journal} {\bibinfo  {journal} {Phys.
  Rev. B}\ }\textbf {\bibinfo {volume} {78}},\ \bibinfo {pages} {235302}
  (\bibinfo {year} {2008})}\BibitemShut {NoStop}%
\bibitem [{\citenamefont {Juska}\ \emph {et~al.}(2011)\citenamefont {Juska},
  \citenamefont {Dimastrodonato}, \citenamefont {Mereni}, \citenamefont
  {Gocalinska},\ and\ \citenamefont {Pelucchi}}]{Juska}%
  \BibitemOpen
  \bibfield  {author} {\bibinfo {author} {\bibfnamefont {G.}~\bibnamefont
  {Juska}}, \bibinfo {author} {\bibfnamefont {V.}~\bibnamefont
  {Dimastrodonato}}, \bibinfo {author} {\bibfnamefont {L.}~\bibnamefont
  {Mereni}}, \bibinfo {author} {\bibfnamefont {A.}~\bibnamefont {Gocalinska}},
  \ and\ \bibinfo {author} {\bibfnamefont {E.}~\bibnamefont {Pelucchi}},\
  }\href {\doibase 10.1186/1556-276X-6-567} {\bibfield  {journal} {\bibinfo
  {journal} {Nanoscale Research Letters}\ }\textbf {\bibinfo {volume} {6}},\
  \bibinfo {pages} {567} (\bibinfo {year} {2011})}\BibitemShut {NoStop}%
\bibitem [{\citenamefont {Schulz}\ \emph
  {et~al.}(2015{\natexlab{a}})\citenamefont {Schulz}, \citenamefont {Tanner},
  \citenamefont {O'Reilly}, \citenamefont {Caro}, \citenamefont {Martin},
  \citenamefont {Bagot}, \citenamefont {Moody}, \citenamefont {Tang},
  \citenamefont {Griffiths}, \citenamefont {Oehler}, \citenamefont {Kappers},
  \citenamefont {Oliver}, \citenamefont {Humphreys}, \citenamefont
  {Sutherland}, \citenamefont {Davies},\ and\ \citenamefont
  {Dawson}}]{SSC2015a}%
  \BibitemOpen
  \bibfield  {author} {\bibinfo {author} {\bibfnamefont {S.}~\bibnamefont
  {Schulz}}, \bibinfo {author} {\bibfnamefont {D.~P.}\ \bibnamefont {Tanner}},
  \bibinfo {author} {\bibfnamefont {E.~P.}\ \bibnamefont {O'Reilly}}, \bibinfo
  {author} {\bibfnamefont {M.~A.}\ \bibnamefont {Caro}}, \bibinfo {author}
  {\bibfnamefont {T.~L.}\ \bibnamefont {Martin}}, \bibinfo {author}
  {\bibfnamefont {P.~A.~J.}\ \bibnamefont {Bagot}}, \bibinfo {author}
  {\bibfnamefont {M.~P.}\ \bibnamefont {Moody}}, \bibinfo {author}
  {\bibfnamefont {F.}~\bibnamefont {Tang}}, \bibinfo {author} {\bibfnamefont
  {J.~T.}\ \bibnamefont {Griffiths}}, \bibinfo {author} {\bibfnamefont
  {F.}~\bibnamefont {Oehler}}, \bibinfo {author} {\bibfnamefont {M.~J.}\
  \bibnamefont {Kappers}}, \bibinfo {author} {\bibfnamefont {R.~A.}\
  \bibnamefont {Oliver}}, \bibinfo {author} {\bibfnamefont {C.~J.}\
  \bibnamefont {Humphreys}}, \bibinfo {author} {\bibfnamefont {D.}~\bibnamefont
  {Sutherland}}, \bibinfo {author} {\bibfnamefont {M.~J.}\ \bibnamefont
  {Davies}}, \ and\ \bibinfo {author} {\bibfnamefont {P.}~\bibnamefont
  {Dawson}},\ }\href {\doibase 10.1103/PhysRevB.92.235419} {\bibfield
  {journal} {\bibinfo  {journal} {Phys. Rev. B}\ }\textbf {\bibinfo {volume}
  {92}},\ \bibinfo {pages} {235419} (\bibinfo {year}
  {2015}{\natexlab{a}})}\BibitemShut {NoStop}%
\bibitem [{\citenamefont {Schulz}\ \emph {et~al.}(2014)\citenamefont {Schulz},
  \citenamefont {Caro},\ and\ \citenamefont {O'Reilly}}]{SSC2014}%
  \BibitemOpen
  \bibfield  {author} {\bibinfo {author} {\bibfnamefont {S.}~\bibnamefont
  {Schulz}}, \bibinfo {author} {\bibfnamefont {M.~A.}\ \bibnamefont {Caro}}, \
  and\ \bibinfo {author} {\bibfnamefont {E.~P.}\ \bibnamefont {O'Reilly}},\
  }\href {\doibase http://dx.doi.org/10.1063/1.4872317} {\bibfield  {journal}
  {\bibinfo  {journal} {Applied Physics Letters}\ }\textbf {\bibinfo {volume}
  {104}},\ \bibinfo {eid} {172102} (\bibinfo {year} {2014})}\BibitemShut
  {NoStop}%
\bibitem [{\citenamefont {Schulz}\ \emph
  {et~al.}(2015{\natexlab{b}})\citenamefont {Schulz}, \citenamefont {Caro},
  \citenamefont {Coughlan},\ and\ \citenamefont {O'Reilly}}]{SSC2015b}%
  \BibitemOpen
  \bibfield  {author} {\bibinfo {author} {\bibfnamefont {S.}~\bibnamefont
  {Schulz}}, \bibinfo {author} {\bibfnamefont {M.~A.}\ \bibnamefont {Caro}},
  \bibinfo {author} {\bibfnamefont {C.}~\bibnamefont {Coughlan}}, \ and\
  \bibinfo {author} {\bibfnamefont {E.~P.}\ \bibnamefont {O'Reilly}},\ }\href
  {\doibase 10.1103/PhysRevB.91.035439} {\bibfield  {journal} {\bibinfo
  {journal} {Phys. Rev. B}\ }\textbf {\bibinfo {volume} {91}},\ \bibinfo
  {pages} {035439} (\bibinfo {year} {2015}{\natexlab{b}})}\BibitemShut
  {NoStop}%
\bibitem [{\citenamefont {Schliwa}\ \emph
  {et~al.}(2009{\natexlab{b}})\citenamefont {Schliwa}, \citenamefont
  {Winkelnkemper},\ and\ \citenamefont {Bimberg}}]{Schliwa2009}%
  \BibitemOpen
  \bibfield  {author} {\bibinfo {author} {\bibfnamefont {A.}~\bibnamefont
  {Schliwa}}, \bibinfo {author} {\bibfnamefont {M.}~\bibnamefont
  {Winkelnkemper}}, \ and\ \bibinfo {author} {\bibfnamefont {D.}~\bibnamefont
  {Bimberg}},\ }\href {\doibase 10.1103/PhysRevB.79.075443} {\bibfield
  {journal} {\bibinfo  {journal} {Phys. Rev. B}\ }\textbf {\bibinfo {volume}
  {79}},\ \bibinfo {pages} {075443} (\bibinfo {year}
  {2009}{\natexlab{b}})}\BibitemShut {NoStop}%
\bibitem [{\citenamefont {Franceschetti}\ and\ \citenamefont
  {Zunger}(1997)}]{Franceschetti}%
  \BibitemOpen
  \bibfield  {author} {\bibinfo {author} {\bibfnamefont {A.}~\bibnamefont
  {Franceschetti}}\ and\ \bibinfo {author} {\bibfnamefont {A.}~\bibnamefont
  {Zunger}},\ }\href {\doibase 10.1103/PhysRevLett.78.915} {\bibfield
  {journal} {\bibinfo  {journal} {Phys. Rev. Lett.}\ }\textbf {\bibinfo
  {volume} {78}},\ \bibinfo {pages} {915} (\bibinfo {year} {1997})}\BibitemShut
  {NoStop}%
\bibitem [{\citenamefont {Franceschetti}\ \emph {et~al.}(1998)\citenamefont
  {Franceschetti}, \citenamefont {Wang}, \citenamefont {Fu},\ and\
  \citenamefont {Zunger}}]{Franceschetti98}%
  \BibitemOpen
  \bibfield  {author} {\bibinfo {author} {\bibfnamefont {A.}~\bibnamefont
  {Franceschetti}}, \bibinfo {author} {\bibfnamefont {L.~W.}\ \bibnamefont
  {Wang}}, \bibinfo {author} {\bibfnamefont {H.}~\bibnamefont {Fu}}, \ and\
  \bibinfo {author} {\bibfnamefont {A.}~\bibnamefont {Zunger}},\ }\href
  {\doibase 10.1103/PhysRevB.58.R13367} {\bibfield  {journal} {\bibinfo
  {journal} {Phys. Rev. B}\ }\textbf {\bibinfo {volume} {58}},\ \bibinfo
  {pages} {R13367} (\bibinfo {year} {1998})}\BibitemShut {NoStop}%
\bibitem [{\citenamefont {Rohlfing}\ and\ \citenamefont
  {Louie}(2000)}]{Rohlfing}%
  \BibitemOpen
  \bibfield  {author} {\bibinfo {author} {\bibfnamefont {M.}~\bibnamefont
  {Rohlfing}}\ and\ \bibinfo {author} {\bibfnamefont {S.~G.}\ \bibnamefont
  {Louie}},\ }\href {\doibase 10.1103/PhysRevB.62.4927} {\bibfield  {journal}
  {\bibinfo  {journal} {Phys. Rev. B}\ }\textbf {\bibinfo {volume} {62}},\
  \bibinfo {pages} {4927} (\bibinfo {year} {2000})}\BibitemShut {NoStop}%
\bibitem [{\citenamefont {Wimmer}\ \emph {et~al.}(2006)\citenamefont {Wimmer},
  \citenamefont {Nair},\ and\ \citenamefont {Shumway}}]{Wimmer}%
  \BibitemOpen
  \bibfield  {author} {\bibinfo {author} {\bibfnamefont {M.}~\bibnamefont
  {Wimmer}}, \bibinfo {author} {\bibfnamefont {S.~V.}\ \bibnamefont {Nair}}, \
  and\ \bibinfo {author} {\bibfnamefont {J.}~\bibnamefont {Shumway}},\ }\href
  {\doibase 10.1103/PhysRevB.73.165305} {\bibfield  {journal} {\bibinfo
  {journal} {Phys. Rev. B}\ }\textbf {\bibinfo {volume} {73}},\ \bibinfo
  {pages} {165305} (\bibinfo {year} {2006})}\BibitemShut {NoStop}%
\bibitem [{\citenamefont {Winkelnkemper}\ \emph {et~al.}(2006)\citenamefont
  {Winkelnkemper}, \citenamefont {Schliwa},\ and\ \citenamefont
  {Bimberg}}]{Bimberg06}%
  \BibitemOpen
  \bibfield  {author} {\bibinfo {author} {\bibfnamefont {M.}~\bibnamefont
  {Winkelnkemper}}, \bibinfo {author} {\bibfnamefont {A.}~\bibnamefont
  {Schliwa}}, \ and\ \bibinfo {author} {\bibfnamefont {D.}~\bibnamefont
  {Bimberg}},\ }\href {\doibase 10.1103/PhysRevB.74.155322} {\bibfield
  {journal} {\bibinfo  {journal} {Phys. Rev. B}\ }\textbf {\bibinfo {volume}
  {74}},\ \bibinfo {pages} {155322} (\bibinfo {year} {2006})}\BibitemShut
  {NoStop}%
\bibitem [{\citenamefont {Kindel}\ \emph {et~al.}(2010)\citenamefont {Kindel},
  \citenamefont {Kako}, \citenamefont {Kawano}, \citenamefont {Oishi},
  \citenamefont {Arakawa}, \citenamefont {H\"onig}, \citenamefont
  {Winkelnkemper}, \citenamefont {Schliwa}, \citenamefont {Hoffmann},\ and\
  \citenamefont {Bimberg}}]{Kindel}%
  \BibitemOpen
  \bibfield  {author} {\bibinfo {author} {\bibfnamefont {C.}~\bibnamefont
  {Kindel}}, \bibinfo {author} {\bibfnamefont {S.}~\bibnamefont {Kako}},
  \bibinfo {author} {\bibfnamefont {T.}~\bibnamefont {Kawano}}, \bibinfo
  {author} {\bibfnamefont {H.}~\bibnamefont {Oishi}}, \bibinfo {author}
  {\bibfnamefont {Y.}~\bibnamefont {Arakawa}}, \bibinfo {author} {\bibfnamefont
  {G.}~\bibnamefont {H\"onig}}, \bibinfo {author} {\bibfnamefont
  {M.}~\bibnamefont {Winkelnkemper}}, \bibinfo {author} {\bibfnamefont
  {A.}~\bibnamefont {Schliwa}}, \bibinfo {author} {\bibfnamefont
  {A.}~\bibnamefont {Hoffmann}}, \ and\ \bibinfo {author} {\bibfnamefont
  {D.}~\bibnamefont {Bimberg}},\ }\href {\doibase 10.1103/PhysRevB.81.241309}
  {\bibfield  {journal} {\bibinfo  {journal} {Phys. Rev. B}\ }\textbf {\bibinfo
  {volume} {81}},\ \bibinfo {pages} {241309} (\bibinfo {year}
  {2010})}\BibitemShut {NoStop}%
\bibitem [{\citenamefont {Schulz}\ \emph
  {et~al.}(2006{\natexlab{b}})\citenamefont {Schulz}, \citenamefont
  {Schumacher},\ and\ \citenamefont {Czycholl}}]{Schulz06}%
  \BibitemOpen
  \bibfield  {author} {\bibinfo {author} {\bibfnamefont {S.}~\bibnamefont
  {Schulz}}, \bibinfo {author} {\bibfnamefont {S.}~\bibnamefont {Schumacher}},
  \ and\ \bibinfo {author} {\bibfnamefont {G.}~\bibnamefont {Czycholl}},\
  }\href {\doibase 10.1103/PhysRevB.73.245327} {\bibfield  {journal} {\bibinfo
  {journal} {Phys. Rev. B}\ }\textbf {\bibinfo {volume} {73}},\ \bibinfo
  {pages} {245327} (\bibinfo {year} {2006}{\natexlab{b}})}\BibitemShut
  {NoStop}%
\bibitem [{\citenamefont {Benchamekh}\ \emph {et~al.}(2015)\citenamefont
  {Benchamekh}, \citenamefont {Raouafi}, \citenamefont {Even}, \citenamefont
  {Ben Cheikh~Larbi}, \citenamefont {Voisin},\ and\ \citenamefont
  {Jancu}}]{Benchamekh2015}%
  \BibitemOpen
  \bibfield  {author} {\bibinfo {author} {\bibfnamefont {R.}~\bibnamefont
  {Benchamekh}}, \bibinfo {author} {\bibfnamefont {F.}~\bibnamefont {Raouafi}},
  \bibinfo {author} {\bibfnamefont {J.}~\bibnamefont {Even}}, \bibinfo {author}
  {\bibfnamefont {F.}~\bibnamefont {Ben Cheikh~Larbi}}, \bibinfo {author}
  {\bibfnamefont {P.}~\bibnamefont {Voisin}}, \ and\ \bibinfo {author}
  {\bibfnamefont {J.-M.}\ \bibnamefont {Jancu}},\ }\href {\doibase
  10.1103/PhysRevB.91.045118} {\bibfield  {journal} {\bibinfo  {journal} {Phys.
  Rev. B}\ }\textbf {\bibinfo {volume} {91}},\ \bibinfo {pages} {045118}
  (\bibinfo {year} {2015})}\BibitemShut {NoStop}%
\bibitem [{\citenamefont {Korkusinski}\ \emph {et~al.}(2011)\citenamefont
  {Korkusinski}, \citenamefont {Voznyy},\ and\ \citenamefont
  {Hawrylak}}]{Korkusinski}%
  \BibitemOpen
  \bibfield  {author} {\bibinfo {author} {\bibfnamefont {M.}~\bibnamefont
  {Korkusinski}}, \bibinfo {author} {\bibfnamefont {O.}~\bibnamefont {Voznyy}},
  \ and\ \bibinfo {author} {\bibfnamefont {P.}~\bibnamefont {Hawrylak}},\
  }\href {\doibase 10.1103/PhysRevB.84.155327} {\bibfield  {journal} {\bibinfo
  {journal} {Phys. Rev. B}\ }\textbf {\bibinfo {volume} {84}},\ \bibinfo
  {pages} {155327} (\bibinfo {year} {2011})}\BibitemShut {NoStop}%
\bibitem [{\citenamefont {Richard}\ \emph {et~al.}(2004)\citenamefont
  {Richard}, \citenamefont {Aniel},\ and\ \citenamefont {Fishman}}]{Fishman}%
  \BibitemOpen
  \bibfield  {author} {\bibinfo {author} {\bibfnamefont {S.}~\bibnamefont
  {Richard}}, \bibinfo {author} {\bibfnamefont {F.}~\bibnamefont {Aniel}}, \
  and\ \bibinfo {author} {\bibfnamefont {G.}~\bibnamefont {Fishman}},\ }\href
  {\doibase 10.1103/PhysRevB.70.235204} {\bibfield  {journal} {\bibinfo
  {journal} {Phys. Rev. B}\ }\textbf {\bibinfo {volume} {70}},\ \bibinfo
  {pages} {235204} (\bibinfo {year} {2004})}\BibitemShut {NoStop}%
\bibitem [{\citenamefont {Lew Yan~Voon}\ and\ \citenamefont
  {Ram-Mohan}(1993)}]{Lew_Yan_Voon}%
  \BibitemOpen
  \bibfield  {author} {\bibinfo {author} {\bibfnamefont {L.~C.}\ \bibnamefont
  {Lew Yan~Voon}}\ and\ \bibinfo {author} {\bibfnamefont {L.~R.}\ \bibnamefont
  {Ram-Mohan}},\ }\href {\doibase 10.1103/PhysRevB.47.15500} {\bibfield
  {journal} {\bibinfo  {journal} {Phys. Rev. B}\ }\textbf {\bibinfo {volume}
  {47}},\ \bibinfo {pages} {15500} (\bibinfo {year} {1993})}\BibitemShut
  {NoStop}%
\end{thebibliography}%
\bibliographystyle{apsrev4-1}

\end{document}